\documentclass[12pt]{article}
\usepackage{amsmath,amssymb}
\usepackage[english]{babel}
\usepackage[cp866]{inputenc}
\usepackage{hyperref}

\numberwithin{equation}{section}

\newcommand{\bq}{\begin{eqnarray}}
\newcommand{\eq}{\end{eqnarray}}
\newcommand{\bbq}{\begin{equation*}}
\newcommand{\eeq}{\end{equation*}}

\newcommand{\ra}{\rightarrow}

\newcommand{\lm}{\Lambda_2}
\newcommand{\la}{\Lambda_Q}

\newcommand{\ov}{\overline}

\newcommand{\lt}{\tilde\Lambda}
\newcommand{\wh}{\widehat}
\newcommand{\wt}{\widetilde}
\newcommand{\wmu}{\widetilde{\mu}_{\rm x}}

\newcommand{\qq}{{\ov Q}Q}
\newcommand{\Qo}{({\ov Q}Q)_1}
\newcommand{\Qt}{({\ov Q}Q)_2}

\newcommand{\nd}{{\ov N}_c}

\newcommand{\mx}{\mu_{\rm x}}

\newcommand{\bb}{2N_c-N_F}
\newcommand{\bt}{{\rm b}_2}
\newcommand{\w}{{\cal W}}
\newcommand{\no}{{\rm n}_1}
\newcommand{\nt}{{\rm n}_2}
\newcommand{\tm}{\widetilde m}

\begin{document}

\begin{center}{\bf \large Mass spectra in softly broken $\mathbf{{\cal N}=2}$ SQCD} \end{center}
\vspace{1cm}
\begin{center}\bf {Victor L. Chernyak $^{a,\, b}$ } \end{center}
\begin{center}(e-mail: v.l.chernyak@inp.nsk.su) \end{center}
\begin{center} a)\,\, Novosibirsk State University, 630090 Novosibirsk, Russian Federation \end{center}
\begin{center} b)\,\, Budker Institute of Nuclear Physics SB RAS, 630090 Novosibirsk, Russian Federation
\end{center}
\vspace{1cm}
\begin{center}{\bf Abstract} \end{center}
\vspace{1cm}

Considered are ${\cal N}=2\,\, SU(N_c)$ or $U(N_c)$ SQCD with $N_F<2N_c-1$ quark flavors with the quark mass term $m{\rm Tr}\,({\overline Q} Q)$ in the superpotential.  ${\cal N}=2$ supersymmetry is softly broken down to ${\cal N}=1$ by the mass term $\mu_{\rm x}{\rm Tr}\,(X^2)$ of colored adjoint scalar partners of gluons, $\mu_{\rm x}\ll\Lambda_2$ (\,$\Lambda_2$ is the scale factor of the $SU(N_c)$ gauge coupling).

There is a large number of different types of vacua in these theories with both unbroken and spontaneously broken flavor symmetry, $U(N_F)\rightarrow U({\rm n}_1)\times U({\rm n}_2)$. We consider in this paper the large subset of these vacua with the unbroken non-trivial $Z_{2N_c-N_F\geq 2}$ discrete symmetry, at different hierarchies between the Lagrangian parameters $m\gtrless\Lambda_2,\,\, \mu_{\rm x}\gtrless m$. The forms of low energy Lagrangians, charges of light particles and mass spectra are described in the main text for all these vacua.

The calculations of power corrections to the leading terms of the low energy quark and dyon condensates are presented in two important Appendices. These calculations confirm independently in a non-trivial way a self-consistency of the whole approach.

Our results differ essentially from corresponding results in recent related papers arXiv:1304.0822, arXiv:1403.6086, and arXiv:1704.06201 of M.Shifman and A.Yung (and in a number of their previous numerous papers on this subject), and we explain in the text the reasons for these differences. (See also the extended critique of a number of results of these authors in section 8 of arXiv:1308.5863).
\vspace*{1cm}

\begin{center} To Arkady Vainshtein on his 75-th Anniversary \end{center}

\newpage

\tableofcontents
\numberwithin{equation}{section}

\section{Introduction}

\hspace*{4mm} The masses and charges of BPS particles in ${\cal N}=2\,\, SU(2)$ SYM and $SU(2)$ SQCD with $N_F=1...4$ quark flavors at small $\mx\neq 0$ were found in seminal papers of N. Seiberg and E. Witten \cite{SW1,SW2}. They presented in particular the corresponding spectral curves from which the masses and charges of these BPS particles can be calculated at $\mx\ra 0$. The forms of these curves have been generalized then, in particular, to ${\cal N}=2\,\, SU(N_c)$ SYM \cite{DS} and ${\cal N}=2\,\, SU(N_c)$ SQCD with $1\leq N_F\leq 2N_c$ quark flavors \cite{KL,AF,HO,Shap,APS}.

In what follows we deal mainly with ${\cal N}=2\,\, SU(N_c)$ SQCD with $N_F<2N_c-1$ flavors of equal mass quarks $Q^i_a,\,{\ov Q}_i^{\,a},\,i=1...N_F,\,\,a=1...N_c$. The Lagrangian of this UV free theory, broken down to ${\cal N}=1$ by the mass term $\mx{\rm Tr}\,(X^{\rm adj}_{SU(N_c)})^2$ of adjoint scalars, can be written at the sufficiently high scale $\mu\gg\max\{\lm, m\}$ as (the gluon exponents in Kahler terms are implied in \eqref{(1.1)} and everywhere below)
\bbq
K=\frac{2}{g^2(\mu)}{\rm Tr}\,\Bigl [(X^{\rm adj}_{SU(N_c)})^\dagger X^{\rm adj}_{SU(N_c)}\Bigr ]_{N_c}+{\rm Tr}\,(Q^\dagger Q+{\ov Q}^{\,\dagger} {\ov Q})_{N_c}\,, \quad X^{\rm adj}_{SU(N_c)}=T^A X^A\,,\quad A=1,\,...,N_c^2-1\,,
\eeq
\bq
{\cal W}_{\rm matter}=\mx{\rm Tr}\,(X^{\rm adj}_{SU(N_c)})_{N_c}^2 +{\rm Tr}\,\Bigl (m\,{\ov Q} Q-{\ov Q}\sqrt{2} X^{\rm adj}_{SU(N_c)} Q \Bigr )_{N_c}\,,\quad {\rm Tr}\, (T^{A_1} T^{A_2})=\frac{1}{2}\,\delta^{A_1 A_2}\,,\label{(1.1)}
\eq
with the scale factor $\lm$ of the gauge coupling $g(\mu)$, and ${\rm Tr}$ in \eqref{(1.1)} is over all colors and flavors. The "softly broken"\, ${\cal N}=2$ theory means $\mx\ll\lm$.

The spectral curve corresponding to \eqref{(1.1)} can be written at $N_F < 2N_c-1$ e.g. in the form \cite{APS}
\bq
y^2=\prod_{i=1}^{N_c}(z+\phi_i)^2-4\lm^{\bb}(z+m)^{N_F}\,,\quad \sum_{i=1}^{N_c}\phi_i=0{\rm\,\,\, in\,\,\, SU(N_c)}\,, \label{(1.2)}
\eq
where $\{\phi_i\}$ is a set of gauge invariant co-ordinates on the moduli space. Note that, as it is, this curve can be used for the calculation of BPS particle masses at $\mx\ra 0$ only.

In principle, it is possible to find  at $\mx\ra 0$ from the curve \eqref{(1.2)} the charges and masses of all massive and massless BPS particles in different multiple vacua of this theory. But in practice it is very difficult to do this for general values of $N_c$ and $N_F$, especially as for charges. The important property of the $SU(N_c)$ curve \eqref{(1.2)} which will be used in the text below is that there are maximum $N_c-1$ double roots. And the vacua we will deal with below at $\mx\neq 0$ are just these vacua. In other words, {\it in all $SU(N_c)$ vacua we will deal with, the curve} \eqref{(1.2)} {\it has $N_c-1$ double roots and two single roots}.

The first detailed attempt has been made in \cite{APS} (only the case $m=0$ was considered) to classify the vacua of \eqref{(1.1)} for general values of  $N_c$ and $N_F$, to find charges of light BPS particles (massless at $\mx\ra 0$) in these vacua  and forms of low energy Lagrangians. It was found in particular that in all vacua the $SU(N_c)$ gauge symmetry is broken spontaneously at the scale $\mu\sim\lm$ by higgsed adjoint scalars, $\langle X^{\rm adj}_{SU(N_c)}\rangle\sim\lm$.

That this should happen can be understood qualitatively as follows \cite{ch6}. The perturbative NSVZ $\beta$-function of the (effectively) massless unbroken ${\cal N}=2$ SQCD is exactly one loop \cite{NSVZ1,NSVZ2}. The theory \eqref{(1.1)} is UV free at $(2N_c-N_F)>0$. At small $\mx$ and $m\ll\lm$, and if the whole matrix $\langle X^{adj}_{SU(N_c)}\rangle\ll\lm$ (i.e. in the effectively massless at the scale $\mu\sim\lm$ unbroken $\,{\cal N}=2$ SQCD), the coupling $g^2(\mu)$ is well defined at $\mu\gg\lm$, has a pole at $\mu=\lm$ and becomes negative at $\mu<\lm$. To avoid this unphysical behavior, {\it the field $X$ is necessarily higgsed breaking the $SU(N_c)$ group, with (at least some) components $\langle X^A\rangle\sim
\lm$}. (This becomes especially clear in the really unbroken ${\cal N}=2$ theory at $\mx=0$ and small $m\ll\lm$, when all chiral quark condensates, i.e. their corresponding mean values $\langle...\rangle$, are zero, so that all quarks are not higgsed in any case, and all particles are very light at $\langle X^{\rm adj}_{SU(N_c)}\rangle\ll\lm$).

Remind that exact values of $\langle X^A\rangle$ account for possible non-perturbative instanton contributions. And it is worth also to note that, in the unbroken ${\cal N}=2$ SQCD with small $m$ and $\mx=0$, if there will be no some components $\langle X^A\rangle\sim\lm$, i.e. all $\langle X^A\rangle
\ll\lm$, the non-perturbative instanton contributions will be also non operative at the scale $\lm$ without both the corresponding fermion masses $\sim\lm$ and the infrared cut off $\,\, \rho\lesssim\, 1/\lm$ supplied by some $\langle X^A\rangle\sim\lm$, so that the problem with $g^2(\mu<\lm)<0$ will survive. Moreover, for the same reasons, if there remains the non-Abelian subgroup unbroken at the scale $\lm$, it has to be IR free (or at least conformal).

It was found in \cite{APS} that (at $m=0$) and $\mx\ra 0$ there are two qualitatively different branches of vacua. On the baryonic  branch the lower energy gauge group at $\mu<\lm$ is $SU(N_F-N_c)\times U^{\bb}(1)$, while on non-baryonic branches these are $SU(\no)\times U^{N_c-\no}(1),\,\, \no\leq N_F/2$.

Besides, at $m\neq 0$, the global flavor symmetry $U(N_F)$ of \eqref{(1.1)} is unbroken or broken spontaneously as $U(N_F)\ra U(\no)\times U(\nt),\, 1\leq\no\leq [N_f/2]$ in various vacua \cite{CKM}.

As was also first pointed out in \cite{APS}, there is the residual $Z_{2N_c-N_F}$ discrete R-symmetry in theory \eqref{(1.1)}. {\it This symmetry will play a crucial role in what follows, as well as a knowledge of multiplicities of all different types of vacua}, see e.g. section 3 in \cite{ch4} and/or section 4 in \cite{ch5}. The charges of fields and parameters in the superpotential of \eqref{(1.1)} under $Z_{\bb}=\exp\{i\pi/(\bb)\}$ transformation are: $q_{\lambda}= q_{\rm\theta}=1,\,\, q_{X}=q_{\rm m}=2,\,\, q_Q=q_{\,\ov Q}=q_{\lm}=0,\,\, q_{\mx}=-2$. If this non-trivial at $\bb\geq 2$ discrete $Z_{\bb}$ symmetry is broken spontaneously in some vacua, there will appear then the factor $\bb$ in their multiplicity. Therefore, if multiplicities of different types of vacua are known, we can see explicitly whether this $Z_{\bb}$ symmetry is broken spontaneously or not in these vacua. As a result, it is seen the following at $0<N_F<2N_c-1,\, m\ll\lm$ (in the language of \cite{ch4,ch6}). -

I) In $SU(N_c)$ theories at $m\ll\lm$ ($\,\langle ({\ov Q} Q)_{1,2}\rangle_{N_c}$ and $\langle S\rangle_{N_c}$ are the quark and gluino condensates summed over all their colors\,). The discrete $Z_{\bb\geq 2}$ symmetry is broken spontaneously. \,{-} \\
1) In L (large) vacua with the unbroken flavor symmetry $U(N_F)$, the multiplicity $2N_c-N_F$ and with $\langle\qq\rangle_{N_c}\sim\mx\lm,\,\,\langle S\rangle_{N_c}\sim\mx\lm^2$.\\
2) In Lt (\,L-type\,) vacua with $U(N_F)\ra U(\no)\times U(\nt),\,\,1\leq\no\leq N_F/2$, the multiplicity $(2N_c-N_F)C^{\,\no}_{N_F},\,C^{\,\no}_{N_F}=
(N_F!/\no!\,\nt!)\,,$ and with $\langle\Qo\rangle_{N_c}\sim\langle\Qt\rangle_{N_c}\sim\mx\lm,\,\,\langle S\rangle_{N_c}\sim\mx\lm^2$.\\
3) In special vacua with $U(N_F)\ra U(\no=N_F-N_c)\times U(\nt=N_c)$, the multiplicity $(2N_c-N_F)C^{\,N_F-N_c}_{N_F}$, and with $\langle\Qo\rangle_{N_c}\sim\mx m,\,\, \langle\Qt\rangle_{N_c}=\mx\lm,\,\, \langle S\rangle_{N_c}\sim\mx m\lm$.

While $Z_{\bb\geq 2}$ is unbroken. - \\
4) In br2 (br=breaking) vacua with $U(N_F)\ra U(\no)\times U(\nt),\,\, \nt>N_c$, the multiplicity $(N_F-N_c-\no)C^{\,\no}_{N_F}$ and with the dominant condensate $\langle\Qt\rangle_{N_c}\sim\mx m\gg \langle\Qo\rangle_{N_c}\sim\mx m (m/\lm)^{(2N_c-N_F)/(\nt-N_c)}$, $\langle S\rangle_{N_c}\sim\mx m^2 (m/\lm)^{(2N_c-N_F)/(\nt-N_c)}$.\\
5) In S (small) vacua with unbroken $U(N_F)$, the multiplicity $N_F-N_c$ and with $\langle\qq\rangle_{N_c}\sim
\mx m,\,\, \langle S\rangle_{N_c}\sim\mx m^2 (m/\lm)^{(2N_c-N_F)/(N_F-N_c)}$ (see section 3 in \cite{ch4} or section 4 in \cite{ch5} for all details).

II). $Z_{\bb\geq 2}$ is unbroken in all multiple vacua of $SU(N_c)$ theories at $m\gg\lm$ (see Part II in the table of contents).

III) There are additional vs (very special) vacua with $\langle S\rangle_{N_c}=0,\, U(N_F)\ra U(\no=N_F-N_c)\times U(\nt=N_c)$ and the multiplicity
$C^{N_F-N_c}_{N_F}$ in $U(N_c)$ theories at all $m\gtrless\lm$ (such vacua are absent in $SU(N_c)$ theories, see Appendix in \cite{ch4}).  $Z_{\bb\geq 2}$ is unbroken in these vs-vacua at all $m\gtrless\lm$.\\

It was stated in \cite{APS} that {\it in all vacua} of $SU(N_c)$ \eqref{(1.1)}, baryonic and non-baryonic, all light particles in low energy Lagrangians with $m=0$ and small $\mx$, i.e. quarks, gluons and scalars in non-Abelian sectors and corresponding particles in Abelian ones, are pure magnetic.

Really, the final purpose of \cite{APS} was an attempt to derive the Seiberg duality for ${\cal N}=1$ SQCD \cite{S2,IS}. The idea was as follows. On the one hand, the electric ${\cal N}=2$ SQCD in \eqref{(1.1)} with small both $\mx$ and $m$ flows clearly to the standard ${\cal N}=1$ electric SQCD with $SU(N_c)$ gauge group, $N_F$ flavors of light quarks and without adjoint scalars $X^{\rm adj}$ at $m={\rm const},\,\,\mx\ra \infty$ and fixed $\la^{3N_c-N_F}=\lm^{2N_c-N_F}\mx^{N_c}$. On the other hand, the authors of \cite{APS} expected that, starting in \eqref{(1.1)} at $\mx\ll\lm$ from vacua of the baryonic branch with the $SU(\nd=N_F-N)$ low energy gauge group {\it by itself} and increasing then $\mx\gg\lm$, they will obtain Seiberg's dual ${\cal N}=1$ magnetic SQCD with $SU(\nd=N_F-N_c)$ gauge group and $N_F$ flavors of massless dual magnetic quarks. (But, in any case, $N^2_F$ light Seiberg's mesons $M^i_j\ra ({\ov Q}_j Q^i)$ of dual ${\cal N}=1\,\, SU(\nd)$ SQCD \cite{S2} were missing in this approach, see section 8 in \cite{ch6} for a more detailed critique of \cite{APS}\,).

The results from \cite{APS} were generalized then in \cite{CKM} to non-zero mass quarks, with emphasis on properties of flavor symmetry breaking in different vacua. Besides, multiplicities of various vacua were discussed in this paper in some details. (A complete detailed description of different vacua multiplicities and values of quark and gluino condensates, $\langle{\ov Q}_j Q^i\rangle_{N_c}$ and $\langle S\rangle_{N_c}$, in various vacua and at different hierarchies between $\mx,\,m$ and $\lm$ (or $\la^{3N_c-N_F}=\lm^{2N_c-N_F}
\mx^{N_c}$) see in \cite{ch4,ch6,ch5}\,).

But further detailed investigations of light particles charges in theory \eqref{(1.1)}, in some simplest examples with small values of $N_c$ and $N_F$, did not confirm the statement \cite{APS} that all these particles, and in particular light quarks, are pure magnetic.

Specifically, the $SU(N_c=3)$ vacua of \eqref{(1.1)}, with $N_F=4,\,\, \no=2$ (in the language of \cite{ch4}, these are Lt vacua with the gluino condensate $\langle S\rangle_{N_c}\sim\mx\lm^2$ and spontaneously broken non-trivial $Z_{\bb=2}$ symmetry at $m\ll\lm$), and with $N_F=5,\,\, \no=2$ (these are special vacua with $\langle S\rangle_{N_c}\sim\mx m\lm$ in the language of \cite{ch4}, $Z_{\bb=1}$ symmetry acts trivially in this case and gives no restrictions on the form of $\langle X^{\rm adj}_{SU(3)}\rangle$), were considered in details at $\mx\ra 0$ and $m\lessgtr\lm$ in \cite{MY}. Besides, the vs (very special) $U(N_c=3)=SU(N_c=3)\times U^{(0)}(1)$ vacua with $N_F=5,\,\, \no=2,\, \langle S\rangle_{N_c}=0$ were considered earlier in \cite{SY3} (these are $r=N_c$ vacua in the language of \cite{SY1}, but the discrete $Z_{2N_c-N_F=1}$ symmetry also acts trivially on $X^{\rm adj}_{SU(3)}$ in this case). It was shown in \cite{SY3,MY} that the low energy non-Abelian gauge group $SU(\no=2)$ at $m\ll\lm$ in all these cases is really dyonic, its {\it flavored} "quarks" (massless at $\mx\ra 0$) form $SU(2)$ doublet of dyons with the $SU(3)$ fundamental electric charges and with the hybrid magnetic charges which are fundamental with respect to $SU(2)$ but adjoint with respect to $SU(3)$.

This e.g. can be imagined for these UV free $SU(N_c=3)$ theories with $N_F=4(5)$ and $m\ll\lm,\,\mx= 0$ as follows. To avoid $g^2(\mu<\lm)<0$ and without restrictions from the non-trivial unbroken $Z_{2N_c-N_F}$ symmetry, the whole electric gauge group $SU(3)$ is higgsed at $\mu\sim\lm$  by $\langle X^{\rm adj}_{SU(3)}\rangle\sim\lm$ as $SU(3)\ra U^{2}(1)$, so that all original pure electrically charged particles acquire masses $\sim\lm$. But, due to strong coupling $a(\mu\sim\lm)=N_c g^2(\mu\sim\lm)/2\pi\sim 1$ and very specific properties of the enhanced ${\cal N}=2$ SUSY, there appear light composite BPS solitons massless at $\mx= 0$. The heavy pure electric flavored quarks $Q^{\,i}_a,\,{\ov Q}^{\,a}_i,\,\, a=1,\,2,\,\,i=1...4(5)$ with masses $\sim\lm$ combine with heavy unflavored pure magnetic monopoles, also with masses $\sim\lm$ and with above described hybrid magnetic charges, and form $SU(2)$ doublet of flavored BPS dyons $D^{\,i}_a,\,{\ov D}^{\,a}_i$, massless at $\mx= 0$. Besides, due to the unbroken at $\mx= 0$ enhanced ${\cal N}=2$ SUSY and BPS properties, also the heavy electric $SU(2)$ gluons $W^{\pm}$ combine similarly with corresponding heavy pure magnetic ones and form massless at $\mx= 0\,$ $W^{\pm}_D$ dyonic gluons (and the same for scalars $X^{\pm}$). As a result, the whole ${\cal N}=2$ lower energy theory (conformal at $N_F=4$ and IR free at $N_F=5$) with {\it the non-Abelian dyonic gauge group} $SU(2)$ emerges, with all its BPS particles massless at $\mx=0$. And this is the most surprising phenomenon.

These $\,SU(2)$ dyonic quarks $D^{\,i}_a, {\ov D}^{\,a}_i$ and all heavy original electrically charged particles with masses $\sim\lm$ are mutually non-local. Therefore, when these dyonic quarks $D^{\,i}_a, {\ov D}^a_i$ are higgsed at small $\mx\neq 0$ as $\langle D^{\,i}_a\rangle=\langle{\ov D}_i^{\,a}\rangle
\sim\delta^i_a\,(\mx\lm)^{1/2}$, this results in a weak confinement of all these heavy pure electrically charged particles, the string tension is $\sigma^{1/2}\sim (\mx\lm)^{1/2}\ll\lm$. (The confinement is weak in a sense that the string tension is much smaller than particle masses). In the case of $U(N_c=3)$ vs-vacua with $\langle S\rangle_{N_c}=0$, there are in addition the massless at $\mx= 0$ {\it unflavored} $SU(2)$ singlet BPS dyons $D_3, {\ov D}_{3}$, with $\langle{\ov D}_{3}\rangle=\langle D_3\rangle\sim (\mx\lm)^{1/2}$ at $\mx\neq 0$ \cite{SY3}. All original pure electric particles are also confined.\\

Besides, it was demonstrated explicitly in \cite{SY3,MY} at $\mx= 0$ on examples considered, how vacua with some massless at $\mx= 0$ pure electric original quarks $Q^i$ at $m\gg\lm$ can evolve at $m\ll\lm$ into vacua with only dyonic massless quarks $D^i$. This is not because at some special value $m=m_0\sim\lm$ the massless pure electric quarks mysteriously transform {\it by themselves} into massless dyonic quarks. The real mechanism is different and not mysterious. In the vicinity of $m=m_0$, there is a number of vacua in which {\it additional} corresponding particles become light. At $m=m_0$ the vacua collide, all these particles become massless, and the low energy regime is conformal. In particular, both some flavored pure electric quarks $Q^i$ and flavored dyonic quarks $D^i$ are massless at this point (as well as some pure magnetic monopoles). The masses of these quarks with {\it fixed charges} behave continuously but non-analytically in $m/\lm$ at this special collision point $m= m_0$. E.g., in selected at $m\lessgtr m_0$ vacua colliding at $m=m_0$ with other vacua, the mass of some light pure electric quarks $Q^i$ stays at zero at $m\geq m_0$, then begins to grow at $m<m_0$ and becomes $\sim\lm$ at $m\ll\lm$, while the mass of light composite dyonic quarks $D^i$ is zero in these vacua at $m\leq m_0$, then begins to grow at $m>m_0$ and becomes large at $m\gg m_0$. {\it Equivalently}, tracing e.g. the evolution with $m/\lm$ in the small vicinity of $m=m_0$ of {\it charges} of, separately, massless or nearly massless flavored quark-like particles at $m\lessgtr m_0$, one can say that their charges jump at the collision point $m=m_0$, i.e.: a) the massless particles at $m > m_0$ are pure electric quarks $Q^i$, while they are dyonic quarks $D^i$ at $m < m_0$; b) the very light (but not massless) flavored quark-like particles are the dyonic quarks $D^i$ at $m > m_0$, while they are  pure electric quarks $Q^i$ at $m < m_0$. These two views of the non-analytic behavior at the collision point $m= m_0$ when traversing the small vicinity of the collision point, i.e. either looking at the non-analytic but continuous evolution of small masses of particles with fixed charges, or looking on the jumps of charges of, separately, massless particles and of nearly massless particles, are clearly {\it equivalent and indistinguishable} because all these particles simultaneously become massless at the collision point $m=m_0$. These are two different projections (or two sides) of the one phenomenon.

The examples of these conformal regimes for the ${\cal N}=2\,\,SU(2)$ theory were described earlier in \cite{APSW,BF,GVY} (but it is clear beforehand that to avoid unphysical $g^2(\mu<\lm)<0$ in UV free ${\cal N}=2\,\,SU(2)$ with $N_F=1,2,3$ and $m\ll\lm$, the whole gauge group is higgsed necessarily at the scale $\mu\sim\lm$ by $\langle X^{\rm adj}_{SU(2)}\rangle\sim\lm,\, SU(2)\ra U(1)$ \cite{SW1,SW2}, so that all original pure electrically charged particles are indeed heavy, with masses~ $\sim\lm$).\\

The properties of light particles with masses $\ll\lm$ in theory \eqref{(1.1)} (really, in the $U(N_c)$ theory with especially added $U^{(0)}(1)$ multiplet) for general values of $N_c,\, N_F$ and possible forms of low energy Lagrangians at scales $\mu<\lm$ in different vacua were considered later by M. Shifman and A. Yung in a series of papers, see e.g. the recent papers \cite{SY1,SY2} and references therein. At $m\ll\lm$, for vacua of baryonic branch with the lower energy ${\cal N}=2\,\,\, SU(\nd=N_F-N_c)\times U^{2N_c-N_F+1}(1)$ gauge group at $\mu<\lm$ \cite{APS}, it was first proposed naturally by these authors in \cite{SY1} that all light charged particles of ${\cal N}=2\,\, SU(\nd)$ are the original pure electric particles, as they have not received masses $\sim\lm$ directly from higgsed $\langle X^{\rm adj}_{SU(N_c)}\rangle$ (this is forbidden in these vacua by the unbroken non-trivial $Z_{\bb\geq 2}$ symmetry \cite{APS}). But they changed later their mind to the opposite
\footnote{\,
Because not $SU(N_c)$ but $U(N_c)$ group was considered in \cite{SY1,SY2}, this their statement concerns really br2 and very special vacua of $U(N_c)$, see table of contents, sections 2.2 and 6 (these are respectively zero vacua and $r=N_c$ vacua in the language of \cite{SY1,SY2}).
}
and, extrapolating freely by analogy their previous results for $r=N_c=3,\, N_F=5,\, \no=2$ very special vacua in \cite{SY3} (see also \cite{SY4}) to all br2 and very special vacua, stated in \cite{SY2} that this $SU(\nd)$ group is not pure electric but dyonic. This means that at $m\ll\lm$ all original pure electrically charged particles of $SU(\nd)$ have received large masses $\sim\lm$, but now not directly from $\langle X^{\rm adj}_{SU(N_c)}\rangle\sim\lm$ (this is forbidden by the non-trivial unbroken $Z_{\bb\geq 2}$ symmetry), but from some mysterious "outside" sources and decoupled at $\mu<\lm$, while the same gauge group $SU(\nd)$ of light composite dyonic solitons was formed. Unfortunately, it was overlooked in \cite{SY2} that, in distinction with their example with $U(N_c=3),\,\,N_F=5,\,\,\no=2$ in \cite{SY3} with the trivial $Z_{2N_c-N_F=1}$ symmetry giving no restrictions on the form of $\langle X\rangle$, the appearance e.g. in the superpotential at $\mu\sim\lm$ of mass terms of original $SU(\nd)$ quarks like $\sim\lm{\rm Tr}\,({\ov Q}Q)_{\nd}$ or terms like $\sim\lm{\rm Tr\,}(X^2_{SU(\nd)})$ for scalars is forbidden in vacua with the non-trivial unbroken $Z_{2N_c-N_F\geq 2}$ symmetry, independently of whether these terms originated directly from $\langle X\rangle$ or from unrecognized "outside".

The authors of \cite{APS} understood that, at $m\ll\lm$ and small $\mx$, the quarks $Q^i$ from $SU(\nd)$
cannot acquire heavy masses $\sim\lm$ in vacua with unbroken non-trivial $Z_{2N_c-N_F\geq 2}$ symmetry and remain light (i.e. with masses $\ll\lm$). But they argued then that a number of vacua with light (massless at $\mx=0$) pure electric quarks at $m\gg\lm$, when evolved to $m\ll\lm$, have only light quarks with non-zero magnetic charges. And they considered this as a literal evidence that phases with light pure electrically and pure magnetically charged quarks are not different and can transform freely into each other by some unidentified mechanism. And they supposed that in vacua of the baryonic branch just this mysterious mechanism transformed at $m\ll\lm$ all light pure electric particles from $SU(\nd)$ into light pure magnetic ones. Remind that, as was shown later in \cite{GVY,SY3,MY} and described above, in vacua with light electric quarks at $m\gg\lm$, {\it these quarks do not literally transform by themselves into dyonic ones} at $m\ll\lm$. The mechanism is different, see above. Therefore, we will consider in what follows that such mysterious mechanism which transforms literally at $m\ll\lm$ light electric quarks into magnetic ones does not exist really.~
\footnote{\,
It is clear from \cite{SY1,SY2} that these authors also disagree with \cite{APS} at this point.
}
\vspace*{5mm}

Our purpose in this paper is similar to \cite{SY1,SY2}, but we limit ourselves at $m\ll\lm$ only to vacua of baryonic branch with the non-trivial unbroken discrete $Z_{\bb\geq 2}$ symmetry. As will be seen from the text below, this $Z_{\bb}$ symmetry is strong enough and helps greatly.

We introduce also the two following assumptions of general character.\,-

{\bf Assumption A}: {\it in unbroken ${\cal N}=2$ (i.e. at $\mx\ra 0$) and at least in considered vacua with the non-trivial unbroken $Z_{\bb\geq 2}$ symmetry, the original pure electrically charged particles receive additional contributions to their masses only from higgsed} $\langle X\rangle\neq 0$ (remind that $\langle X\rangle$ includes in general all non-perturbative instanton contributions). In other words, at $\mx\ra 0$ {\it they are BPS particles} in vacua considered. For instance, if light quarks $Q^i$ with $m\ll\lm$ do not receive contributions to their masses from some $\langle X^{A}\rangle\neq 0$, they remain as they were, i.e. the light electric quarks with masses $m$. The opposite is not true in this very special ${\cal N}=2$ theory in the strong coupling regime $a(\mu\sim\lm)=N_c g^2(\mu\sim\lm)/2\pi=O(1)$. The original pure electric quarks which have received masses $\sim\lm$ from some $\langle X^{A}\rangle\sim\lm$, can combine e.g. with some heavy composite pure magnetic monopoles, also with masses $\sim\lm$, and form massless at $\mx\ra 0$ composite BPS dyons. And all that.

{\bf Assumption B}: {\it there are no massless particles at $\mx\neq 0,\, m\neq 0$ in considered vacua of the $SU(N_c)$ theory} \eqref{(1.1)} (except for Nambu-Goldstone multiplets originating from spontaneously broken global symmetry $U(N_F)\ra U(\no)\times U(\nt)$\,). This more particular assumption will help us to clarify the charges of massless at $\mx\ra 0$ charged BPS dyons $D_j$, see section 2.1\,.

Besides, to calculate mass spectra in some ${\cal N}=1$ SQCD theories obtained from ${\cal N}=2$ after increasing masses $\mx$ of adjoint scalar multiplets and decoupling them as heavy ones at scales $\mu<\mx^{\rm pole}$, we use the dynamical scenario introduced in \cite{ch3}. {\it But it is really needed in this paper only for those few cases with $3N_c/2<N_F<2N_c-1$ when so obtained ${\cal N}=1$ SQCD enters at lower energies the strong coupling conformal regime with} $a_*=O(1)$, see section 2.4 below. Remind that this scenario \cite{ch3} assumes for these cases that at those even lower scales where this conformal regime is broken when quarks are either higgsed or decouple as heavy, no {\it additional} parametrically light composite solitons are formed in this ${\cal N}=1$ SQCD without colored scalar fields $X^{\rm adj}$.
\footnote{\,
It is worth noting that the appearance of additional light solitons will influence the 't Hooft triangles of this ${\cal N}=1$ theory. \label{(f3)}
}

Let us emphasize that this dynamical scenario satisfies all those checks of Seiberg's duality hypothesis
for ${\cal N}=1$ SQCD which were used in \cite{S2}. In other cases, if this ${\cal N}=1$ SQCD stays in the IR free weak coupling logarithmic regime, its dynamics is simple and clear and there is no need for any additional dynamical assumptions.\\

Below in sections 2-8 are presented our results (see the table of contents) for charges of light particles, the forms of low energy Lagrangians and mass spectra in various vacua of \eqref{(1.1)} with the unbroken discrete $Z_{\bb\geq 2}$ symmetry, at different hierarchies between lagrangian parameters $m,\,\mx$ and $\lm$.

For further convenience, we present here also the form of the Konishi anomalies \cite{Konishi} in the $SU(N_c)$ theory \eqref{(1.1)} in vacua with the spontaneously broken global flavor symmetry $U(N_F)\ra U(\no)\times U(\nt)$, following from the effective superpotential $\w^{\rm eff}_{\rm tot}$ which accounts for all anomalies and contains {\it only} quark bilinears $\Pi^i_j=({\ov Q}_j Q^i)_{N_c}$. This $\w^{\rm eff}_{\rm tot}$ is used here only as a most convenient tool for finding the Konishi anomalies for the quark condensates $\langle\Qo\rangle_{N_c}$ and $\langle\Qt\rangle_{N_c}$ in \eqref{(1.4)},
\bq
\w^{\,\rm eff}_{\rm tot}(\Pi)=m\,{\rm Tr}\,({\ov Q} Q)_{N_c}-\frac{1}{2\mx}\Biggl [ \,\sum_{i,j=1}^{N_F} ({\ov Q}_j Q^i)_{N_c}({\ov Q}_{\,i} Q^j)_{N_c}-\frac{1}{N_c}\Bigl ({\rm Tr}\,({\ov Q} Q)_{N_c}\Bigr )^2\Biggr ]-\nd S_{N_c} \,, \label{(1.3)}
\eq
\bbq
S_{N_c}=\Bigl(\frac{\det (\qq)_{N_c}}{\la^{3N_c-N_F}=\lm^{2N_c-N_F}\mx^{N_c}}\Bigr )^{\frac{1}{N_F-N_c}}\,,
\quad \langle S\rangle_{N_c} =\Bigl(\frac{\langle\det\qq\rangle=\langle\Qo\rangle^{\no}\langle\Qt
\rangle^{\nt}}{\lm^{2N_c-N_F}\mx^{N_c}}\Bigr )^{\frac{1}{N_F-N_c}}_{N_c}\,,
\eeq
see e.g. section 3 in \cite{ch4} and/or section 4 in \cite{ch5}, and section 8 in \cite{ch6},
\, $N_F=\no+\nt,\,\,1\leq\no\leq [N_F/2],\,\, \nt\geq\no$.
\bq
\langle \Qo+\Qt-\frac{1}{N_c}{\rm Tr}\,(\qq)\rangle_{N_c}=\mx m\,,\quad \langle S \rangle_{N_c}=\frac{\langle \Qo\rangle_{N_c}\langle\Qt\rangle_{N_c}}{\mx}\,,\label{(1.4)}
\eq
\bbq
\langle\,\sum_{a=1}^{N_c}{\ov Q}^{\,a}_j Q^i_a\,\rangle=\delta^i_j\,\langle\Qo\rangle_{N_c}\,,\,\,i,j=
1\,...\,\no,\,\, \langle\,\sum_{a=1}^{N_c}{\ov Q}^{\,a}_j Q^i_a\,\rangle=\delta^i_j\,
\langle\Qt\rangle_{N_c}\,,\,\,i,j=\no+1\,...\,N_F\,,
\eeq
\bbq
\langle {\rm Tr}\,({\ov Q} Q \rangle_{N_c}=\no\langle\Qo\rangle_{N_c}+\nt\langle\Qt\rangle_{N_c}\,,
\eeq
\bbq
\mx\langle {\rm Tr}\, (\sqrt{2}\,X^{\rm adj}_{SU(N_c)})^2\rangle_{N_c}=(2N_c-N_F)\langle S \rangle_{N_c}+m\,\langle {\rm Tr}\,({\ov Q} Q) \rangle_{N_c}\,,
\eeq
where $\langle S \rangle_{N_c}$ is the gluino condensate summed over all $N_c^2-1$ colors.\\

The organization of this paper is as follows (see also the table of contents). In $SU(N_c)$ gauge theories with $m\ll\lm$, the br2 vacua with the broken global flavor symmetry, $U(N_F)\ra U(\no)\times U(\nt),\, \nt>N_c$, and the unbroken non-trivial discrete symmetry $Z_{\bb\geq 2}$ are considered in section 2.1\,. We discuss in detail in this section 2.1 the charges of massless at $\mx\ra 0$ BPS particles in these vacua, the low energy Lagrangians and mass spectra at smallest $0<\mx\ll\Lambda^{SU(\nd-{\rm n}_1)}_{{\cal N}=2\,\, SYM}$. The mass spectra in these br2 vacua at larger values of $\mx$ are described in sections 2.3 and 2.4. These results serve then as a basis for a description of similar regimes in sections 6-8. The br2 vacua
of the $U(N_c)$ theory at smallest $0<\mx\ll\Lambda^{SU(\nd-\no)}_{{\cal N}=2\,\, SYM}$ are considered in section 2.2. Besides, the S vacua with the unbroken flavor symmetry $U(N_F)$ (i.e. $\no=0$) in the $SU(N_c)$ theory at $m\ll\lm$ are considered in section 3.

The light particle charges, low energy Lagrangians and mass spectra in a large number of vacua of $SU(N_c)$ theory with $m\gg\lm$ are considered in sections 4 and 5.

Mass spectra in specific additional vs (very special) vacua with $\langle S\rangle_{N_c}=0,\, N_c\leq N_F< 2N_c-1$ present in $U(N_c)$ theory (but absent in $SU(N_c)$, see Appendix in \cite{ch4}\,), are considered in section 6 at $m\lessgtr\lm$ and various values of $0<\mx\ll\lm$.

The mass spectra are calculated in section 7 in br2 vacua of $SU(N_c)$ theory with $U(N_F)\ra U(\no)\times U(\nt),\, \nt<N_c,\,\,m\gg\lm$ at various values of $\mx$.

And finally, the mass spectra are described in section 8 for vacua with $0<N_F<N_c$ and $m\gg\lm\,$.\\

Calculations of power corrections to the leading terms of the low energy quark and dyon condensates are presented in two important Appendices. These calculations give additional {\it independent} confirmation of a self-consistency of the whole approach.

\addcontentsline{toc}{section}
{\hspace*{4cm}\bf Part I.\,\, Small quark masses, $\Large\mathbf{m\ll\lm}$ }

\vspace*{3mm}

\begin{center}{\hspace*{1cm}\bf\Large Part I.\,\, Small quark masses, $\Large\mathbf{m\ll\lm}$ } \end{center}

\section{ Broken flavor symmetry,\,\, br2 vacua}
\numberwithin{equation}{section}

From \eqref{(1.3)},\eqref{(1.4)} for the $SU(N_c)$ theory, the quark condensates (summed over all $N_c$ colors) at $N_c+1<N_F<2N_c-1$ in the considered br2 vacua (i.e. breaking 2, with the dominant condensate $\langle({\ov Q}Q)_2\rangle_{N_c}$) with $\nt>N_c\,,\, \no<\nd=(N_F-N_c)$ in theory \eqref{(1.1)} (these are vacua of the baryonic branch in the language \cite{APS} or zero vacua in \cite{SY1,SY2}\,) look as, see section 3 in \cite{ch4} or section 4 in \cite{ch5}, (the leading terms only)
\footnote{\,
Here and everywhere below\,: $A\approx B$ has to be understood as an equality neglecting smaller power corrections, and $A\ll B$ has to be understood as $|A|\ll |B|$. \label{(f4)}
}
\bq
\langle({\ov Q}Q)_2\rangle_{N_c}=\mx m_1+\frac{N_c-\no}{\nt-N_c}\langle({\ov Q}Q)_1\rangle_{N_c}\approx \mx m_1\,,\quad \langle({\ov Q}Q)_1\rangle_{N_c}\approx\mx m_1\Bigl (\frac{m_1}{\lm}
\Bigr)^{\frac{\bb}{{\rm n}_2-N_c}},\label{(2.1)}
\eq
\bbq
\frac{\langle ({\ov Q}Q)_1\rangle_{N_c}}{\langle({\ov Q}Q)_2\rangle_{N_c}}\approx\Bigl (\frac{m_1}{\lm}
\Bigr)^{\frac{2N_c-N_F}{{\rm n}_2-N_c}}\ll 1\,,\,\,  N_c+1\leq N_F\leq 2N_c-2\,,\,\,  m_1=\frac{N_c}{N_c-n_2}\,m\,,\,\, \nd\equiv N_F-N_c,
\eeq
\bbq
\langle S\rangle_{N_c}=\frac{\langle\Qo\rangle_{N_c}\langle\Qt\rangle_{N_c}}{\mx}\approx\mx m_1^2\Bigl (\frac{m_1}{\lm}\Bigr)^{\frac{\bb}{{\rm n}_2-N_c}}\ll\mx m^2,\quad \nt > N_c\,,
\eeq
and this shows that {\it the multiplicity of these vacua} is $N_{\rm br2}=(\nt-N_c)\, C^{\,\nt}_{N_F}=(\nd-n_1)\,C^{\,\no}_{N_F}$, the factor $C^{\,\no}_{N_F}=N_F!/[\no!\,\nt!]$ originates from the spontaneous breaking of the global $U(N_F)$ flavor symmetry of \eqref{(1.1)}.

Because the multiplicity $N_{\rm br2}$ of these br2 vacua in \eqref{(2.1)} does not contain the factor $\bb$, this shows that the non-trivial at $\bb\geq 2$ discrete symmetry $Z_{\bb}$ {\it is not broken spontaneously in these vacua}.

\numberwithin{equation}{subsection}
\subsection{$SU(N_c)$, small $\mx\,,\,\,\,\mx\ll\Lambda^{SU(\nd-{\rm n}_1)}_{{\cal N}=2\,\,SYM}\ll m$}

We describe first the overall qualitative picture of various stages (in order of decreasing energy scale) of the gauge and flavor symmetry breaking in these br2 vacua. As will be shown below, it leads in a practically unique way to a right multiplicity of these vacua with unbroken discrete $Z_{\bb\geq 2}$ symmetry, and it is non-trivial to achieve this.\\

{\bf 1)}\, As was explained in Introduction, at small $m$ and $\mx$, the adjoint field $X^{\rm adj}_{SU(N_c)}$ higgses necessarily the UV free $SU(N_c)$ group at the largest scale $\mu\sim\lm$, and in the case considered this first stage looks as: $\,SU(N_c)\ra SU(N_F-N_c)\times U^{(1)}(1)\times U(1)^{\bb-1}$ \cite{APS}, while all other components of $\langle X^A \rangle\lesssim m$ are much smaller, see \eqref{(2.1.3)},\eqref{(2.1.6)} below,
\bq
\langle\sqrt{2}\, X^{\rm adj}_{SU(\bb)}\,\rangle\sim\lm\,{\rm diag}(\,\underbrace{\,0}_{\nd}\,; \underbrace{\,\omega^0,\,\,\omega^1,\,...\,,\,\omega^{\bb-1}}_{\bb}\,)\,,\quad \omega=\exp\{\frac{2\pi i}{\bb}\}\,,  \label{(2.1.1)}
\eq
\bbq
\sqrt{2}\, X^{(1)}_{U(1)}=a_{1}\,{\rm diag}(\,\underbrace{\,1}_{\nd}\,;\, \underbrace{\,c_1}_{\bb}\,),\quad c_1=-\,\frac{\nd}{\bb}\,,\quad \langle a_{1}\rangle=\frac{1}{c_1}\, m\,.
\eeq
This pattern of symmetry breaking is required by the unbroken $Z_{\bb\geq 2}$ discrete symmetry. Remind that  charges under $Z_{\bb}=\exp\{\pi i/(\bb)\}$ transformation are: $q_{\lambda}=q_{\theta}=1,\,\,q_{X}=q_{\rm m}=2,\,\,q_Q=q_{\ov Q}=q_{\lm}=0,\,\,\mx=-2$, so that $\langle a_{1}\rangle$ respects $Z_{\bb}$ symmetry, while the form of $\langle\sqrt{2}\, X^{\rm adj}_{SU(\bb)}\,\rangle$ ensures the right behavior under $Z_{\bb }$ transformation up to interchanging terms in \eqref{(2.1.1)} (the Weyl symmetry).

As a result, all original $SU(2N_c-N_F)$ charged electric particles acquire large masses $\sim\lm$ and decouple at lower energies $\mu<\lm/(\rm several)$. But due to strong coupling, $a(\mu\sim\lm)=N_c g^2(\mu\sim\lm)/2\pi\sim 1$ and the enhanced ${\cal N}=2$ SUSY, the light composite BPS dyons $D_j, {\ov D}_j,\,\, j=1...2N_c-N_F$ (massless at $\mx\ra 0$) are formed in this sector at the scale $\mu\sim\lm$, their number is required by the unbroken $Z_{\bb\geq 2}$ symmetry which operates interchanging them with each other. This is seen also from the spectral curve \eqref{(1.2)}: it has $\bb$ unequal double roots $e_j\approx \omega^{\,j-1}\lm,\,\, j=1...\bb,$ corresponding to these $\bb$ massless BPS solitons \cite{APS,CKM}. The charges of these dyons $D_j$ are discussed below in detail. They are all flavor singlets, their $U^{(1)}(1)$ and $SU(N_c)$ baryon charges which are $SU(\bb)$ singlets are pure electric, while they are coupled with the whole set of $U^{\bb-1}(1)$ independent light (massless at $\mx=0$) multiplets in \eqref{(2.1.2)}.

{\it All original electric particles with $SU(\nd)$ colors remain light}, i.e. with masses $\ll\lm$. This is a consequence of the non-trivial unbroken $Z_{\bb\geq 2}$ symmetry and their BPS properties. As for the quarks with $SU(\nd)$ colors, the appearance in the superpotential of the mass terms like $\sim\lm {\rm Tr\,}({\ov Q} C_Q Q)_{\nd}$ with at least some non-zero constants $C_{Q}=O(1)$ will break spontaneously $Z_{\bb\geq 2}$, independently of the source of such terms (while higgsed some $\langle Q^i_a\rangle\sim\lm$ are forbidden by the unbroken continuous $U(1)$ R-symmetry). Therefore, the unbroken $Z_{\bb\geq 2}$ is sufficient by itself to forbid so large masses. This is clear. But we will need below stronger assumption ${\bf "A"}$ formulated in Introduction that these $SU(\nd)$ quarks are BPS particles, i.e. they receive additional contributions to their masses only directly from $\langle{\sqrt 2} X^{(1)}_{U(1)}\rangle$ in \eqref{(2.1.1)} (and from $\langle X^{\rm adj}_{SU(\nd}\rangle$, see below). This forbids also the appearance of {\it any additional}\, smaller "outside" contributions to their masses, e.g. $c_{a_1}\langle a_1\rangle {\rm Tr\,}({\ov Q} Q)_{\nd},\, c_{a_1}=O(1)$ or $c_m m {\rm Tr\,}({\ov Q} Q)_{\nd},\, c_m=O(1)$, although these are not forbidden by the unbroken $Z_{\bb\geq 2}$. Besides, this BPS assumption ${\bf "A"}$ concerns all original electric particles of $SU(\nd)$. (The absence of such additional contributions is confirmed by the {\it independent} calculations of quark condensates, see \eqref{(2.1.14)} and the text just below it).

The $SU(\nd)$ gauge symmetry can still be considered as unbroken at scales $m\ll\mu<\lm$ (it is broken only at the scale $\sim m$ due to higgsed $X^{\rm adj}_{SU(\nd)}$, see below). Therefore, after integrating out all heavy particles with masses $\sim\lm$ and at sufficiently small $\mx$, the superpotential at the scale $\mu=\lm/{(\rm several)}$ can be written as
\bq
\wh{\w}_{\rm matter}=\w_{SU(\nd)}+\w_D+\w_{a_1}+\dots\,,\label{(2.1.2)}
\eq
\bbq
\w_{SU(\nd)}={\rm Tr}\,\Bigl [{\ov Q}\Bigl (m-a_1-\sqrt{2}X^{\rm adj}_{SU(\nd)}\Bigr ) Q\Bigr ]_{\nd}+\wmu {\rm Tr}\,(X^{\rm adj}_{SU(\nd)})^2\,,\quad \wmu=\mx(1+\delta_2)\,,
\eeq
\bbq
\w_{D}=(m-c_1 a_1)\sum_{j=1}^{\bb}{\ov D}_j D_j-\sum_{j=1}^{\bb} a_{D,j}{\ov D}_j D_j-
\mx\lm\sum_{j=1}^{\bb}\omega^{j-1}a_{D,j}+
\eeq
\bbq
+\mx L \sum_{j=1}^{\bb} a_{D,j}+\mx\sum_{i,j} C_{i,j} a_{D,i}a_{D,j}+O\Bigl (\mx\frac{a_D^3}{\lm} \Bigr )\,,
\eeq
\bbq
\w_{a_1}=\frac{\mx}{2}(1+\delta_1)\frac{\nd N_c}{\bb}\,a^2_1+\mx N_c \delta_3\,a_1(m-c_1 a_1)+\mx N_c\delta_4 (m-c_1 a_1)^2\,,
\eeq
where $a_{D,j}$ are the light neutral scalars, $L$ is the Lagrange multiplier, $\langle L\rangle=O(m)$, all $\delta_i=O(1)$. The dots in \eqref{(2.1.2)} denote smaller power corrections.

The way the parameter "m" enters \eqref{(2.1.2)} can be understood as follows. This parameter "m" in the original superpotential $\w_{\rm matter}$ in \eqref{(1.1)} can be considered as a soft scalar $U^{(0)}(1)$ background field coupled universally to the $SU(N_c)$ singlet scalar electric baryon current, $\delta\w_B=m J_B$, and this current looks at the scale $\mu\gg\lm$ as $J_B={\rm Tr}\,({\ov Q} Q)_{N_c}$. But at the lower energy scale $\mu\ll\lm$ only lighter particles with masses $\ll\lm$ and with a non-zero electric baryon charge contribute to $J_B$, these are ${\rm Tr}\,({\ov Q} Q)_{\nd}$ and $\sum_{j=1}^{\bb}{\ov D}_j D_j$ (see bellow). Moreover, the field $a_{1}$  plays a similar role, but its couplings are different for two different color sectors of $SU(N_c)$, see \eqref{(2.1.1)},\eqref{(2.1.2)}.

The terms with $\delta_i,\,\, i=1\,...\,4\,,\,\,\delta_i=O(1)$, arose in \eqref{(2.1.2)} from integrating out heaviest original fields with masses $\sim \lm$ in the soft background of lighter $X^{\rm adj}_{SU(\nd)},\,\, a_1$ and $(m-c_1 a_1)$ fields. These heavy fields are\,: charged $SU(\bb)$ adjoints with masses
$ (\Lambda_j-\Lambda_i),\,\, \Lambda_j=\omega^{\,j-1}\lm$, the heavy quarks with $SU(\bb)$ colors and masses $\Lambda_j-(m-c_1 a_1)$, and heavy adjoint hybrids $SU(N_c)/[SU(\nd)\times SU(\bb)\times U^{(1)}(1)]$ with masses $\Lambda_j-[\sqrt{2}X^{\rm adj}_{SU(\nd)}+(1-c_1)a_1]$. Note that terms like $\mx\lm^2,\,\,\mx a_1\lm$ and $\mx (m-c_1 a_1)\lm$ cannot appear in \eqref{(2.1.2)}, they are forbidden by the unbroken $Z_{\bb\geq 2}$ symmetry. Moreover, the term $\sim\mx m^2$ also cannot appear in \eqref{(2.1.2)} as there are no heavy fields with masses $\Lambda_j-m$, the soft background fields include "m" only in the combination $(m-c_1 a_1)$.\\

The above described picture is more or less typical, the only really non-trivial point concerns the charges of $2N_c-N_F$ dyons $D_j$. Therefore, we describe now some necessary requirements to their charges.

a) First, the $SU(\nd)$ part of \eqref{(2.1.2)}, {\it due its own internal dynamics}, ensures finally the right pattern of the spontaneous flavor symmetry breaking $U(N_F)\ra U(\no)\times U(\nt)$ and the right multiplicity of vacua, $N_{\rm br2}=(\nd-\no) C^{\,\no}_{N_F}$, see below. Therefore, the charges of these $\bb$ dyons $D_j$ should be such as not to spoil the right dynamics in the $SU(\nd)$ sector.

b) These charges should be such that the lower energy Lagrangian respects the discrete $Z_{\bb\geq 2}$ symmetry. This fixes, in particular, the number $\bb$ of these dyons because $Z_{\bb\geq 2}$ transformations interchanges them between each other.

c) These dyons have to be flavor singlets. Otherwise, because they are all higgsed at small $\mx\neq 0$, this will result in a wrong pattern of the spontaneous flavor symmetry breaking. Besides, the charges of these dyons $D_j$ formed at the scale $\mu\sim\lm$ cannot know about properties of further color breaking in $SU(\nd)$ at much lower scales $\mu\sim m\ll\lm$, i.e. they do not know the number $~\no$.

d) None of $U^{\bb-1}(1)$ charges of these light BPS dyons $D_j$ cannot be pure electric because all $SU(\bb)$ adjoint $\langle a_j\rangle\sim\lm$ in \eqref{(2.1.1)}, so that all BPS solitons with some pure electric $U^{\bb-1}(1)$ charges will have large masses $\sim\lm$ in this case.

e) Their magnetic charges should be such that they will be mutually local with respect to all original electric particles in the $SU(\nd)$ sector. Otherwise, in particular, because all these $2N_c-N_F$ dyons and all original pure electric quarks from $SU(\nd)$ have non-zero $U^{(1)}(1)$ charges, and because all dyons are higgsed, $\langle{\ov D}_j\rangle=\langle D_j\rangle\sim (\mx\lm)^{1/2}$, this will lead to confinement of all quarks from the $SU(\nd)$ sector with the string tension $\sigma^{1/2}\sim (\mx\lm)^{1/2}$ if the  magnetic $U^{(1)}(1)$ charge of these dyons will be non-zero. E.g., see below, all $N_F$ flavors of confined light electric quarks with $SU(\no)$ colors will decouple then in any case at the scale $\mu<(\mx\lm)^{1/2}$, the flavor symmetry $U(N_F)$ will remain unbroken in the whole $SU(\nd)$ sector, while the remained ${\cal N}=2 \,\,SU(\no)$ SYM will give the additional wrong factor $\no$ in the multiplicity. All this is clearly unacceptable. This excludes the variant with the non-zero $U^{(1)}(1)$ magnetic charge, and also with non-zero $SU(\nd)$ magnetic charges of these dyons. Similarly, when considering the $U(N_c)=SU(N_c)\times U^{(0)}(1)$ theory in section 2.2 below, this requires also that the $SU(N_c)$ baryon charge of these dyons has to be pure electric. On the other hand, to be mutually local with the whole $SU(\nd)$ part, the magnetic parts of all their $U^{\bb-1}(1)$ charges have to be $SU(\bb)$ root-like (i.e. $SU(\bb)$ adjoints most naturally).

f) If these $\bb$ dyons $D_j$ will have zero $U^{(1)}(1)$ charge, they will be not coupled then with the $U^{(1)}(1)$ multiplet \eqref{(2.1.1)}, this will correspond to the first term of $\w_D$ in \eqref{(2.1.2)},\eqref{(2.1.5)} of the form $m\sum_{j}^{\bb}{\ov D}_j D_j$. This will lead to extra massless particles at $\mx\neq 0$ and even to the internal inconsistency, see below.

The case with zero $SU(N_c)$ baryon charge of dyons, i.e. with the first term of $\w_{D}$ of the form $[\,-c_1 a_1\sum_{j=1}^{\bb}{\ov D}_j D_j\,]$ in \eqref{(2.1.2)}, \eqref{(2.1.5)}, will lead to wrong values of quark condensates, see below.

Finally, about the case when this first term of $\w_{D}$ in \eqref{(2.1.2)} equals zero. All $\bt=\bb$ dyons ${\ov D}_j, D_j$ are higgsed at $\mx\neq 0$ (this is a consequence of the unbroken $Z_{\bb\geq 2}$ symmetry which operates interchanging them with each other), but there will be only $\bb-1$ independent multiplets $U^{\bb-1}(1)$ they are coupled with. There will remain then at $\mx\neq 0$ {\it additional} massless Nambu-Goldstone bosons resulting in the additional (infinite) factor in the multiplicity of these vacua. Moreover, counting degrees of freedom in the $SU(\nd)\times U^{(1)}(1)$ sector (see below), it is seen that one ${\cal N}=1$ photon multiplet will remain exactly massless (while its ${\cal N}=2$ scalar partner will have very small mass $\sim\mx$ only due to breaking ${\cal N}=2\ra {\cal N}=1$ at small $\mx\neq 0$).

Clearly, these considerations with extra massless particles also concern as well the cases when even one out of $U^{\bb-1}(1)$ Abelian photon multiplets will be not coupled with a whole set of $\bb$ dyons. Due to the unbroken $Z_{\bb}$ discrete symmetry, all $\bb-1$ Abelian multiplets will be not coupled then. In any case, at least, all $U^{\bb-1}(1)$ photons will remain massless at $\mx\neq 0$, while because all dyons will be higgsed, there will be a large number of extra Numbu-Goldstone particles. At least, all these cases do not look realistic. According to our assumption "{\bf B}" (see Introduction), all variants with extra massless particles at $\mx\neq 0,\, m\neq 0$ in $SU(N_c)$ theories (in addition to normal Nambu-Goldstone particles originating from the spontaneous global flavor symmetry breaking) in $SU(N_c)$ theory are excluded.

g) All $\bb$ these BPS dyons have to be mutually local and massless at $\mx=0$ (so that $\langle a_{D,j}\rangle=0$, see \eqref{(2.1.5)},\eqref{(2.1.6)} below).

Therefore, as a result, these mutually local BPS dyons $D_j, {\ov D}_j$, massless at $\mx=0$, should have non-zero both {\it pure electric} $U^{(1)}(1)$ and $SU(N_c)$ baryon charges and, on the whole, all $\bb-1$  independent charges of the broken $SU(\bb)$ subgroup, with their magnetic parts being the appropriate $SU(\bb)$ adjoints and all non-zero. They will be mutually local then with the whole $SU(\nd)$ sector, coupled with $"m"$, with the electric $U^{(1)}(1)$ multiplet, and with all $U^{\bb-1}(1)$ multiplets, see\eqref{(2.1.1)},\eqref{(2.1.2)}. There will be no massless particles at $\mx\neq 0,\, m\neq 0$ (except for $2\nt\no$ Nambu-Goldstone multiplets), and the multiplicity and pattern of flavor symmetry breaking will be right, see below.

On the other hand, the whole set of $\bb$ these dyons with such charges, all higgsed at $\mx\neq 0$, will be mutually non-local to all heavy original $SU(\bb)$ charged pure electric particles with masses $\sim\lm$, so that these last will be weakly confined, the string tension is $\sigma^{1/2}_D\sim (\mx\lm)^{1/2}\ll\lm$.\\

{\bf 2)}\, The second stage is the breaking $SU(\nd)\ra SU(n_1)\times  U^{(2)}(1)\times SU(\nd-n_1)$ at the lower scale $\mu\sim m\ll\lm$ {\it in the weak coupling regime}, see \eqref{(2.1.6)} below, $1\leq \no<\nd=N_F-N_c$, (this stage is qualitatively similar to those in section 4.1)
\bq
\sqrt{2}\, X^{(2)}_{U(1)}=a_{2}\,{\rm diag}(\,\underbrace{\,1}_{\no}\,;\, \underbrace{\,c_2}_{\nd-\no}\,;\,\underbrace{0}_{\bb}\,),\quad c_2=-\,\frac{\no}{\nd-\no}\,,\quad \langle a_{2}\rangle=\frac{N_c}{\nd}\, m\,.\label{(2.1.3)}
\eq

As a result, all original electric quarks with $SU(\nd-\no)$ colors and $N_F$ flavors have masses $m_2=\langle m- a_1-c_2 a_2\rangle=m\,N_c/(\nd-\no)$, and all original electrically charged hybrids $SU(\nd)/[SU(\no)\times SU(\nd-\no)\times U(1)]$ also have masses $(1-c_2)\langle a_2\rangle= m_2$, and so all these particles decouple as heavy at scales $\mu\lesssim m$. But the original electric quarks with $SU(\no)$ colors and $N_F$ flavors have masses $\langle m- a_1- a_2\rangle=0$, they remain massless at $\mx\ra 0$ and survive at $\mu\lesssim m$. Therefore, there remain two lighter non-Abelian sectors $SU(\no)\times SU(\nd-\no)$. As for the first one, it is the IR free ${\cal N}=2\,\, SU(\no)$ SQCD with $N_F$ flavors of original electric quarks massless at $\mx\ra 0\,,\,\, N_F>2\no,\,\, \no<\nd\,$. As for the second one, it is the ${\cal N}=2\,\, SU(\nd-\no)$ SYM with the scale factor $\Lambda^{SU(\nd-{\rm n}_1)}_{{\cal N}=2\,\, SYM}$ of its gauge coupling, compare with $\langle S\rangle_{N_c}$ in \eqref{(2.1)},
\bq
\Bigl (\Lambda^{SU(\nd-\no)}_{{\cal N}=2\,\, SYM}\Bigr )^{2(\nd-\no)}=\frac{\Lambda_{SU(\nd)}^{2\nd-N_F}
(m-a_1-c_2 a_2)^{N_F}}{[\,(1-c_2)a_2\,]^{\,2\no}}\,,\label{(2.1.4)}
\eq
\bbq
m_2=\langle m-a_1-c_2 a_2\rangle=\langle\, (1-c_2) a_2\,\rangle=\frac{N_c}{\nt-N_c}\,m=-m_1\,,
\eeq
\bbq
\langle\Lambda^{SU(\nd-\no)}_{{\cal N}=2\,\, SYM}\rangle^2=m^2_2\Bigl (\frac{m_2}{\Lambda_{SU(\nd)}}\Bigr )^{\frac{2N_c-N_F}{\nd-\no}}\ll m^2\,,\quad\langle S\rangle_{\nd-\no}=\mx(1+\delta_2)\langle\Lambda^{SU(\nd-\no)}_{{\cal N}=2\, SYM}\rangle^2\,,
\eeq
where $\Lambda_{SU(\nd)}$ is the scale factor of the $SU(\nd)$ gauge coupling at the scale $\mu=\lm/(\rm several)$ in \eqref{(2.1.2)}, after integrating out heaviest particles with masses $\sim\lm$ (it will be determined below in \eqref{(2.1.15)}).\\

{\bf 3)}\, At the third stage, to avoid unphysical $g^2(\mu<\Lambda^{SU(\nd-\no)}_{{\cal N}=2\,\, SYM})<0$ of  UV free $SU(\nd-\no)\,\, {\cal N}=2$ SYM, the field $X^{adj}_{SU(\nd-\no)}$ is higgsed, $\langle X^{adj}_{SU(\nd-\no)}\rangle\sim \langle\Lambda^{SU(\nd-{\rm n}_1)}_{{\cal N}=2\,\, SYM}\rangle$, breaking $SU(\nd-\no)$ in a well known way \cite{DS}, $SU(\nd-\no)\ra U^{\nd-\no-1}(1)$. All original pure electrically charged adjoint gluons and scalars of ${\cal N}=2\,\,SU(\nd-\no)$ SYM acquire masses $\sim\langle\Lambda^{SU(\nd-{\rm n}_1)}_{{\cal N}=2\,\, SYM}\rangle$ and decouple at lower energies $\mu<\langle\Lambda^{SU(\nd-{\rm n}_1)}_{{\cal N}=2\,\, SYM}\rangle$. Instead, $\nd-\no-1$ light composite pure magnetic monopoles (massless at $\mx\ra 0$) are formed, $M_i,\,{\ov M}_i,\,\, i=1,...,\nd-\no-1$, with their $SU(\nd-\no)$ adjoint magnetic charges. The factor $\nd-\no$ in the overall multiplicity of considered br2 vacua originates from the multiplicity of vacua of this ${\cal N}=2\,\,SU(\nd-\no)$ SYM. Besides, the appearance of two single roots with $(e^{+}-e^{-})\sim \langle\Lambda^{SU(\nd-\no)}_{{\cal N}=2\,\, SYM}\rangle$, of the curve \eqref{(1.2)} in these br2 vacua is connected just with this ${\cal N}=2\,\,SU(\nd-\no)$ SYM \cite{DS}. Other $\nd-\no-1$ double roots originating from this ${\cal N}=2$ SYM sector are unequal double roots corresponding to $\nd-\no-1$ pure magnetic monopoles $M_{\rm n}$ massless at $\mx\ra 0$.\\

{\bf 4)}\,  All $\bb$ dyons ${\ov D}_j, D_j$ are higgsed at the scale $\sim (\mx\lm)^{1/2},\,\,\langle D_j\rangle=\langle {\ov D}_j\rangle\sim (\mx\lm)^{1/2}\gg (\mx m)^{1/2}$. As a result, $\bb$ long ${\cal N}=2\,\, U(1)$ multiplets of massive photons are formed (including  $U^{(1)}(1)$ with its scalar $a_1$), all with masses $\sim (\mx\lm)^{1/2}$. No massless particles remain in this sector at $\mx\neq 0$.

The set $\bb$ of these higgsed dyons with the $SU(2N_c-N_F)$ adjoint magnetic charges is mutually non-local with all original $SU(2N_c-N_F)$ pure electrically charged particles with largest masses $\sim\lm$. Therefore, all these heaviest electric particles are confined, the string tension is $\sigma^{1/2}_D\sim (\mx\lm)^{1/2}$. All these heaviest confined particles will form a large number of hadrons with masses $\sim\lm$. But this confinement is weak, in the sense that the tension of the confining string is much smaller than particle masses, $\sigma^{1/2}_{D}\sim (\mx\lm)^{1/2}\ll\lm\,$.\\

{\bf 5)}\, $\no$ out of $N_F$ electric quarks of IR free ${\cal N}=2\,\, SU(\no)$ SQCD are higgsed at the scale $\mu\sim (\mx m)^{1/2}$ in the weak coupling region, with $\langle{\ov Q}_k^{\,a}\rangle=\langle Q^k_a\rangle\sim \delta^k_a\,(\mx m)^{1/2},\,\, a=1,...,\no,\,\, k=1,...,N_F$. As a result, the whole electric group $SU(\no)$ is broken and $\no^2$ long ${\cal N}=2$ multiplets of massive electric gluons are formed (including  $U^{(2)}(1)$ with its scalar $a_2$), all with masses $\sim (\mx m)^{1/2}\ll (\mx\lm)^{1/2}$ (for simplicity, here and everywhere below in similar cases we ignore logarithmic factors $\sim g$, these are implied where needed). The global flavor symmetry is broken spontaneously at this scale, $U(N_F)\ra U(\no)\times U(\nt)$, and $2\no\nt$ massless Nambu-Goldstone multiplets are formed in this sector (in essence, these are quarks $Q^k_a,\, {\ov Q}^{\,a}_{k},\,a=1,...,\no,\,\, k=\no+1,...,N_F$ ). This is a reason for the origin of the factor $C^{\,\no}_{N_F}$ in multiplicity of these br2 vacua. No additional massless particles remain in this sector at $\mx\neq 0$.\\

{\bf 6)}\,  $\nd-\no-1$ magnetic monopoles are higgsed at the lowest scale, $\langle{\ov M}_{\rm n}\rangle=\langle M_{\rm n}\rangle\sim (\mx\langle\Lambda^{SU(\nd-{\rm n}_1)}_{{\cal N}=2\,\, SYM}\rangle)_{,}^{1/2}$ and $\nd-{\rm n}_1-1$ long ${\cal N}=2$ multiplets of massive $U^{\nd-\no-1}(1)$ photons are formed, all with masses $\sim (\mx\langle\Lambda^{SU(\nd-{\rm n}_1)}_{{\cal N}=2\,\, SYM}\rangle)^{1/2}\ll (\mx m)^{1/2}$. All original pure electrically charged particles with non-singlet $SU(\nd-{\rm n}_1)$ charges and masses either $\sim m$ or $\sim\langle\Lambda^{SU(\nd-{\rm n}_1)}_{{\cal N}=2\,\, SYM}\rangle$ are weakly confined, the string tension is $\sigma^{1/2}_{\rm SYM}\sim (\mx\langle\Lambda^{SU(\nd-{\rm n}_1)}_{{\cal N}=2\,\, SYM}\rangle)^{1/2}$. No massless particles remain in this sector at $\mx\neq 0,\, m\neq 0$.\\

As a result of all described above, the lowest energy superpotential at the scale $\mu=\langle\Lambda^
{SU(\nd-\no)}_{{\cal N}=2\,\, SYM}\rangle/(\rm several)$ (and e.g. at very small $\mx\ll\langle\Lambda^
{SU(\nd-\no)}_{{\cal N}=2\,\, SYM}\rangle^2/\lm$) can be written in these br2 vacua as 
\bq
\w^{\,\rm low}_{\rm tot}=\w^{\,(SYM)}_{SU(\nd-\no)}+\w^{\,\rm low}_{\rm matter}+\dots\,,\quad
\w^{\,\rm low}_{\rm matter}=\w_{SU(\no)}+\w_{D}+\w_{a_1,\,a_2}\,,\label{(2.1.5)}
\eq
\bbq
\w^{\,(SYM)}_{SU(\nd-\no)}=(\nd-\no)\wmu\Bigl (\Lambda^{SU(\nd-{\rm n}_1)}_{{\cal N}=2\,\, SYM}\Bigr )^2+\w^{\,(M)}_{SU(\nd-\no)}\,,\quad \wmu=\mx(1+\delta_2)\,,
\eeq
\bbq
\w^{\,(M)}_{SU(\nd-\no)}= - \sum_{n=1}^{\nd-\no-1} {\tilde a}_{M,\,\rm n}\Biggl [\, {\ov M}_{\rm n} M_{\rm n}+\wmu\langle\Lambda^{SU(\nd-\no)}_{{\cal N}=2\,\,SYM}\rangle\Biggl (1+O\Bigl (\frac{\langle\Lambda^{SU(\nd-\no)}_{{\cal N}=2\,\,SYM}\rangle}{m}\Bigr )\Biggr )\,d_{\rm n}\,\Biggr ]\,,
\eeq
\bbq
\w_{SU(\no)}=(m-a_1-a_2)\,{\rm Tr}\,({\ov Q} Q)_{\no}-\,{\rm Tr}\,({\ov Q}\sqrt{2} X^{\rm adj}_{SU(\no )} Q)_{\no}+\wmu{\rm Tr}\,(X^{\rm adj}_{SU(\no)})^2\,,
\eeq
\bbq
\w_{D}=(m-c_1 a_1)\sum_{j=1}^{\bb}{\ov D}_j D_j-\sum_{j=1}^{\bb} a_{D,j}\,{\ov D}_j D_j\,-
\,\mx\lm\sum_{j=1}^{\bb}\omega^{j-1}\,a_{D,j}+\mx L \sum_{j=1}^{\bb} a_{D,j}\,,
\eeq
\bbq
\w_{a_1,\,a_2}=\frac{\mx}{2}(1+\delta_1)\frac{\nd N_c}{\bb} a_1^2+\frac{\wmu}{2}\frac{\no \nd}{\nd-\no} a_2^2+\mx N_c \delta_3 a_1(m-c_1 a_1)+\mx N_c\delta_4 (m-c_1 a_1)^2+O\Bigl (\mx a_{D,j}^2 \Bigr ),
\eeq
where coefficients $d_i=O(1)$ in \eqref{(2.1.5)} are known from \cite{DS}), and dots in \eqref{(2.1.5)} denote smaller power suppressed corrections (these are always implied and dots are omitted below in the text), . It is important to notice that the unbroken $Z_{2N_c-N_F\geq 2}$ symmetry restricts strongly their possible values. The matter is that e.g. power suppressed corrections in $\w_{a_1}$ like $\sim\mx a_1^2(a_1/\lm)^{N}$ originate from $SU(N_c)\ra SU(\nd)\times U^{(1)}(1)\times SU(2N_c-N_F)$ at the scale $\sim\lm$ and they do not know the number $\no$ originating only at much lower scales $\mu\sim m\ll\lm$. Then the unbroken $Z_{2N_c-N_F\geq 2}$ symmetry requires that $N=(2N_c-N_F) k,\, k=1,2,3...$.

Remind that charges of fields and parameters entering \eqref{(2.1.5)} under $Z_{\bb}=\exp\{i\pi/(2N_c-N_F)\}$ transformation are\,: $q_{\lambda}=q_{\theta}=1,\,\, q_{X_{SU(\no)}^{\rm adj}}=q_{a_1}=q_{a_2}=q_{a_{D,j}}=q_{{\tilde a}_{M,\rm n}}=q_{\rm m}=q_{L}=2,\,\,q_{Q}=q_{\ov Q}=q_{D_j}=q_{{\ov D}_j}=q_{M_{\rm n}}=q_{{\ov M}_{\rm n}}=q_{\lm}=0,\,\, q_{\mx}=-2$. The non-trivial $Z_{\bb\geq 2}$ transformations change only numerations of dual scalars $a_{D,j}$ and dyons $D_j, {\ov D}_j$ in \eqref{(2.1.5)}, so that $\int d^2\theta\,\w_{\rm tot}^{\,\rm low}$ is $Z_{\bb}$-invariant.

All (massless at $\mx\ra 0$) $\bt=2N_c-N_F$ dyons $D_j$, $\no<\nd$ quarks $Q^k$ and $\nd-\no-1$ monopoles $M_{\rm n}$ in \eqref{(2.1.5)} are higgsed at $\mx\neq 0$. As a result, we obtain from \eqref{(2.1.5)} (neglecting power corrections)
\bq
\langle a_1\rangle=\frac{1}{c_1}\,m=-\frac{\bb}{\nd}\,m\,,\,\, \langle a_2\rangle=\langle m-a_1\rangle=
\frac{N_c}{\nd}\,m\,,\,\, \langle X^{\rm adj}_{SU({\rm n}_1)}\rangle=\langle a_{D,\,j}\rangle=\langle{\tilde a}_{M,\rm n}\rangle=0\,,\label{(2.1.6)}
\eq
\bbq
\langle{\ov M}_{\rm n} M_{\rm n}\rangle=\langle{\ov M}_{\rm n}\rangle\langle M_{\rm n}\rangle\approx - \wmu\langle\Lambda^{SU(\nd-\no)}_{{\cal N}=2\,\, SYM}\rangle\,d_{\rm n}\,,\,\, d_{\rm n}=O(1),\quad \langle{\ov D}_j D_j\rangle=\langle{\ov D}_j\rangle\langle D_j\rangle\approx -\mx\lm\,\omega^{j-1}\,,
\eeq
\bbq
\langle\Qo\rangle_{\no}=\langle{\ov Q}^1_1\rangle\langle Q^1_1\rangle\approx\wmu\frac{\nd}{\nd-\no}\langle a_2\rangle\approx \wmu\frac{N_c}{\nd-\no}\,m,\,\, \langle\Qt\rangle_{\no}=\sum_{a=1}^{\no}\langle{\ov Q}^{\,a}_2\rangle\langle Q^2_a\rangle=0,\,\,\langle S\rangle_{\no}=0
\eeq
(the dyon condensates are dominated by terms $\sim\mx\lm$ plus smaller terms $\sim\mx m$, see Appendix B).

Now, in a few words, it is not difficult to check that the variant of \eqref{(2.1.5)} with the first term of $\w_D$ of the form $m\sum_{j=1}^{\bb}{\ov D}_j D_j$ (i.e. zero $U^{(1)}(1)$ charge of dyons) will result in $\langle\partial\w^{\,\rm low}_{\rm matter}/\partial L\rangle=\mx m (\bb)/2\neq 0$, and this is the internal inconsistency.\\

The $\bb$ unequal double roots $e^{(D)}_j\approx\omega^{j-1}\lm$ of the curve \eqref{(1.2)} correspond to these BPS dyons formed at the scale $\mu\sim\lm$. Together with $\no$ equal double roots $e^{(Q)}_k= - m$ of original pure electric quarks (higgsed at small $\mx\neq 0$) from ${\cal N}=2 \,\, SU(\no)$ SQCD, and $\nd-\no-1$ unequal double roots of $SU(\nd-\no)$ adjoint pure magnetic monopoles from ${\cal N}=2 \,\,  SU(\nd-\no)$ SYM (formed at the scale $\mu\sim\langle\Lambda^{SU(\nd-\no)}_{{\cal N}=2\,\, SYM}\rangle$), they constitute the total set of $N_c-1$ double roots of the curve \eqref{(1.2)}.

The short additional explanations will be useful at this point. As was pointed out in Introduction, the spectral curve \eqref{(1.2)} can be used, in general, to determine the masses of BPS charged particles (and the multiplicities and charges of massless charged BPS particles in particular) in the formal limit $\mx\ra 0$ only. But in cases considered in this paper when there are exactly massless Nambu-Goldstone particles due to the spontaneous breaking of the global $U(N_F)$ flavor symmetry in many vacua, some additional infrared regularization is needed to really have only $N_c-1$ exactly massless at $\mx=0$ charged BPS particles and corresponding them $N_c-1$ well defined unequal double roots of the $SU(N_c)$ curve \eqref{(1.2)}. We have in mind that, e.g. in br2 vacua considered, all charged particles of the ${\cal N}=2\,\, SU(\no)$ subgroup will be massless at $\mx=0$ for equal mass quarks, this corresponds to $\no$ equal double roots $e^{(Q)}_k= - m,\,\, k=1...\no$ of the curve \eqref{(1.2)}. To have really only $\no$ exactly massless at $\mx=0$ charged BPS particles in this $SU(\no)$ sector, we have to split slightly the quark masses. For instance, in br2 vacua of this section it is sufficient to split slightly the masses of quarks from $SU(\no)$ with the first $\no$ flavors. I.e.\,: $m_k= m+{\delta m}_k,\,\, k=1...N_F,\,\, {\delta m}_k\neq 0,\,\,  {\delta m}_k\sim {\delta m}_n, \,\, \sum_{k=1}^{\no}{\delta m}_k=0$, for $k, n=1...\no$\,; while ${\delta m}_k=0$ for $k=\no+1\, ...\, N_F$. The mass splittings ${\delta m}_k,\,\, k=1...\no$ are arbitrary small but fixed to be unequal and nonzero. In this case, $\langle X^{\rm adj}_{SU({\rm n}_1)}\rangle$ in \eqref{(2.1.6)} will be nonzero, $\langle{\sqrt 2} X^{\rm adj}_{SU({\rm n}_1)}\rangle={\rm diag}(\,{\delta m}_{1}\,...\,{\delta m}_{\no}\,)$, and $SU(\no)\ra U^{\no-1}(1)$. The exactly massless charged particles at $\mx=0$ will be only $\no$ quarks $Q^{i=a}_a,\, {\ov Q}^{\,a}_{j=a},\, a=1...\no$ (higgsed at $0<\mx\ll {\delta m}_k$), while all other charged particles of $SU(\no)$ (including all previously massless Nambu-Goldstone particles) will acquire nonzero masses $O({\delta m})$. And the $SU(N_c)$ curve \eqref{(1.2)} will have $\no$ slightly split quark double roots $e^{(Q)}_k= - m_k,\,\, k=1...\no$, and $N_c-1$ unequal double roots on the whole. And this shows now most clearly that {\it there are no other charged BPS solitons massless at $\mx\ra 0$}. This infrared regularization will be always implied in the text below when we will speak about only $N_c-1$ double roots of the $SU(N_c)$ curve \eqref{(1.2)} at $\mx\ra 0$ (or about $N_c$ double roots of the $U(N_c)$ curve \eqref{(1.2)} in vs vacua of section 6).\\

Our purpose now is to calculate $\delta_2$ in order to find the leading terms of quark condensates $\langle{\rm Tr}\,{\ov Q} Q\rangle_{\no}$ and the monopole condensates $\langle M_i\rangle$ in \eqref{(2.1.6)}, and to calculate $\Lambda_{SU(\nd)}$ in \eqref{(2.1.4)}. For this, on account of leading terms $\sim\mx m^2$ and the leading power correction $\sim\langle S\rangle_{SU(\nd-\no)}$, we write, see \eqref{(1.1)},\eqref{(2.1.5)},\eqref{(2.1.4)},
\bq
\langle\w^{\,\rm low}_{\rm tot}\rangle=\langle\w^{\,(SYM)}_{SU(\nd-\no)}\rangle+\langle\w^{\,\rm low}_{\rm matter}\rangle\,,\quad \wmu=\mx(1+\delta_2)\,,\label{(2.1.7)}
\eq
\bbq
\langle\w^{\,(SYM)}_{SU(\nd-\no)}\rangle=\wmu\langle{\rm Tr\,}(X^{adj}_{SU(\nd-\no)})^2\rangle=(\nd-\no)
\langle S\rangle_{SU(\nd-\no)}=(\nd-\no)\wmu\langle\Lambda^{SU(\nd-\no)}_{{\cal N}=2\,\, SYM}\rangle^2\,,
\eeq
where $\langle\w^{\,(SYM)}_{SU(\nd-\no)}\rangle$ is the contribution of the ${\cal N}=2\,\,SU(\nd-\no)$ SYM.

First, to determine $\delta_{1,2}=O(1)$ in \eqref{(2.1.5)}, it will be sufficient to keep only the leading term $\langle\w^{\,\rm low}_{\rm matter}\rangle\sim\mx m^2$ in \eqref{(2.1.7)},\eqref{(2.1.5)} and to neglect even the leading power correction $\langle\w^{\,(SYM)}_{SU(\nd-\no)}\rangle$,
\bq
\langle\w^{\,\rm low}_{\rm tot}\rangle\approx\langle\w^{\,\rm low}_{\rm matter}\rangle=\frac{\mx}{2}(1+\delta_1)\frac{\nd N_c}{\bt}\Bigl [ \langle a_1\rangle^2=\frac{\bt^2}{\nd^{\,2}}\,m^2\Bigr ]+\frac{\mx}{2}(1+\delta_2)\frac{{\rm n}_1\nd}{\nd-{\rm n}_1}\Bigl [\langle a_2\rangle^2=\frac{N_c^{\,2}}{\nd^{\,2}}\,m^2 \Bigr ].\quad\,\,\label{(2.1.8)}
\eq

On the other hand, the exact total effective superpotential $\w^{\,\rm eff}_{\rm tot}$ which accounts explicitly for all anomalies and contains {\it only} quark bilinears $\Pi^i_j=({\ov Q}_j Q^i)_{N_c}$ looks as, see \cite{ch4,ch6} and \eqref{(1.3)},\eqref{(1.4)},
\bq
\w^{\,\rm eff}_{\rm tot}(\Pi)=m\,{\rm Tr}\,({\ov Q} Q)_{N_c}-\frac{1}{2\mx}\Biggl [ \,\sum_{i,j=1}^{N_F} ({\ov Q}_j Q^i)_{N_c}({\ov Q}_{\,i} Q^j)_{N_c}-\frac{1}{N_c}\Bigl ({\rm Tr}\,({\ov Q} Q)_{N_c}\Bigr )^2\Biggr ]-\nd S_{N_c} \,, \label{(2.1.9)}
\eq
\bbq
\langle S\rangle_{N_c}=\Bigl(\frac{\langle\det\qq\rangle=\langle\Qo\rangle^{\no}\langle\Qt
\rangle^{\nt}}{\lm^{2N_c-N_F}\mx^{N_c}}\Bigr )^{\frac{1}{N_F-N_c}}_{N_c}\,.
\eeq

Kipping only the leading terms $\sim\mx m^2$ in \eqref{(2.1.9)} and using \eqref{(2.1)} obtained from \eqref{(2.1.9)},\eqref{(1.4)},
\bq
\langle\w^{\,\rm eff}_{\rm tot}\rangle\approx\frac{1}{2}\frac{\nt N_c}{N_c-\nt}\mx m^2\approx -\,\frac{1}{2}\frac{\nt N_c}{\nd-\no}\mx m^2\,.\label{(2.1.10)}
\eq

From $\langle\w^{\,\rm low}_{\rm tot}\rangle=\langle\w^{\,\rm eff}_{\rm tot}\rangle$ in \eqref{(2.1.8)},\eqref{(2.1.10)},
\bq
-2\nd N_c=\nd\bt\delta_1+\no(N_c\delta_2-\bt\delta_1)\,.\label{(2.1.11)}
\eq
But all coefficients $\delta_{i},\,\, i=1...4,$ in \eqref{(2.1.2)} originate from the large scale $\sim\lm$, from integrating out the heaviest fields with masses $\sim\lm$, and they do not know the number $\no$ which originates only from the behavior of $X^{\rm adj}_{SU(\nd)}$ in \eqref{(2.1.2)} at the much lower scale $\sim\langle a_2\rangle\sim m\ll\lm$, so that
\bq
N_c\delta_2=\bt\delta_1\,,\quad \delta_1=-\,\frac{2N_c}{\bt}\,,\quad \delta_2= -2\,,\quad \bt=2N_c-N_F\,.\label{(2.1.12)}
\eq
(Besides, it is shown in Appendix B that $\delta_3=0$ in \eqref{(2.1.2)},\eqref{(2.1.5)}, see \eqref{(B.4)}).\\

Therefore, we obtain from $\langle\partial\w^{\,\rm low}_{\rm matter}/\partial a_2\rangle=0$, see \eqref{(2.1.5)},\eqref{(2.1.6)},\eqref{(2.1.12)}, for the leading term of $\langle\Qo\rangle_{\no}\rangle$ in br2 vacua, compare with \eqref{(2.1)},
\bq
\langle{\rm Tr}\,({\ov Q} Q)\rangle_{\no}=\no\langle\Qo\rangle_{\no}+\nt\langle\Qt\rangle_{\no}=
\no\langle\Qo\rangle_{\no}\approx\frac{{\rm n}_1 N_c}{N_c-\nt}\,\mx m\approx\no\mx m_1\,,\label{(2.1.13)}
\eq
\bbq
\langle\Qo\rangle^{SU(N_c)}_{\no}=\sum_{a=1}^{\no}\langle{\ov Q}_1^{\,a} Q^1_a\rangle=\langle{\ov Q}_1^{\,1} \rangle\langle Q^1_1\rangle\approx\mx m_1\approx\langle\Qt\rangle^{SU(N_c)}_{N_c},\,\, \langle\Qt\rangle_{\no}=\sum_{a=1}^{\no}\langle{\ov Q}_2^{\,a} \rangle\langle Q^2_a\rangle=0.
\eeq

The whole group $SU(\no)\times U^{(2)}(1)$ is higgsed by $\no$ electric quarks (the power correction from the SYM part in \eqref{(2.1.5)} is accounted for in \eqref{(2.1.14)}, see \eqref{(A.24)})
\bq
\frac{1}{\mx}\langle{\ov Q}_k^{\,a}\rangle\langle Q^k_a\rangle\approx\delta^k_a\, m_1 \Bigl [\, 1+\frac{\bb}{\nt-N_c}\Bigl (\frac{m_1}{\lm}\Bigr )^{\frac{2N_c-N_F}{\nt-N_c}}\,\Bigr ]\,,\quad \wmu\langle S\rangle_{\no}=\langle\Qo\rangle_{\no}\langle\Qt\rangle_{\no}=0,\quad \label{(2.1.14)}
\eq
\bbq
a=1...\no\,,\quad k=1...N_F\,,\quad m_1=\frac{N_c}{N_c-\nt}\, m\,,
\eeq
and $\no^2$ long ${\cal N}=2$ multiplets of massive gluons with masses $\sim (\mx m)^{1/2}$ are formed (including $U^{(2)}(1)$ with its scalar $a_2$). It is worth to remind that at sufficiently small  $\mx$ the ${\cal N}=2$ SUSY remains unbroken at the level of particle masses $O(\sqrt{\mx})$, and only small corrections $O(\mx)$ to masses of corresponding scalar superfields break ${\cal N}=2$ to ${\cal N}=1$, see e.g. \cite{VY2}.
\footnote{\,
In comparison with the standard ${\cal N}=1$ masses $\sim (\mx m_1)^{1/2}$ of $\no^2$ gluons and $\no^2$ of their ${\cal N}=1$ (real) scalar superpartners originating from D-terms, the same masses of additional $2\no^2$ (complex) scalar superpartners of long ${\cal N}=2$ gluon multiplets originate here from F-terms in \eqref{(2.1.5)}: $\sim\sum_{a,b=1}^{\no} X^a_b \Pi^b_a$, where $\no^2$ of $X^a_b,\, \langle X^a_b\rangle=0$, include $(\no^2-1)$ of $X^{adj}_{SU(\no)}$ and one ${\hat a}_2=a_2-\langle a_2\rangle$, and $\no^2$ "pions" are $\Pi^b_a=\sum_{i=1}^{\no}({\ov Q}^{\,b}_i Q_a^i),\, \langle\Pi^b_a\rangle\approx \delta^b_a\,\mx m_1$. The Kahler terms of all $\no^2$ pions $\Pi^b_a$ look as $K_{\Pi}=2\,{\rm Tr}\,\sqrt{\Pi^\dagger \Pi}\,$.
}

As it should be, there remain $2\no\nt$ {\it exactly massless} Nambu-Goldstone multiplets $({\ov Q}_2 Q^1)
_{\no}, ({\ov Q}_1 Q^2)_{\no}$ (these are in essence the quarks ${\ov Q}$ and  $Q$ with $\nt=N_F-\no$ flavors and $SU(\no)$ colors). The factor $\nd-\no$ in the multiplicity $N_{\rm br2}=(\nd-\no)C^{\,\no}_{N_F}$ of these vacua originates from $SU(\nd-\no)$ SYM, while $C^{\,\no}_{N_F}$ is due to the spontaneous flavor symmetry breaking in the $SU(\no)$ sector.\\

The value of the quark condensate in \eqref{(2.1.14)} is verified by two independent calculations using
values of double roots of the curve \eqref{(1.2)} and \eqref{(2.2.10)}, see \eqref{(A.31)},\eqref{(B.14)}, and {\it these calculations are valid only for charged BPS particles massless at $\mx = 0$}. This right value in \eqref{(2.1.14)} is really of importance because it is a check of the main assumption "A" formulated in Introduction, i.e. the BPS properties of original quarks in \eqref{(1.1)} at $\mx\ra 0$. Indeed, if original quarks were not BPS particles, then their mass terms in \eqref{(2.1.2)},\eqref{(2.1.5)} will receive e.g. the additional contributions $c_{a_1} a_1 {\rm Tr}(\qq)_{\nd},\,\, c_{a_1}=O(1)$ and/or $c_m m{\rm Tr}(\qq)_{\no},\,\, c_m=O(1)$, from integrated out heavier particles (or from elsewhere). The value of $\langle a_2\rangle$ in \eqref{(2.1.6)} will be then changed, and this will spoil the correct value of the quark condensate in \eqref{(2.1.14)}. (And the same for the $U(N_c)$ theory in section 2.2 below, see \eqref{(2.2.5)},\eqref{(2.2.8)},\eqref{(A.27)},\eqref{(A.29)}). On the other hand, the additional contributions $\sim\mx\delta_i$ into masses of adjoint scalars in \eqref{(2.1.2)} do not
contradict the ${\cal N}=2$ BPS properties of these scalars as these additional contributions to their masses are only due to the explicit breaking ${\cal N}=2\ra {\cal N}=1$ by $\mx\neq 0$.\\

Now, in short about the variant of \eqref{(2.1.5)} with the first term of $\w_D$ of the form $[\,-c_1 a_1\sum_{j=1}^{\bt}{\ov D}_j D_j\,]$, i.e. with zero $SU(N_c)$ baryon charge of dyons $D_j$. With such a $\w_D$ term in \eqref{(2.1.5)}, this will result in $\langle a_1\rangle=0,\,\,\langle a_2\rangle=m$. Then, instead of \eqref{(2.1.12)},\eqref{(2.1.13)} we will obtain
\bbq
(1+\delta_2)=-\frac{N^2_c}{\nd^{\,2}}\,\, \ra\,\,  \langle\Qo\rangle^{SU(N_c)}_{\no}\approx\mx(1+\delta_2)
\frac{\nd}{\nd-\no}\langle a_2\rangle\approx -\frac{N_c^2}{\nd(\nd-\no)}\mx m\approx\frac{N_c}{\nd}\mx m_1\,,
\eeq
and this disagrees with the {\it independent} calculation of $\langle\Qo\rangle^{SU(N_c)}_{\no}\approx \mx m_1$ in this $SU(N_c)$ theory using the roots of the curve \eqref{(1.2)}, see \eqref{(A.31)} in Appendix A or \eqref{(B.14)} in Appendix B.

And finally, to determine $\Lambda_{SU(\nd)}$ in \eqref{(2.1.4)}, we account now for the leading power corrections $\langle\,\delta \w^{\,\rm eff}_{\rm tot}\rangle\sim\langle S\rangle_{N_c}\sim\langle\,\delta \w_{\rm tot}^{\,\rm low}\rangle\sim\langle S\rangle_{\nd-\no}$, see \eqref{(2.1)},\eqref{(2.1.4)},\eqref{(2.1.9)},\eqref{(2.1.12)},
\bbq
\langle\,\delta \w^{\,\rm eff}_{\rm tot}\rangle=(N_c-\nt)\langle S\rangle_{N_c}\approx (N_c-\nt)\mx\, m^2_1\Bigl (\frac{m_1}{\Lambda_2}\Bigr )^{\frac{2 N_c-N_F}{\nt-N_c}}\,,\quad m_1=\frac{N_c}{N_c-\nt}\,m\,,
\eeq
\bbq
\langle\,\delta \w^{\,\rm low}_{\rm tot}\rangle=\langle\w^{\,(SYM)}_{SU(\nd-\no)}\rangle=\wmu\langle\,{\rm Tr\,} ( X^{adj}_{SU(\nd-\no)} )^2\,\rangle=(\nd-\no)\langle S\rangle_{SU(\nd-\no)}= (\nd-\no)\wmu\langle\Lambda^{SU(\nd-\no)}_{{\cal N}=2\,\, SYM}\rangle^2
\eeq
\bbq
\approx(N_c-\nt)\mx m^{2}_2\Bigl (\frac{m_2}{\Lambda_{SU(\nd)}}\Bigr )^{\frac{2N_c-N_F}{\nt-N_c}},\quad \wmu=\mx(1+\delta_2)=-\mx\,,\quad m_2=\frac{N_c}
{\nd-\no}\,m = - m_1\,,
\eeq
\bq
\langle\,\delta \w^{\,\rm eff}_{\rm tot}\rangle=\langle\,\delta \w^{\,\rm low}_{\rm tot}\rangle\quad\ra\quad \Lambda_{SU(\nd)}=-\lm\,.\label{(2.1.15)}
\eq

\vspace*{1mm}

The whole spectrum of masses in the br2 vacua considered and, in particular, the spectrum of non-zero masses arising from \eqref{(2.1.5)} at small $\mx\ll\langle\Lambda^{SU(\nd-\no)}_{{\cal N}=2\,\, SYM}\rangle$ was described above in this section. It is in accordance with $N_c-1$ double roots of the curve \eqref{(1.2)} in these br2-vacua at $\mx\ra 0$. Of them: $\,2N_c-N_F$ unequal roots corresponding to $2N_c-N_F$ dyons $D_j$, then $\no$ equal roots corresponding to $\no$ pure electric quarks $Q^k$ of $SU(\no)$ (higgsed at $\mx\neq 0$), and finally $\nd-\no-1$ unequal roots corresponding to $\nd-\no-1$ pure magnetic monopoles $M_i$ of $SU(\nd-\no)\,\, {\cal N}=2$ SYM. Remind that two single roots $(e^{+}-e^{-})\sim\langle\Lambda^{SU(\nd-
{\rm n}_1)}_{{\cal N}=2\,\, SYM}\rangle$ of the curve \eqref{(1.2)} originate in these br2 vacua from the ${\cal N}=2\,\, SU(\nd-\no)$ SYM sector.

As it is seen from the above, {\it the only exactly massless at $m\neq 0,\, \mx\neq 0$ particles} in the Lagrangian \eqref{(2.1.5)} are $2 \no\nt$ (complex) Nambu-Goldstone multiplets originating from the spontaneous breaking of the global flavor symmetry, $U(N_F)\ra U(\no)\times U(\nt)$.

The calculation of the leading power correction $\sim\mx\langle\Lambda^{SU(\nd-{\rm n}_1)}_{{\cal N}=2\,\, SYM}\rangle^2/m$ to the value of $\langle\Qo\rangle^{SU(N_c)}_{\no}$ in \eqref{(2.1.14)} is presented in Appendices A and B, and power corrections to the dyon condensates in \eqref{(2.1.6)} are calculated in Appendix B. These calculations give an additional {\it independent} check of a self-consistency of the whole approach.\\

According to reasonings given above in this section, the quarks with all $N_F$ flavors, $SU(2N_c-N_F)$ colors, and masses $\sim\lm$ are too heavy and {\it not higgsed}, i.e. $\langle Q^i_a\rangle=\langle{\ov Q}^{\,b}_j\rangle=0,\,\, a,b=\nd+1...N_c$. Indeed, at small $\mx$, at least all {\it light} $\,U^{\bb}(1)$ physical phases of all these quark fields $Q^i_a, {\ov Q}^{\,b}_j, a,b=\nd+1...N_c,\, i,j=1...N_F$, still fluctuate {\it freely and independently} in the whole interval from higher energies down to $\mu^{\rm low}_{\rm cut}=(\rm several)g(\mx\lm)^{1/2}\ll\lm$, so that the mean values of quarks fields {\it integrated over this energy range} look as
\bq
\langle{\ov Q}^{\,b}_j Q_a^i\rangle_{\mu^{\rm low}_{\rm cut}}=\langle{\ov Q}^{\,b}_j\rangle_{\mu^{\rm low}_{\rm cut}}\langle Q_a^i\rangle_{\mu^{\rm low}_{\rm cut}}=0,\quad a,b=\nd+1...N_c,
\quad \mu^{\rm low}_{\rm cut}=(\rm several)g(\mx\lm)^{1/2},\,\,\,\label{(2.1.16)}
\eq

For this reason, at the scale $\mu=\lm/(\rm several)$, {\it after all particles with masses $\sim\lm$ decoupled as heavy, there are no any additional contributions to particle masses in the Lagrangian \eqref{(2.1.2)}} from mean values of heavy quark fields. (Really, because $\nt>N_c,\, \no<\nd$ in these br2 vacua, it is clear beforehand that the quarks $Q^2, {\ov Q}_2$ in $\nt>N_c$ equal condensates $\langle\Qt\rangle_{N_c}$ are definitely not higgsed at all due to the rank restrictions, higgsed are the light quarks $Q^i_a, {\ov Q}^{\,a}_i,\, \, i,a=1...\no$ only, see \eqref{(2.1.14)} and the end of section 2.4).

Nevertheless, the {\it total mean values} of some heavy quark bilinears, $\langle{\ov Q}^{\,b}_j Q_a^i\rangle,\, a,b=\nd+1...N_c$ are non-zero (remind that, by definition, for any operator $O$, the expression $\langle O\rangle$ denotes its {\it total mean value  integrated from sufficiently high energies down to} $\mu^{\rm lowest}_{\rm cut}=0$). The largest among them in this section are the total mean values of $SU(\bb)$ adjoint but $U(N_F)$ singlet parts of heavy quark bilinears, see \eqref{(2.1.1)}
\bq
2\sum_{i=1}^{N_F}\sum_{A=1}^{(\bb)^{2}-1}T^{A}\langle{\rm Tr\,} {\ov Q}_i T^{A} Q^i\rangle=\mx\langle
\sqrt{2} X^{adj}_{SU(\bb)}\rangle\sim\mx\lm{\,\rm diag}(\,\underbrace{\,0}_{\nd}\,;\underbrace{\omega^0,\,
\omega^1...\,\omega^{\bb-1}}_{\bb})\,.\quad\,\,\,\, \label{(2.1.17)}
\eq

But non-zero mean values of heavy quark bilinears in \eqref{(2.1.17)} do not mean that these quarks are higgsed (\, i.e. $\langle Q^i_a\rangle=\langle{\ov Q}^{\,a}_i\rangle\neq 0\,$), because these total mean values of bilinears are {\it not factorizable}. The non-zero values of quark bilinears in \eqref{(2.1.17)} originate really, on the first stage, from effects of loops integrated over the energy range from high energies down to $\mu=\lm/(\rm several)$ (in the strong coupling regime at the scale $\mu\sim\lm$), transforming such bilinears operators of heavy quarks into corresponding $SU(\bb)$ adjoint parts of bilinears operators of light dyons in \eqref{(2.1.2)}, i.e. $\sim [\,{\ov D}_j D_j-(\bb)^{-1}\Sigma_D\,],\,\,\,\, \Sigma_D=\sum_{j=1}^{\bb} {\ov D}_j D_j$.  And the largest non-zero mean values $\sim\mx\lm$ in
\eqref{(2.1.17)} are formed finally not at the scale $\mu\sim\lm/(\rm several)$, because at least all $U^{\bb-1}(1)\times U^{(1)}(1)$ physical phases of ${\ov D}_j$ and $D_j$ fluctuate freely and independently in the whole energy interval $\mu^{\rm low}_{\rm cut} < \mu<\lm/(\rm several)$, so that $\langle{\ov D}_j\rangle_{\mu^{\rm low}_{\rm cut}}=\langle D_j\rangle_{\mu^{\rm low}_{\rm cut}}=\langle\Sigma_D\rangle_{\mu^{\rm low}_{\rm cut}}=0$. They are formed only at the much lower scale $\sim g(\mx \lm)^{1/2}\ll\lm$, in the weak coupling regime, from $SU(\bb)$ adjoint parts of {\it coherent condensates} of these {\it higgsed light dyons} (resulting in appearance of $\sim g(\mx\lm)^
{1/2}$ masses of all $U^{\bb-1}(1)\times U^{(1)}(1)$ ${\cal N}=2$ multiplets): $\,[\langle{\ov D}_j D_j-(\bb)^{-1}\Sigma_D\,\rangle ]=[\langle{\ov D}_j\rangle\langle D_j\rangle -(\bb)^{-1}\sum_{j=1}^{\bb}
\langle{\ov D}_j\rangle\langle D_j\rangle ]= - \mx\lm\,\omega^{j-1}$,  see \eqref{(2.1.6)},\eqref{(B.8)},\eqref{(B.15)}.\\

Let us emphasize also the following point. The largest {\it numerical values} $\langle X^{adj}_{SU(\bb)}
\rangle\sim\lm$ in \eqref{(2.1.1)} (remind that {\it these numbers are the total mean values, i.e. integrated down to $\mu^{\rm lowest}_{\rm cut}=0$}) giving the main contributions $\sim\lm$ to particle masses, do not mean really that they originate from the region of scales $\mu\sim\lm$ only. As it is seen from \eqref{(2.1.17)}, these numerical values originate from contributions of {\it the same energy regions which contribute to the total mean values} $\langle{\rm Tr\,} {\ov Q}_i T^{A} Q^i\rangle/\mx\sim\lm$ of heavy quarks. As was explained above, two energy regions are responsible for the result. The first is the region $\mu\sim\lm$ in which the loop effects transform the bilinear heavy quark operators into bilinear operators of light dyons. But this is not sufficient by itself, see \eqref{(2.1.16)}. The final non-zero answer originates only from the region of much lower energies $\mu\sim g(\mx\lm)^{1/2}$, from the contribution of the coherent condensate of higgsed dyons.\\

As it is seen from \eqref{(2.1)},\eqref{(2.4.1)}, the $SU(\bb)$ singlet parts of heavy quark condensates, $\langle{\ov Q}_j Q^i\rangle_{\bb}=\sum_{a=\nd+1}^{N_c}\langle{\ov Q}^{\,a}_j Q^i_{a}\rangle\sim\,
\delta^i_j\,\mx m\ll\mx\lm$, although much smaller, are also non-zero and of the same size as condensates of light higgsed quarks $\langle\Qo\rangle_{\nd}\approx\langle\Qo\rangle_{\no}=\langle{\ov Q}^1_1\rangle\langle Q^1_1\rangle\sim\mx m$ ( the condensates $\langle{\ov Q}_i Q^i\rangle_{\nd}$ of all flavors $i=1...N_F$ of quarks are holomorphic in $\mx$ and their values in \eqref{(2.4.1)},\eqref{(2.4.5)},\eqref{(2.4.6)} are valid at smaller $\mx\ll m$ as well),
\bq
\langle\Qt\rangle_{\bb}=\langle\Qt\rangle_{N_c}-\langle\Qt \rangle_{\nd}\approx\mx m_1\Biggl [1+\frac{\bb}{\nt-N_c}\Bigl (\frac{m_1}{\lm}\Bigr )^{\frac{\bb}{\nt-N_c}}\Biggr ) \Biggr ]\,,\label{(2.1.18)}
\eq
\bbq
\langle\Qo\rangle_{\bb}=\langle\Qo\rangle_{N_c}-\langle\Qo\rangle_{\nd}\approx - \mx m_1\Biggl [1+\frac{\bb}{\nt-N_c}\Bigl (\frac{m_1}{\lm}\Bigr )^{\frac{\bb}{\nt-N_c}}\Biggr ) \Biggr ]\,,
\eeq
\bbq
\langle{\rm Tr\,}\qq\rangle_{\bb}=\nt\langle\Qt\rangle_{2N_c-N_F}+\no\langle\Qo\rangle_{2N_c-N_F}\approx  (\nt-\no)\mx m_1\Biggl [1+\frac{\bb}{\nt-N_c}\Bigl (\frac{m_1}{\lm}\Bigr )^{\frac{\bb}{\nt-N_c}}\Biggr ) \Biggr].
\eeq

Besides, the value of the flavor singlet part $\langle{\rm Tr\,}\qq\rangle_{\bb}$ in \eqref{(2.1.18)} can be connected with the $SU(\bb)$ singlet part of the dyon condensate, $\langle\Sigma_D\rangle$. Considering 'm' as the background field, from \eqref{(1.1)},\eqref{(2.1.2)} and using $\delta_3=0$ from \eqref{(B.4)},
\bq
\langle\,\frac{\partial}{\partial m}\w_{\rm matter}\rangle=\langle\,\frac{\partial}{\partial m}
{\widehat\w}_{\rm matter}\rangle \ra\langle{\rm Tr\,}\qq\rangle_{\bb}=\langle\Sigma_D\rangle
\equiv\sum_{j=1}^{\bb}\langle{\ov D}_j D_j\rangle=\sum_{j=1}^{\bb}\langle{\ov D}_j\rangle\langle D_j\rangle.   \quad\,\,\,\,\label{(2.1.19)}
\eq

As for the value of $\langle\Sigma_D\rangle$ in br2 vacua of the $SU(N_c)$ theory, it is obtained {\it independently} in Appendix B using \eqref{(2.2.10)} and the values of roots of the curve \eqref{(1.2)}, see \eqref{(B.7)},\eqref{(B.15)}
\bq
\langle\Sigma_D\rangle=\sum_{j=1}^{\bb}\langle{\ov D}_j\rangle\langle D_j\rangle\approx (\nt-\no)\mx m_1 \Biggl [1+\frac{\bb}{\nt-N_c}\Bigl (\frac{m_1}{\lm}\Bigr )^{\frac{\bb}{\nt-N_c}}\Biggr ) \Biggr]\,,\label{(2.1.20)}
\eq
this agrees with \eqref{(2.1.18)},\eqref{(2.1.19)}.\\

The non-zero values of the heavy quark condensates in left hand sides of \eqref{(2.1.18)} originate finally: a) from the quantum loop effects at the scale $\mu\sim\lm$, transforming, on the first "preliminary" stage, the bilinear operators of heavy quark fields with masses $\sim\lm$ into bilinear operators of light particles, \, b) finally, from formed at lower energies $\mu\ll\lm$ {\it non-zero "genuine" (i.e. coherent) condensates of light higgsed dyons and quarks}. I.e.: 1) the dyon condensates $\langle D_j\rangle=\langle {\ov D}_j\rangle\approx (\mx\lm)^{1/2},\, j=1...\bb$, formed at $\mu\sim g(\mx\lm)^{1/2}$, and 2) the quark condensates $\langle Q^i_a\rangle=\langle{\ov Q}^{\,a}_i\rangle\sim (\mx m)^{1/2},\,\, i,a=1...\no$, formed at $\mu\sim g (\mx m)^{1/2}\ll g(\mx\lm)^{1/2}$.\\

In connection with \eqref{(2.1.19)},\eqref{(2.1.20)}, it is worth asking the following question. Because the difference between $\w_{\rm matter}$ in \eqref{(1.1)} and $\widehat\w_{\rm matter}$ in \eqref{(2.1.2)} originate from the color symmetry breaking $SU(N_c)\ra SU(\nd)\times U^{\bb}(1)$ in the high energy region $\mu\sim\lm$, how e.g. the condensate $\langle\Sigma_D\rangle$ in \eqref{(2.1.20)} of dyons which are formed at the scale $\mu\sim\lm$ knows about the number $\no$ which originates only at much lower energies ? (By the way, put attention that even the leading term $\sim\mx m$ of $\langle\Sigma_D\rangle$ in \eqref{(2.1.20)} is much smaller than the separate terms $\langle{\ov D}_j\rangle\langle D_j\rangle\sim\mx\lm$, but these largest terms $\sim\mx\lm$ cancel in the sum over j).

The answer is that, indeed, both ${\rm Tr\,}(\qq)_{2N_c-N_F}$ and $\Sigma_D$ in \eqref{(2.1.19)} are by itself $U(N_F)_{\rm flavor}\times U(\nd)_{\rm color}$ singlets. The explicit dependence on $\no$ appears in $\langle\Sigma_D\rangle$ in \eqref{(2.1.20)} {\it only from the region of much lower energies $\sim g(\mx m)^{1/2}$, after taking the total vacuum averages $\langle...\rangle$ in br2 vacua}. Remind that, by definition, $\langle O\rangle$ means the mean value of any operator $O$ integrated down to $\mu^{\rm lowest}_{\rm cut}=0$. This explicit dependence of leading terms $\sim\mx m$ in $\langle\Sigma_D\rangle$ \eqref{(2.1.20)} on $\no$ originates finally in \eqref{(2.1.5)},\eqref{(2.1.6)} at energies $\mu<(\rm several)g(\mx m)^{1/2}$ in the following way. a) At the first "preliminary" stage, from the spontaneous color symmetry breaking $SU(\nd)\ra U(\no)\times SU(\nd-\no)$ at the scale $\mu\sim m\ll\lm$ by $\langle X^{adj}_{SU(\nd)}\rangle\sim \langle a_2\rangle\sim\langle\Qo\rangle_{\no}/\mx\sim m$, see \eqref{(2.1.3)},\eqref{(2.1.5)}, giving masses $m_3\sim m$ to all hybrid adjoints of $SU(\nd)$ and quarks with $SU(\nd-\no)$ colors.  (But the {\it number} $\langle a_2\rangle\sim m$ is really formed not at the scale $\mu\sim m$, but only at lower energies $\mu\sim g(\mx m)^{1/2}$, due to higgsed quarks from $SU(\no)$). b) Finally, from the coherent condensate formed by light higgsed quarks $\langle{\ov Q}^{\,a}_i\rangle=\langle Q_a^i\rangle\sim\delta^i_a (\mx m)^{1/2},\,\, a=1...\no,\, i=1...N_F$, from the color subgroup $SU(\no)$ at the scale $\sim g(\mx m)^{1/2}\ll m$, giving masses $\sim g(\mx m)^{1/2}$ to all $SU(\no)\times U^{(2)}(1)$ gluons, see \eqref{(2.1.14)}. (While the total mean value $\langle a_1\rangle= - m (\bb)/\nd$ is formed at the higher scale $\sim g(\mx\lm)^{1/2}$ due to higgsed dyons, see \eqref{(2.1.5)},
\eqref{(2.1.6)}, but it is independent of $\no$). The number $\no$ penetrates then into $\langle\Sigma_D\rangle$ from {\it the low energy relation between the vacuum averages} following from \eqref{(2.1.5)} (see also \eqref{(A.23)})
\bq
\langle\frac{\partial\w^{\,\rm low}_{\rm tot}}{\partial a_{1}}\rangle=\Biggl [\, - \langle{\rm Tr\,}\qq\rangle_{\no}+\langle\frac{\partial\w^{(SYM)}_{SU(\nd-\no)}}{\partial a_{1}}\rangle\,\Biggr ]
-c_1\langle\Sigma_D\rangle+\langle\frac{\partial\w_{\rm a_{1}}}{\partial a_{1}}\rangle=0\,,\label{(2.1.21)}
\eq
because all $N_F$ flavors of quarks from the color $SU(\nd)$ and all dyons interact with $a_1$. (Moreover, the non-leading term of $\langle\Sigma_D\rangle$ in \eqref{(2.1.20)} originates at even lower energies from the SYM part in \eqref{(2.1.21)}, and from the corresponding SYM correction in $\langle{\rm Tr\,}\qq\rangle_{\no}$, see \eqref{(A.23)},\eqref{(A.24)}).\\

The heavy quark condensates \eqref{(2.1.18)} include both the $SU(N_F)$ adjoint and singlet in flavor parts. As for the singlet part, its numerical value is given in \eqref{(2.1.19)},\eqref{(2.1.20)}. As for the adjoint parts, they look as, see \eqref{(2.1.19)}
\bq
\langle\Qt-\frac{1}{N_F}{\rm Tr\,}(\qq)\rangle_{\bb}=A\langle\Qt-\frac{1}{N_F}{\rm Tr\,}(\qq)\rangle_{\nd},
\label{(2.1.22)}
\eq
\bq
\langle\Qt\rangle_{\bb}=A\langle\Qt-\frac{1}{N_F}{\rm Tr\,}(\qq)\rangle_{\nd}+
\frac{1}{N_F}\langle\Sigma_D\rangle.\label{(2.1.23)}
\eq
\bq
\langle\Qo-\frac{1}{N_F}{\rm Tr\,}(\qq)\rangle_{\bb}=A\langle\Qo-\frac{1}{N_F}{\rm Tr\,}(\qq)\rangle_{\nd}  \label{(2.1.24)}
\eq
\bq
\langle\Qo\rangle_{\bb}=A\langle\Qo-\frac{1}{N_F}{\rm Tr\,}(\qq)\rangle_{\nd}+ \frac{1}{N_F}\langle\Sigma_D\rangle\,,\label{(2.1.25)}
\eq
where $A=O(1)$ is some constant originating from integrating out in \eqref{(2.1.22)},\eqref{(2.1.24)} the loop effects from the Lagrangian \eqref{(1.1)} over the energy region $\mu\sim\lm$. It is seen from
\eqref{(2.1.22)},\eqref{(2.1.24)},\eqref{(2.4.1)},\eqref{(2.4.6)} that the main contributions $\sim\mx m$ to the "non-genuine" (i.e. non-factorizable) adjoint bilinear condensates of heavy non-higgsed quarks are induced finally by the "genuine" (i.e. coherent) factorizable condensate $\langle\Qo\rangle_{\nd}\approx
\langle\Qo\rangle_{\no}=\langle{\ov Q}^1_1\rangle\langle Q^1_1\rangle\approx\mx m_1$ of light higgsed quarks through the quantum loop effects.

Unfortunately, we cannot calculate 'A' directly, but we can claim that it does not know the number $\no$ appearing only at much lower scale $\mu\sim m\ll\lm$. Besides, we can find 'A' e.g. from \eqref{(2.1.23)} using \eqref{(2.4.1)},\eqref{(2.1.20)} for the right hand part and $\langle\Qt\rangle_{\bb}$ in \eqref{(2.1.18)} for the left one, and then predict $\langle\Qo\rangle_{\bb}$ in \eqref{(2.1.25)}. As a result,
\bq
A= -2\,,\quad \langle\Qo\rangle_{\bb}\approx - \mx m_1\Biggl [1+\frac{\bb}{\nt-N_c}\Bigl (\frac{m_1}{\lm}\Bigr )^{\frac{\bb}{\nt-N_c}} \Biggr]\,, \label{(2.1.26)}
\eq
this agrees with \eqref{(2.1.18)}, and $A=\,-2$ in independent of $\no$ as it should be.\\

On the whole, the total decomposition of quark condensates $\langle (\qq)_{1,2}\rangle_{N_c}$ over their separate color parts look in these br2 vacua as follows.\\
I) The condensate $\langle\Qo\rangle_{N_c}$.\\
a) From \eqref{(2.1.14)}, the (factorizable) condensate of higgsed quarks in the $SU(\no)$ part
\bq
\langle\Qo\rangle_{\no}=\langle{\ov Q}^1_1\rangle\langle Q^1_1\rangle\approx\mx m_1 \Bigl [\, 1+\frac{\bb}{\nt-N_c}\Bigl (\frac{m_1}{\lm}\Bigr )^{\frac{2N_c-N_F}{\nt-N_c}}\,\Bigr ]\,.\label{(2.1.27)}
\eq
b) The (non-factorizable) condensate $\langle\Qo\rangle_{\nd-\no}$ is determined by the one-loop Konishi anomaly for the heavy non-higgsed quarks with the mass $m_2$ in the $SU(\nd-\no)$ SYM sector, see \eqref{(2.1.4)},\eqref{(2.1.15)}
\bq
\langle\Qo\rangle_{\nd-\no}=\frac{\langle S\rangle_{\nd-\no}}{m_2}\approx\mx m_1\Bigl (\frac{m_1}{\lm}\Bigr )^{\frac{\bb}{\nt-N_c}}\,.\label{(2.1.28)}
\eq
c) From \eqref{(2.1.18)}, the (non-factorizable) condensate of heavy non-higgsed quarks with masses $\sim\lm$
\bq
\langle\Qo\rangle_{\bb}\approx - \mx m_1\Biggl [1+\frac{\bb}{\nt-N_c}\Bigl (\frac{m_1}{\lm}\Bigr )^{\frac{\bb}{\nt-N_c}} \Biggr ]\,.\label{(2.1.29)}
\eq
Therefore, on the whole
\bq
\langle\Qo\rangle_{N_c}=\langle\Qo\rangle_{\no}+\langle\Qo\rangle_{\nd-\no}+\langle\Qo\rangle_{\bb}\approx
\mx m_1\Bigl (\frac{m_1}{\lm}\Bigr )^{\frac{\bb}{\nt-N_c}}\,,\,\,\quad \label{(2.1.30)}
\eq
as it should be, see \eqref{(2.1)}.\\
II) The condensate $\langle\Qt\rangle_{N_c}$.\\
a) From \eqref{(2.1.6)} the (factorizable) condensate of the non-higgsed massless Nambu-Goldstone particles in the $SU(\no)$ part
\bq
\langle\Qt\rangle_{\no}=\sum_{a=1}^{\no}\langle{\ov Q}^{\,a}_2\rangle\langle Q^2_{a}\rangle=0\,.\label{(2.1.31)}
\eq
b) The (non-factorizable) condensate $\langle\Qt\rangle_{\nd-\no}$ is determined by the same one-loop Konishi anomaly for the heavy non-higgsed quarks with the mass $m_2$ in the $SU(\nd-\no)$ SYM sector,
\bq
\langle\Qt\rangle_{\nd-\no}=\frac{\langle S\rangle_{\nd-\no}}{m_2}\approx\mx m_1\Bigl (\frac{m_1}{\lm}\Bigr )^{\frac{\bb}{\nt-N_c}}\,.\label{(2.1.32)}
\eq
c) From \eqref{(2.1.18)}, the (non-factorizable) condensate of heavy non-higgsed quarks with masses $\sim\lm$
\bq
\langle\Qt\rangle_{\bb}\approx \mx m_1\Biggl [1+\frac{\bb}{\nt-N_c}\Bigl (\frac{m_1}{\lm}\Bigr )^{\frac{\bb}{\nt-N_c}} \Biggr ]\,.\label{(2.1.33)}
\eq
Therefore, on the whole
\bq
\langle\Qt\rangle_{N_c}=\langle\Qt\rangle_{\no}+\langle\Qt\rangle_{\nd-\no}+\langle\Qt\rangle_{\bb}\approx
\mx m_1 \Biggl [1+\frac{N_c-\no}{\nt-N_c}\Bigl (\frac{m_1}{\lm}\Bigr )^{\frac{\bb}{\nt-N_c}} \Biggr ],\,\,\,\quad \label{(2.1.34)}
\eq
as it should be, see \eqref{(2.1)}.\\

And finally, remind that the mass spectra in these br2 vacua depend essentially on the value of $m/\lm$ and all these vacua with $\nt>N_c$ evolve at $m\gg\lm$ to br1 vacua of sections 4.1 or 4.3 below, see section 3 in \cite{ch4} or section 4 in \cite{ch5}.

\numberwithin{equation}{subsection}
\subsection{$U(N_c)$\,,\,\,small $\mx\,,\,\,\mx\ll\Lambda^{SU(\nd-\no)}_{{\cal N}=2\,\, SYM}$}

$U(N_c)$ theory is obtained by adding one $SU(N_c)$ singlet $U^{(0)}(1)$ ${\cal N}=2$ multiplet, with its scalar field $\sqrt{2} X^{(0)}=a_0 I$, where $I$ is the unit $N_c\times N_c$ matrix. Instead of \eqref{(1.1)}, the superpotential looks now as
\bq
{\cal W}_{\rm matter}=\frac{\mu_{0}}{2} N_c a^2_0+\mx{\rm Tr}\,(X^{\rm adj}_{SU(N_c)})^2 +{\rm Tr}\,\Bigl [ (m-a_0)\,{\ov Q} Q-{\ov Q}\sqrt{2} X^{\rm adj}_{SU(N_c)} Q \Bigr ]_{N_c}\,,\quad \mu_0=\mx\,,\label{(2.2.1)}
\eq
and we consider in this paper only the case with $\mu_0=\mx$. The only change in the Konishi anomalies \eqref{(1.3)} and in the value of $\langle\Qt\rangle_{N_c}$ in \eqref{(2.1)} will be that now
\bq
\langle \Qo+\Qt\rangle_{N_c}=\mx m\,,\quad  \langle\Qt\rangle_{N_c}\approx \mx m\gg \langle\Qo\rangle_{N_c}\approx\mx m\Bigl (\frac{m}{\lm}
\Bigr)^{\frac{\bb}{{\rm n}_2-N_c}},  \label{(2.2.2)}
\eq
while in br2 vacua the small ratio $\langle \Qo\rangle_{N_c}/\langle\Qt\rangle_{N_c}\ll 1$ and small $\langle S\rangle_{N_c}\ll\mx m^2$ in \eqref{(2.1)} remain parametrically the same. Besides, from \eqref{(2.2.1)},\eqref{(2.2.2)} (neglecting power corrections)
\bq
\langle a_0\rangle=\frac{\langle\,{\rm Tr}\,(\qq)\rangle_{N_c}}{N_c\mx}=\frac{\no \langle\Qo\rangle_{N_c}+\nt \langle\Qt\rangle_{N_c}}{N_c \mx}\approx\frac{\nt}{N_c}\,m \,,\quad \langle m-a_0\rangle\approx -\,\frac{\nd-\no}{N_c}\,m\,.\label{(2.2.3)}
\eq

The changes in \eqref{(2.1.2)} and \eqref{(2.1.5)} are also very simple\,: a)\, the term "$\mx N_c a^2_{0}/2$"\, is added,\,\, b) "$m$"\, is replaced by "$m-a_0$" in all other terms. So, instead of \eqref{(2.1.5)},\eqref{(2.1.6)}, we have now
\bq
\w^{\,\rm low}_{\rm tot}=\w^{\,(SYM)}_{SU(\nd-\no)}+\w^{\,\rm low}_{\rm matter}+\dots\,,\quad
\w^{\,\rm low}_{\rm matter}=\w_{SU(\no)}+\w_{D}+\w_{a}\,,
\label{(2.2.4)}
\eq
\bbq
\w^{\,(SYM)}_{SU(\nd-\no)}=(\nd-\no)\,\wmu\Bigl (\Lambda^{SU(\nd-\no)}_{{\cal N}=2\,\, SYM}\Bigr )^2+\w^{\,(M)}_{SU(\nd-\no)}\,,\quad\langle\Lambda^{SU(\nd-\no)}_{{\cal N}=2\,\,SYM}\rangle^2
\approx m^2\Bigl (\frac{m}{\lm}\Bigr )^{\frac{2N_c-N_F}{\nt-N_c}}\,,
\eeq
\bbq
\w_{SU(\no)}+\w_{D}=(m-a_0-a_1-a_2)\,{\rm Tr}\,({\ov Q} Q)_{\no}+(m-a_0-c_1 a_1)\sum_{j=1}^{2N_c-N_F}{\ov D}_j D_j+\dots\,,
\eeq
\bbq
\w_{a}=\frac{\mx}{2}\Biggl [N_c a^2_0+(1+\delta_1)\frac{\nd N_c}{\bt} a_1^2+(1+\delta_2)\frac{\no \nd}{\nd-\no} a_2^2+2 N_c \delta_3 a_1(m-a_0-c_1 a_1)+2 N_c \delta_4(m-a_0-c_1 a_1)^2\Biggr ],
\eeq
where dots denote here terms which are either of no importance for us below in this section or too small, see also \eqref{(2.1.5)}.

From \eqref{(2.2.3)},\eqref{(2.2.4)} (the leading terms only)
\bq
\langle a_1\rangle= -\,\frac{\bt}{\nd}\langle m-a_0\rangle\approx \frac{\bt(\nd-\no)}{\nd N_c}\,m\,,\quad \langle a_2\rangle=\langle m-a_0-a_1\rangle\approx -\,\frac{\nd-\no}{\nd}\,m\,, \label{(2.2.5)}
\eq
while, according to reasonings given in section 2.1, all $\delta_i$ should remain the same. Of course, this can be checked directly. Instead of \eqref{(2.1.8)} we have now for the leading terms
\bq
\langle\w^{\,\rm low}_{\rm tot}\rangle\approx\langle\w^{\,\rm low}_{\rm matter}\rangle\approx\frac{\mx}{2}\Biggl [\,N_c\langle a_0\rangle^2+\frac{\nd N_c}{\bt}(1+\delta_1)\langle a_1\rangle^2+\frac{{\rm n}_1\nd}{\nd-{\rm n}_1}(1+\delta_2)\langle a_2\rangle^2\,\Biggr ]\,, \label{(2.2.6)}
\eq
where $\langle a_i\rangle$ are given in \eqref{(2.2.3)},\eqref{(2.2.5)}. Instead of \eqref{(2.1.9)},\eqref{(2.1.10)} we have now (with the same accuracy)
\bq
\w^{\,\rm eff}_{\rm tot}(\Pi)=m\,{\rm Tr}\,({\ov Q} Q)_{N_c}-\frac{1}{2\mx}\Biggl [ \,\sum_{i,j=1}^{N_F} ({\ov Q}_j Q^i)_{N_c}({\ov Q}_{\,i} Q^j)_{N_c}\Biggr ]-\nd S_{N_c}\,\,\ra\,\, \langle\w^{\,\rm eff}_{\rm tot}\rangle\approx\frac{1}{2}\,\nt\,\mx m^2\,.\label{(2.2.7)}
\eq
It is not difficult to check that values of $\delta_{1,2}$ obtained from \eqref{(2.2.6)}=\eqref{(2.2.7)} are the same as in ~\eqref{(2.1.12)}.~
\footnote{\, Besides, one can check that \eqref{(2.2.4)} with all $\delta_{i}=0,\,i=1...4$, will result in the wrong value: $\langle a_0\rangle\approx [\,m\,(\bt+\no)/N_c\,]$, this differs from \eqref{(2.2.3)}.
}

Therefore, instead of \eqref{(2.1.13)}, we obtain now  finally for the leading term, see \eqref{(2.2.4)},\eqref{(2.2.5)},
\bbq
\langle{\rm Tr}\,({\ov Q} Q)\rangle_{\no}=\no\langle\Qo\rangle_{\no}+\nt\langle\Qt\rangle_{\no}= \no\langle\Qo\rangle_{\no}\approx \wmu \,\frac{{\rm n}_1\nd}{\nd-\no}\langle a_2\rangle\approx\no \mx m\,,
\eeq
\bq
\langle\Qo\rangle^{U(N_c)}_{\no}=\sum_{a=1}^{\no}\langle{\ov Q}_1^{\,a}Q^1_a\rangle=\langle{\ov Q}^1_1\rangle\langle Q^1_1\rangle\approx\mx m\approx \langle\Qt\rangle^{U(N_c)}_{N_c}\,,\,\, \langle\Qt\rangle_{\no}=0\,,\,\, \langle S\rangle_{\no}=0, \label{(2.2.8)}
\eq
while as before in $SU(N_c)$
\bq
\langle{\ov D}_j\rangle\langle D_j\rangle= -\mx\lm\omega^{j-1}+O(\mx m)\,.\label{(2.2.9)}
\eq

It is worth to remind that the number of double roots of the curve \eqref{(1.2)} in this case is still $N_c-1$, as two single roots with $e^{+}=-e^{-}\approx 2\langle\Lambda^{SU(\nd-{\rm n}_1)}_{{\cal N}=2\,\, SYM}\rangle$ still originate here from $SU(\nd-\no)\,\, {\cal N}=2$ SYM. In other words, the number of charged higgsed particles remains $N_c-1$ as in the $SU(N_c)$ theory, i.e.\,:  $2N_c-N_F$ dyons $D_j$, $\nd-\no-1$ pure magnetic monopoles $M_i$, and $\no$ electric quarks $Q^k$. But now, in comparison with the $SU(N_c)$ theory, one extra $U^{(0)}(1)$ gauge multiplet is added. As a result, in addition to $2\no\nt$ massless Nambu-Goldstone multiplets, there remains now one exactly massless ${\cal N}=1$ photon multiplet, while the corresponding scalar ${\cal N}=1$ multiplet has now smallest non-zero mass $\sim\mx$ due to breaking ${\cal N}=2\ra {\cal N}=1$.

In so far as we know from \eqref{(2.1.5)},\eqref{(2.2.4)} the charges and numbers of all particles massless at $\mx\ra 0$, we can establish the definite correspondence with the roots of the curve \eqref{(1.2)}: $2N_c-N_F$ unequal double roots $e_j\sim\lm,\, j=1...2N_c-N_F$ correspond to our dyons $D_j$,\, $\nd-\no-1$ unequal double roots $e_i\sim\langle\Lambda^{SU(\nd-{\rm n}_1)}_{{\cal N}=2\,\, SYM}\rangle,\, i=1...N_c-\no-1$ correspond to pure magnetic monopoles from $SU(\nd-\no)$ SYM, and $\no$ equal double roots $e_k\sim m,\, k=1...\no$ correspond to $\no$ higgsed original pure electric quarks from $SU(\no)$. Besides, we know from \eqref{(2.2.8)},\eqref{(2.2.9)} the leading terms of their condensates (see \eqref{(2.1.6)} for the monopole condensates). Then, as a check, we can compare the values of condensates in \eqref{(2.2.8)},\eqref{(2.2.9)} with the formulas proposed in \cite{SY5} where these condensates are expressed through the roots of the $U(N_c)$ curve \eqref{(1.2)}.
\footnote{\,
Remind that these formulas were derived in \cite{SY5} from and checked then on a few simplest examples only, and proposed after this to be universal.

Besides, in general, the main problem with the use of the expressions like \eqref{(2.2.10)} from \cite{SY5} for finding the condensates of definite light BPS particles (massless at $\mx\ra 0$) is that one needs to find first the values of all roots of the curve \eqref{(1.2)}. But for this, in practice, one has to understand, at least, the main properties of the color and flavor symmetry breaking in different vacua, the corresponding mass hierarchies, and multiplicities of each type roots. \label{(f7)}
}

Specifically, these roots look in these $U(N_c)$ br2 vacua as: 1) the quark double roots $e^{(Q)}_k= - m,\,\, k=1...\no$,\,\, 2) the dyon double roots $e^{(D)}_j\approx \omega^{j-1}\lm,\,\,j=1...(2N_c-N_F)$,\,\,
3) two single roots \cite{DS,CIV,CSW} which we know originate from $SU(\nd-\no)\,\,{\cal N}=2$ SYM,\,\, $e^{+}=-e^{-}\approx 2\langle\Lambda^{SU(\nd-{\rm n}_1)}_{{\cal N}=2\,\, SYM}\rangle$. Therefore, with this knowledge, the formulas from \cite{SY5} look as
\bq
\langle{\ov Q}_k Q^k\rangle^{U(N_c)}_{\no}= - \mx\sqrt{(e^{(Q)}_k-e^+)(e^{(Q)}_k-e^-)}\approx -\mx e^{(Q)}_k\approx\mx m\,,\quad k=1...\no, \label{(2.2.10)}
\eq
\bbq
\langle{\ov D}_j D_j\rangle= - \mx\sqrt{(e^{(D)}_k-e^+)(e^{(D)}_k-e^-)}\approx -\mx e^{(D)}_j\approx -\mx\omega^{j-1}\lm\,,\quad j=1...2N_c-N_F\,.
\eeq
( Both $e^{\pm}$ are neglected here because they are non-leading). It is seen that the leading terms of \eqref{(2.2.8)},\eqref{(2.2.9)} and \eqref{(2.2.10)} agree, this is non-trivial as the results were obtained very different methods. The calculation of power corrections to \eqref{(2.2.8)},\eqref{(2.2.9)}, and \eqref{(2.2.10)} is presented in Appendices A and~ B.

\subsection{$SU(N_c)$\,,\,\,larger $\mx\,,\,\,\,\langle\Lambda^{SU(\nd-\no)}_{{\cal N}=2\,\, SYM}\rangle\ll\mx\ll\, m$}

Significant changes for such values of $\mx$ occurs only in the ${\cal N}=2\,\, SU(\nd-\no)$ SYM sector. All $SU(\nd-\no)$ adjoint scalars $X^{adj}_{SU(\nd-\no)}$ are {\it too heavy} now, their $SU(\nd-\no)$ phases fluctuate freely at all scales $\mu\gtrsim\langle\Lambda^{SU(\nd-\no)}_{{\cal N}=1\,\,SYM}\rangle$
in this case and they are not higgsed, i.e. $\langle X^{adj}_{SU(\nd-\no)}\rangle=0$. Instead, they all decouple as heavy at scales $\mu<\mx^{\rm pole}=|g\mx|/(\rm several)\ll\, m$ {\it in the weak coupling region}. There remains ${\cal N}=1\,\,SU(\nd-\no)$ SYM with the scale factor of its gauge coupling $\langle\Lambda^{SU(\nd-\no)}_{{\cal N}=1\,\, SYM}\rangle=[\,\wmu\langle\Lambda^{SU(\nd-\no)}_{{\cal N}=2\,\, SYM}\rangle^2\,]^{1/3},\,\,\,\langle\Lambda^{SU(\nd-\no)}_{{\cal N}=2\,\, SYM}\rangle\ll\langle\Lambda^{SU(\nd-\no)}_{{\cal N}=1\,\, SYM}\rangle\ll\mx\ll~ m$, see \eqref{(2.1.4)},\eqref{(2.1.15)}. The small non-zero (non-factorizable) value $\langle {\,\rm Tr\,} (X^{adj}_{SU(\nd-\no)})^2\rangle=(\nd-\no)\langle S\rangle_{\nd-\no}/\wmu=(\nd-\no)\langle\Lambda^
{SU(\nd-\no)}_{{\cal N}=2\,\, SYM}\rangle^2\ll\langle\Lambda^{SU(\nd-\no)}_{{\cal N}=1\,\, SYM}\rangle^2$ arises here only due to the Konishi anomaly, i.e. from one-loop diagrams with heavy scalars $X^{adj}_{SU(\nd-\no)}$ and their fermionic superpartners with masses $\sim\mx$ inside, not because $X^{adj}_{SU(\nd-\no)}$ are higgsed.

The multiplicity of vacua of this ${\cal N}=1 \,\, SU(\nd-\no)$ SYM is also $\nd-\no$, as it should be.. There appears now a large number of strongly coupled ${\cal N}=1$ gluonia with the mass scale $\sim\langle\Lambda^{SU(\nd-\no)}_{{\cal N}=1\,\,SYM}\rangle$. All heavier $SU(\nd-\no)$ charged original pure electric quarks and hybrids with masses $\sim m$ and scalars $X^{adj}_{SU(\nd-\no)}$ with masses $\sim\mx\gg\langle\Lambda^{SU(\nd-\no)}_{{\cal N}=1\,\,SYM}\rangle$ are still weakly confined, but the tension of the confining string is larger now, $\sigma^{1/2}_{SU(\nd-\no)}\sim\langle\Lambda^{SU(\nd-\no)}_{{\cal N}=1\,\,SYM}\rangle\ll\mx\ll m$.

\subsection{$SU(N_c)$\,,\,\, even larger $\mx\,,\,\,\,m\ll\mx\ll\lm$}

\hspace*{4mm} The case with $m\ll\mx\ll\lm$ and $N_c+1<N_F<3N_c/2$ is described in section 8.1 of  \cite{ch6}. Therefore, we consider here in addition the region $3N_c/2<N_F<2N_c-1$.

Remind, that the whole dyonic $\bt=2N_c-N_F$ sector, including the $U^{(1)}(1)\,\, {\cal N}=2$ photon multiplet with its scalar partner $a_1$, acquires masses $\sim (\mx\lm)^{1/2}\gg\mx$ and decouples at lower energies. But finally, the trace of the heavier scalar $a_1$ remains in the lower energy ${\cal W}_{SU(\nd)}$ superpotential at scales $\mu<(\mx\lm)^{1/2},\,\, \mx\ll(\mx\lm)^{1/2}\ll\lm$, in the form: $\,m\ra{\wt m}= (m-\langle a_1\rangle)=m N_c/\nd\,$. Remind also that, see \eqref{(2.1.2)},\eqref{(2.1.12)}, $\mx\ra\wmu=(1+\delta_2)\mx= - \mx$.

Therefore, after integrating out all particles with masses $\sim (\mx\lm)^{1/2}$, the lower energy superpotential at the scale $\mu= (\mx\lm)^{1/2}/{(\rm several)}\gg\mx$ looks as, see also \eqref{(2.4.5)},\eqref{(2.4.6)}
\bq
\w^{({\cal N}=2,\,\nd)}_{\rm matter}={\rm Tr}\,\Bigl [\,{\ov Q}\,(\tm-\sqrt{2}X^{\rm adj}_{SU(\nd)})\,Q\,
\Bigr ]_{\nd}+\wmu{\rm Tr}\,(X^{\rm adj}_{SU(\nd)})^2\,,\quad  \tm=\frac{N_c}{\nd}m\,,\quad \wmu=-\mx\,,\label{(2.4.1)}
\eq
\bbq
\langle\Qo\rangle_{\nd}= [\,\frac{\nd\,\wmu \tm}{\nd-\no}=\wmu m_2=\mx m_1\,]+\frac{N_c-\no}{\nd-\no}\langle\Qt\rangle_{\nd}\approx\mx m_1\approx\langle\Qt\rangle_{N_c}\,,
\eeq
\bbq
\langle\Qt\rangle_{\nd}\approx\mx m_1\Bigl (\frac{m_1}{\lm}\Bigr)^{\frac{\bb}{{\rm n}_2-N_c}}\approx\langle\Qo\rangle_{N_c}\,,\quad m_2= - m_1=\frac{N_c}{\nt-N_c}\,m\,,
\eeq
\bbq
\langle{\rm Tr}(\sqrt{2}X^{\rm adj}_{SU(\nd)})^2\rangle=\frac{1}{\wmu}\Bigl [(2\nd-N_F)\langle S\rangle_{\nd}+\tm\langle{\rm Tr}({\ov Q} Q)\rangle_{\nd}\Bigr ]\approx \frac{\no\nd}{\nd-\no}\tm^2\,.
\eeq

All $\nd^{\,2}-1\,\,X^{\rm adj}_{SU(\nd)}$ {\it decouple as heavy} at scales $\mu<\mx^{\rm pole}/(\rm several)=g(\mu=\mx^{\rm pole})\mx/(\rm several)$ {\it in the weak coupling region} and can be integrated out. Therefore, the whole unbroken ${\cal N}=1\,\, SU(\nd)$ SQCD with $N_F$ flavors of quarks emerges already at the scale $\mu=\mx^{\rm pole}/(\rm several)$. The scale factor $\lt$ of its gauge coupling is, see \eqref{(2.1.15)},
\bq
(\,\lt\,)^{3\nd-N_F}=\Bigl [\Lambda_{SU(\nd)}= -\lm\Bigr ]^{2\nd-N_F}\,\wmu^{\,\nd}\,,\quad  \frac{\lt}{\wmu}=\Bigl (\frac{\mx}{\lm}\Bigr )^{\frac{2N_c-N_F}{2N_F-3N_c}}\ll 1 \,.\label{(2.4.2)}
\eq

This ${\cal N}=1$ theory is not IR free at $3N_c/2<N_F<2N_c-1$ and so its logarithmically small coupling $g^2(\mu\sim\mx)$ begins to grow logarithmically at $\mu<\mx$. There is a number of variants of the mass spectrum. To deal with them it will be convenient to introduce colorless but flavored auxiliary (i.e. sufficiently heavy, with masses $\sim\mx$, and dynamically irrelevant at $\mu<\mx$) fields $\Phi_i^j$ \cite{ch5}, so that the Lagrangian at the scale $\mu=\mx^{\rm pole}/(\rm several)$ instead of
\bq
K={\rm Tr}\,(Q^\dagger Q+{\ov Q}^\dagger {\,\ov Q}),\,\, \w^{\,({\cal N}=1,\,\nd)}_{\rm matter}=\tm\, {\rm Tr}\,({\ov Q} Q)_{\nd}-\frac{1}{2\wmu}\Biggl ( {\rm Tr}\,({\ov Q} Q)_{\nd}^2-\frac{1}{\nd}\Bigl ({\rm Tr}({\ov Q} Q)_{\nd}\Bigr )^2  \Biggr ) \label{(2.4.3)}
\eq
can be rewritten as
\footnote{\,
all factors of the ${\cal N}=1$ RG evolution in Kahler terms are ignored below for simplicity if they are logarithmic only, as well as factors of $g$ if $g$ is either constant or logarithmically small \label{(f8)}
}
\bq
K={\rm Tr}\,(\Phi^\dagger \Phi)+{\rm Tr}\,(Q^\dagger Q+{\ov Q}^\dagger {\,\ov Q})\,,\quad
\w^{\,({\cal N}=1,\,\nd)}_{\rm matter}=\w_{\Phi}+{\rm Tr}\,\Bigl ({\ov Q}\,\tm^{\rm tot} Q \Bigr )\,,\label{(2.4.4)}
\eq
\bbq
\w_{\Phi}=\frac{\wmu}{2}\Bigl ( {\rm Tr}\,(\Phi^2)-\frac{1}{N_c}({\rm Tr}\,\Phi)^2 \Bigr )\,,\quad {\tm}
=\frac{N_c}{\nd}\,m\,,\quad (\tm^{\rm tot})^j_i=\tm\,\delta^j_i -\Phi^j_i\,,\quad \wmu= - \mx\,,
\eeq
and {\it its further evolution at lower energies is determined now by the dynamics of the genuinely ${\cal N}=1$ theory \eqref{(2.4.4)}}.

The Konishi anomalies for \eqref{(2.4.4)} look as, see \eqref{(2.4.2)}, compare with \eqref{(1.3)},\eqref{(1.4)},
\bq
\langle\Qo+\Qt-\frac{1}{\nd}{\rm Tr}\,\qq\rangle_{\nd}=\wmu \tm\,,\label{(2.4.5)}
\eq
\bbq
\langle S\rangle_{\nd}=\frac{\langle\Qo\rangle_{\nd}\langle\Qt\rangle_{\nd}}{\wmu}=\Biggl(\frac{\langle
\det\qq\rangle_{\nd}}{\lt^{3\nd-N_F}}\Biggr)^{1/N_c}=\Biggl(\,\frac{[\langle\Qo\rangle_
{\,\nd}]^{\no}[\langle\Qt\rangle_{\,\nd}]^{\nt}}{\Lambda_{SU(\nd)}^{2\nd-N_F}\,\wmu^{\,\nd}}\,\Biggr )^{1/N_c}\,,
\eeq
\bbq
\langle\tm^{\rm tot}_1=\tm-\Phi^1_1\rangle=\frac{\langle\Qt\rangle_{\nd}}{\wmu}\,,\quad \langle\tm^{\rm tot}_2=\tm-\Phi^2_2\rangle=\frac{\langle\Qo\rangle_{\nd}}{\wmu}\,.
\eeq
\bbq
\langle\Phi^i_j\rangle=\frac{1}{\wmu}\Biggl [\langle{\ov Q}_{j} Q^i\rangle_{\nd}-\delta^i_j\,\frac{1}{\nd}\langle{\rm Tr\,}\qq\rangle_{\nd} \,\Biggr ]\,.
\eeq

From \eqref{(2.4.5)}, see \eqref{(2.4.1)},\eqref{(2.1.15)} and section 8.1.1 in \cite{ch6} (neglecting all power corrections for simplicity),
\bq
\langle\Qo\rangle_{\nd}= [\,\frac{\nd\,\wmu \tm}{\nd-\no}=\wmu m_2=\mx m_1\,]+\frac{N_c-\no}{\nd-\no}
\langle\Qt\rangle_{\nd}\approx\mx m_1\approx\langle\Qt\rangle_{N_c}\,,\label{(2.4.6)}
\eq
\bbq
\langle\Qt\rangle_{\nd}\approx\wmu m_2\Bigl (\frac{m_2}{\Lambda_{SU(\nd)}}\Bigr)^{\frac{\bb}{\nt-N_c}}
\approx\mx m_1\Bigl (\frac{m_1}{\lm}\Bigr)^{\frac{\bb}{{\rm n}_2-N_c}}\approx\langle\Qo\rangle_{N_c}\,,
\quad \Lambda_{SU(\nd)}= -\lm\,,
\eeq
\bbq
\langle S\rangle_{\nd}=\langle S\rangle_{\,SU(\nd-\no)}^{\,(SYM)}= -\langle S\rangle_{N_c}\,,\quad
m_2=\frac{N_c}{\nt-N_c}\,m= -m_1\,.
\eeq

Besides, it is not difficult to check that accounting for both the leading terms $\sim \mx m^2$ and the leading power corrections $\sim \langle S\rangle_{N_c}= -\langle S\rangle_{\nd}$, see
\eqref{(2.4.3)}-\eqref{(2.4.6)},\, \eqref{(2.1)},\eqref{(2.1.9)},\eqref{(2.1.12)}
\bq
\Bigl [\langle\w^{\,({\cal N}=1,\,\nd)}_{\rm tot}\rangle=\langle\w^{\,({\cal N}=1,\,\nd)}_{\rm matter}\rangle-N_c\langle S\rangle_{\nd}\Bigr ]+\Bigl [\langle\w_{a_1}\rangle=\frac{\mx}{2}(1+\delta_1)\frac{\nd N_c}{\bt}\langle a_1\rangle^2\Bigr ]=\langle \w^{\,\rm eff}_{\rm tot}\rangle\,,\label{(2.4.7)}
\eq
as it should be.\\

{\bf A)} If $m$ is not too small so that $(\mx m)^{1/2}\gg\lt$. Then $\no<\nd$ quarks are higgsed still {\it in the weak coupling regime}
at the scale $\mu\sim(\mx m)^{1/2}\gg\lt,\,\, SU(\nd)\ra SU(\nd-\no)$, and $\no(2\nd-\no)$ gluons acquire masses $\mu_{\rm gl,1}\sim (\mx m)^{1/2}$. There remains at lower energies ${\cal N}=1$ SQCD with the unbroken $SU(\nd-\no)$ gauge group, $\nt>N_c$ flavors of still active lighter quarks $Q^2,\,{\ov Q}_2$ with $SU(\nd-\no)$ colors, $\no^2$ pions $\Pi^1_{1^\prime}\ra \sum_{a=1}^{\nd}({\ov Q}^{\,a}_{i^\prime} Q^i_a),\,\,i^\prime,i=1,\,...,\,\no,\,\, \langle\Pi^1_1\rangle=\langle \Qo\rangle_{\nd}\sim \mx m$, and $2\no\nt$ hybrids $\Pi^2_1,\,\Pi^1_2$ (these are in essence the quarks $Q^2,\,{\ov Q}_2$ with broken colors). The scale factor $\widehat\Lambda$ of the gauge coupling is, see \eqref{(2.4.2)},
\bq
{\wh\Lambda}^{\,\wh{\rm b}_{\rm o}}=\frac{\lt^{3\nd-N_F}}{\det \Pi^1_1}\,,\quad
\Bigl (\frac{{\langle\widehat\Lambda}\rangle^{2}}{\mx m}\Bigr )^{\wh{\rm b}_{\rm o}}\sim\Bigl (\frac{\lt^{2}}{\mx m} \Bigr )^{2N_F-3N_c}\ll 1\,.\,\quad \widehat{\rm b}_{\rm o}= 3(\nd-\no)-\nt\,.\label{(2.4.8)}
\eq

{\bf a1)} If $\wh{\rm b}_{\rm o}<0$. The $SU(\nd-\no)$ theory at $\mu<\mu_{\rm gl,1}$ is then IR free, its coupling is small and still decreases logarithmically with diminishing energy. The quarks $Q^2,\,{\ov Q}_2$ with $\nd-\no$ still active colors and $\nt>N_c$ flavors have masses $m^{\rm pole}_{Q,2}\sim m$ and decouple as heavy at $\mu<\,m$. Notice that {\it all this occurs in the weak coupling regime} in this case, so that {\it there is no need to use the assumed dynamical scenario from} \cite{ch3}.

There remains at lower energies ${\cal N}=1\,\, SU(\nd-\no)$ SYM with the scale factor of its coupling
\bq
\Bigl (\Lambda^{SU(\nd-\no)}_{{\cal N}=1\,\,SYM}\Bigr )^{3(\nd-\no)} = \frac{{\lt}^{\,3\nd-N_F}\,\det\tm^{\rm tot}_2}{\det \Pi^1_1},\quad (\tm^{\rm tot}_2)^{2^\prime}_2=\tm\,\delta^{2^\prime}_2-\Phi^{2^\prime}_2\,, \,\, \frac{\langle\Lambda^{SU(\nd-\no)}_{{\cal N}=1\,\,SYM}\rangle}{m}\ll 1.\label{(2.4.9)}
\eq

After integrating this ${\cal N}=1$ SYM via the VY-procedure \cite{VY}, the Lagrangian looks as, see \eqref{(2.1.15)},\eqref{(2.4.9)},\eqref{(2.4.2)},
\bbq
K={\rm Tr}\,(\Phi^\dagger\Phi)+  {\rm Tr}\,\Biggl [2\sqrt{(\Pi^1_1)^{\dagger}\Pi^1_1}+\Pi_2^1\frac{1}
{\sqrt{(\Pi^1_1)^{\dagger}\Pi^1_1}}\,(\Pi_2^1)^{\dagger}+(\Pi_1^2)^{\dagger}\frac{1}{\sqrt{
(\Pi^1_1)^{\dagger}\Pi^1_1}}\,\Pi_1^2 \Biggr ]\,,
\eeq
\bq
\w^{\,(\nd)}=\w_{\rm non-pert}+\w_{\Phi}+{\rm Tr}\,\Bigl (\tm^{\rm tot}_1\,\Pi^1_1\Bigr )+\w_{\rm hybr}\,, \label{(2.4.10)}
\eq
\bbq
\w_{\rm non-pert}=(\nd-\no)\Bigl (\Lambda^{SU(\nd-\no)}_{{\cal N}=1\,\,SYM}\Bigr )^3= (\nd-\no)\Biggl [\frac{(\Lambda_{SU(\nd)}=-\lm)^{2\nd-N_F}\wmu^{\,\nd}\,\det\tm^{\rm tot}_2}{\det \Pi^1_1}\Biggr ]^{1/(\nd-\no)}\,,
\eeq
\bbq
\w_{\rm hybr}={\rm Tr}\,\Bigl (\tm^{\rm tot}_2\,\Pi_2^1\frac{1}{\Pi^1_1}\Pi_1^2-\Phi_1^2\Pi_2^1-
\Phi_2^1\Pi_1^2\Bigr )\,.
\eeq

The masses of $\no^{\,2}$ pions $\Pi^1_1$ and hybrids from \eqref{(2.4.10)} are, see \eqref{(2.4.4)} for $\w_{\Phi}$,
\bq
\mu(\Pi^1_1)\sim\frac{\langle\Pi^1_1\rangle=\langle\Qo\rangle_{\nd}}{\mx}\sim m\sim m^{\rm pole}_{Q,2}\,,\quad\langle\Pi^1_{1^\prime}\rangle=\langle({\ov Q}_{1^\prime} Q^1)\rangle_{\nd}\approx - \delta^1_{1^\prime} \frac{N_c}{\nd-\no}\,\mx m\,,\label{(2.4.11)}
\eq
\bbq
\mu(\Pi^1_2)=\mu(\Pi^2_1)=0\,,
\eeq
(the main contribution to $\mu(\Pi^1_1)$ gives the term $\sim (\Pi^1_1)^2/\mx$ originating from  ${\rm Tr}\, (\tm^{\rm tot}_1\,\Pi^1_1 )$ in \eqref{(2.4.10)} after integrating out heavier $\Phi^1_1$ with masses $\sim\mx\gg m$ ).\\

On the whole, the mass spectrum of this ${\cal N}=1\,\,SU(\nd)$ theory look in this case as follows.\\
1) There are $\no(2\nd-\no)\,\,{\cal N}=1$ multiplets of massive gluons, $\mu_{\rm gl,1}\sim (\mx m)^{1/2}$.\\
2) A large number of hadrons made from non-relativistic and weakly confined ${\ov Q}_2, Q^2$ quarks with $\nt>N_c$ flavors and $\nd-\no$ colors, their mass scale is $\mu_H\sim m,\,\, \Lambda^{SU(\nd-\no)}_{{\cal N}=1\,\,SYM}\ll \mu_H\ll\mu_{\rm gl,1}$ (the tension of the confining string is $\sigma^{1/2}\sim\Lambda^{SU(\nd-\no)}_{{\cal N}=1\,\,SYM}\ll m$).\\
3) $\no^2$ pions $\Pi^1_1$ with masses $\sim m$.\\
4) A large number of ${\cal N}=1\,\,SU(\nd-\no)$ SYM strongly coupled gluonia with the mass scale $\sim\Lambda^{SU(\nd-\no)}_{{\cal N}=1\,\,SYM}$.\\
5) $2\no\nt$ massless (complex) Nambu-Goldstone ${\cal N}=1$ multiplets $\Pi^1_2,\,\Pi^2_1$.\\
6) All $N_F^2$ fields $\Phi$ have masses $\mu(\Phi)\sim\mx\gg m$ and are dynamically irrelevant at scales $\mu<\mx$.\\
Comparing with the mass spectrum at $N_c+1<N_F<3N_c/2$ in "$\bf A$" of section 8.1.1 in \cite{ch6} it is seen that it is the same (up to different logarithmic factors).

The multiplicity of these vacua is $N_{\rm br2}=(\nd-\no)C^{\,\no}_{N_F}$, as it should be. The factor $\nd-\no$ originates from ${\cal N}=1\,\,SU(\nd-\no)$ SYM and the factor $C^{\,\no}_{N_F}$ - from spontaneous flavor symmetry breaking $U(N_F)\ra U(\no)\times U(\nt)$ due to higgsing $\no$ $\, Q^1,\,{\ov Q}_1$ quarks.\\

{\bf a2)} If $\wh{\rm b}_{\rm o}>0$. Then, because $\wh\Lambda\ll m$, the quarks $Q^2,\,{\ov Q}_2$ decouple as heavy at $\mu\sim m$, still in the weak coupling region. As a result, the mass spectrum will be the same as in "{\bf a1}" above (up to different logarithmic factors).\\

{\bf B)} If $(\mx m)^{1/2}\ll\lt$. Then ${\cal N}=1$ SQCD with $SU(\nd)$ colors and $3N_c/2<N_F<2N_c-1$ quark flavors enters first at $\mu<\lt$ the strongly coupled conformal regime with the frozen gauge coupling $a_*=\nd g^2_*/2\pi=O(1)$.

{\it And only now we use for the first time the assumed dynamical scenario from} \cite{ch3} {\it to calculate the mass spectrum}. Really, what is only assumed in this scenario in the case of the standard ${\cal N}=1$ SQCD conformal regime is the following.

In this ${\cal N}=1$ theory without colored adjoint scalars, unlike the very special ${\cal N}=2$ theory with its additional colored scalar fields $X^{\rm adj}$ and enhanced supersymmetry, {\it no parametrically lighter solitons} (in addition to the ordinary mass spectrum described below) {\it are formed at those scales where the standard ${\cal N}=1$ conformal regime is broken explicitly by non-zero particle masses} (see also the footnote \ref{(f3)}).

The potentially competing masses look then as
\bbq
{\wt m}_{Q,2}^{\rm pole}\sim\frac{m}{z_Q(\lt,m_{Q,2}^{\rm pole})}\sim\lt\Bigl (\frac{m}{\lt}\Bigr )^{N_F/3\nd},\,\, z_Q(\lt,\mu\ll\lt)=\Bigl (\frac{\mu}{|\lt|}\Bigr )^{\gamma_Q}\ll 1,\,\, 0<\gamma_Q=\frac{3\nd-N_F}{N_F}<\frac{1}{2}\,,
\eeq
\bq
\mu^2_{\rm gl,1}\sim z_Q(\lt,\mu_{\rm gl,1})\Bigl [\langle\Qo\rangle_{\nd}\sim\mx m\Bigr ]\,,\quad \mu_{\rm gl,1}\sim\lt\Bigl (\frac{\mx m}{\lt^2}\Bigr )^{N_F/3N_c}\ll\lt\,, \label{(2.4.12)}
\eq
\bbq
m\ll\mu_{\rm gl,1}\ll\lt\ll\mx\ll\lm\,,\quad \frac{{\wt m}_{Q,2}^{\rm pole}}{\mu_{\rm gl,1}}\ll 1\,.
\eeq
Therefore, the quarks $Q^i, {\ov Q}_j,\,\, i,j=1...\no,$ are higgsed at $\mu=\mu_{\rm gl,1}$, the flavor symmetry is broken spontaneously, $U(N_F)\ra U(\no)\times U(\nt)$, and there remains at lower energy ${\cal N}=1$ SQCD with $SU(\nd-\no)$ colors and $\nt$ flavors of still active quarks $Q^2, {\ov Q}_2\,$ with $SU(\nd-\no)$ colors.

{\bf b1)} If $\wh{\rm b}_{\rm o}<0$. Then this theory is IR free and the gauge coupling becomes logarithmically small at $\Lambda^{SU(\nd-\no)}_{{\cal N}=1\,\,SYM}\ll\mu\ll\mu_{\rm gl,1}$. The RG evolution at $\Lambda^{SU(\nd-\no)}_{{\cal N}=1\,\,SYM}<\mu<\mu_{\rm gl,1}$ is logarithmic only (and neglected for simplicity). The pole mass of $Q^2, {\ov Q}_2$ quarks with still active unbroken $\nd-\no$ colors looks really as
\bq
m_{Q,2}^{\rm pole}\sim \frac{m}{z_Q(\lt,\mu_{\rm gl,1})}\sim\lm\Bigl (\frac{\mx}{\lm}\Bigr )^{1/3}\Bigl (\frac{m}{\lm} \Bigr )^{\frac{2(3N_c-N_F)}{3N_c}}\ll\mu_{\rm gl,1}\,.\label{(2.4.13)}
\eq
They decouple then as heavy at $\mu<m_{Q,2}^{\rm pole}$ {\it in the weak coupling region} and there remains ${\cal N}=1\,\, SU(\nd-\no)$ SYM with $\Lambda^{SU(\nd-\no)}_{{\cal N}=1\,\,SYM}\ll m_{Q,2}^{\rm pole}$, see \eqref{(2.4.9)}. Integrating it via the VY procedure \cite{VY}, the lower energy Kahler terms looks now as \cite{ch5},
\bbq
K=z_{\Phi}(\lt,\mu_{\rm gl,1}){\rm Tr} (\Phi^\dagger\Phi )+z_Q(\lt,\mu_{\rm gl,1})
{\rm Tr}\,\Biggl [\,2\,\sqrt{(\Pi^1_1)^{\dagger}\Pi^1_1}+\Pi_2^1\frac{1}
{\sqrt{(\Pi^1_1)^{\dagger}\Pi^1_1}}\,(\Pi_2^1)^{\dagger}+(\Pi_1^2)^{\dagger}\frac{1}{\sqrt{
(\Pi^1_1)^{\dagger}\Pi^1_1}}\,\Pi_1^2 \Biggr ]\,,
\eeq
\bq
z_{\Phi}(\lt,\mu_{\rm gl,1})= \Bigl (\frac{\mu_{\rm gl,1}}{|\lt|}\Bigr )^{\gamma_{\Phi}}\,,\quad
-1<\gamma_{\Phi}=-2\gamma_{Q}<0\,,\quad z_{\Phi}(\lt,\mu_{\rm gl,1})=1/z^2_{Q}(\lt,\mu_{\rm gl,1})\gg 1\,,\label{(2.4.14)}
\eq
while the superpotential is as in \eqref{(2.4.10)}. Therefore, in comparison with \eqref{(2.4.11)}, only the pion masses have changed and are now
\bq
\mu(\Pi^1_1)\sim\frac{\langle\Pi^1_1\rangle=\langle\Qo\rangle_{\nd}}{\mx}\frac{1}{z_Q(\lt,\mu_{\rm gl,1})}\sim\lm\Bigl (\frac{\mx}{\lm}\Bigr )^{1/3}\Bigl (\frac{m}{\lm} \Bigr )^{\frac{2(3N_c-N_F)}{3N_c}}\sim m_{Q,2}^{\rm pole}\,, \label{(2.4.15)}
\eq
while, because $0<\gamma_Q<1/2$, all $N_F^2$ fions $\Phi$ remain too heavy, dynamically irrelevant and not observable as real particles at $\mu<\mx$.\\

{\bf b2)} If $\wh{\rm b}_{\rm o}>0$. Then the ${\cal N}=1\,\, SU(\nd-\no)$ theory remains in the conformal window at $\mu<\mu_{\rm gl,1}$ with the frozen gauge coupling $O(1)$.

The anomalous dimensions at $\mu<\mu_{\rm gl,1}$ look now as, see \eqref{(2.4.8)},
\bq
0<{\wh\gamma}_Q=\frac{{\rm {\wh b}}_{\rm o}}{\nt}<\gamma_Q<\frac{1}{2}\,,\quad {\wh\gamma}_{\Phi}=-2{\wh\gamma}_Q\,,\quad  {\wh z}_Q(\mu_{\rm gl,1},\mu\ll\mu_{\rm gl,1})=\Bigl (\frac{\mu}{\mu_{\rm gl,1}}\Bigr )^{{\wh\gamma}_Q}\ll 1\,.\label{(2.4.16)}
\eq
The pole mass of $Q^2, {\ov Q}_2$ quarks with unbroken colors looks now as, see \eqref{(2.4.12)},
\bq
{\wh m}_{Q,2}^{\rm pole}\sim\frac{m}{z_Q(\lt,\mu_{\rm gl,1})}\frac{1}{{\wh z}_Q(\mu_{\rm gl,1},m_{Q,2}^{\rm pole})}\sim\lm\Bigl (\frac{\mx}{\lm}\Bigr )^{1/3}\Bigl (\frac{m}{\lm}\Bigr )^{\frac{\nt-\no}{3(\nd-\no)}}\sim (\rm several)\,\Lambda^{SU(\nd-\no)}_{{\cal N}=1\,\,SYM}\,,\label{(2.4.17)}
\eq
while the masses of $\no^2$ pions $\Pi^1_1$ remain the same as in \eqref{(2.4.15)} and all $N_F^2\,\, \Phi^j
_i$ also remain dynamically irrelevant. The overall hierarchies of non-zero masses look in this case as
\bq
\mu(\Pi^1_1)\ll\Lambda^{SU(\nd-\no)}_{{\cal N}=1\,\,SYM}\sim {\wh m}_{Q,2}^{\rm pole}\ll\mu_{\rm gl,1}\ll\lt\ll\mx\ll\lm\,.\label{(2.4.18)}
\eq

On the whole for this section 2.4 with $m\ll\mx\ll\lm$. -

In all cases considered the overall phase $\rm\mathbf{Higgs_1-HQ_2}$ (HQ=heavy quark) remains the same in the non-Abelian $SU(\nd)$ sector. At scales $\mx\ll\mu\ll\lm$ it behaves as the effectively massless IR free ${\cal N}=2$ theory. At the scale $\mu\sim \mx^{\rm pole}=g\mx$ all $X^{adj}_{SU(\nd)}$ decouple as heavy and its dynamics becomes those of the ${\cal N}=1$ theory. The $2\no^2$ quarks $Q^1, {\ov Q}_1$ with $SU(\no)$ colors and $\no$ flavors are higgsed while $2\nt (\nd-\no)$ quarks $Q^2, {\ov Q}_2$ with $SU(\nd-\no)$ colors and $\nt$ flavors are in the HQ phase and confined by the ${\cal N}=1\, SU(\nd-\no)$ SYM. The $2\no\nt$ massless Nambu-Goldstone particles in all cases are in essence the hybrid quarks with $\nt$ flavors and broken $SU(\no)$ colors. And the overall qualitative picture is also the same in all variants considered, see e.g. the text after \eqref{(2.4.11)}. Changes only the character of the ${\cal N}=1$ RG evolution, i.e. it is logarithmic or power-like, and this influences the values of non-zero masses and mass hierarchies.\\

Finally, we think it will be useful to make the following short additional comments.\\
In this section, the largest masses $\mx^{\rm pole}=g(\mu=\mx^{\rm pole})\mx\gg (\mx m)^{1/2}\gg m$ in the whole $SU(\nd)$ sector have  $\nd^{\,2}-1$ adjoint scalars $X^{\rm adj}_{SU(\nd)}$. They all are {\it too heavy} now and their light physical $SU(\nd)$ phases fluctuate freely in the whole energy region $\mu> (\rm several) (\mx m)^{1/2}$. Therefore, the mean value of $X^{\rm adj}_{SU(\nd)}$, integrated not only over the interval from the high energy down to $\mx^{\rm pole}/({\rm several})$ as in the Lagrangian \eqref{(2.4.3)},\eqref{(2.4.4)}, but even down to much lower energies $\mu_{\rm cut}^{\rm low}= (\rm several) (\mx m)^{1/2}\ll\mx$ is zero,
\bbq
\langle\sqrt{2} X^{\rm adj}_{SU(\nd)}\rangle_{\mu^{\rm low}_{\rm cut}}=2\sum_{i=1}^{N_F}\sum_{A=1}^{\nd^{\,2}-1}T^{A}\langle{\rm Tr\,}{\ov Q}_i
T^{A} Q^i\rangle_{\mu^{\rm low}_{\rm cut}}/\wmu=
\eeq
\bq
= 2\sum_{i=1}^{\no}\sum_{A=1}^{\nd^{\,2}-1}T^{A} {\rm Tr}\Biggl [\langle{\ov Q}_i\rangle_{\mu^{\rm low}_{\rm cut}} T^{A} \langle Q^i\rangle_{\mu^{\rm low}_{\rm cut}}\Biggr ]/\wmu=0\,,\label{(2.4.19)}
\eq
because all light quark fields $Q^i_a, {\ov Q}^{\,a}_i,\, a=1...\nd,\, i=1...N_F$ (and all $SU(\nd)$ gluons) {\it still fluctuate independently and freely} in this higher energy region.

But its formal total mean value $\langle X^{\rm adj}_{SU(\nd)}\rangle$, integrated {\it by definition} down to $\mu^{\rm lowest}_{\rm cut}=0$, contains the {\it small} but non-zero, $\sim m\ll (\mx m)^{1/2}$, contribution originating only in the lower energy region $\mu\sim (\mx m)^{1/2}\ll\mx$ from the classical coupling of heavy $X^{\rm adj}_{SU(\nd)}$ with {\it lighter higgsed quarks} ${\ov Q}^{\,a}_i, Q^i_a,\, a,i=1...\no$, see e.g. \eqref{(2.4.1)} or \eqref{(2.1.14)},
\bbq
\langle\sqrt{2}X^{\rm adj}_{SU(\nd)}\rangle=\langle\sqrt{2}\sum_{A=1}^{\nd^{\,2}-1} X^{A} T^{A}\rangle = 2\sum_{i=1}^{N_F}\sum_{A=1}^{\nd^{\,2}-1}T^{A}\langle{\rm Tr\,}{\ov Q}_i T^{A} Q^i\rangle/\wmu \approx
\eeq
\bq
\approx 2\sum_{i=1}^{\no}\sum_{A=1}^{\nd^{\,2}-1}T^{A}{\rm Tr\,}[\,\langle{\ov Q}\rangle_i T^{A} \langle Q^i\rangle\,]/\wmu\approx m_2\,{\rm diag}\,(\,\underbrace{\,1-\frac{\no}{\nd}}_{\no}\,;\, \underbrace{\,-\frac{\no}{\nd}}_{\nd-\no}\,)\,.\label{(2.4.20)}
\eq
This leads to the main contribution $\sim m^2$ to ${\rm Tr\,}[\langle (\sqrt{2} X^{\rm adj}_{SU(\nd)})^2\rangle ]$, which in the case considered looks as
${\rm Tr\,}[\langle (\sqrt{2} X^{\rm adj}_{SU(\nd)})^2\rangle ]\approx{\rm Tr\,}[\langle \sqrt{2} X^{\rm adj}_{SU(\nd)}\rangle\langle \sqrt{2} X^{\rm adj}_{SU(\nd)}\rangle]$, see \eqref{(2.4.20)},
\bq
{\rm Tr\,} [\langle (\sqrt{2} X^{\rm adj}_{SU(\nd)})^2\rangle ]\approx \Bigl [\frac{\no (\nd-\no)}{\nd}\, m_2^2=\frac{\no\nd}{\nd-\no}\,\tm^2\,\Bigr ], \,\, \tm=\frac{N_c}{\nd},\,\, m_2=\frac{N_c}{\nd-\no} m\,, \,\,\,\,\label{(2.4.21)}
\eq
this agrees with the exact relation in the last line of \eqref{(2.4.1)} following from the Konishi anomaly (up to small power corrections $\sim m^2_1  (m_1/\lm)^{(\bb)/(\nt-N_c)}$ originating finally at even lower energies from the SYM part), and explains the origin of a smooth holomorphic behavior of $\langle{\rm Tr\,}(\sqrt{2}X^{\rm adj}_{SU(\nd)})^2\rangle$ when $\mx$ is increased from $\mx\ll m$ in section 2.1 to $\mx\gg m$ in this section.

But, in the case considered, this small formal non-zero total mean value of $\langle X^{\rm adj}_{SU(\nd)}
\rangle\sim m\ll (\mx m)^{1/2}$ does not mean that these heavy adjoint scalars are {\it higgsed}, in the sense that {\it they give the additional contributions} $\sim m$ {\it to particle masses} (those of hybrid gluons and $SU(\nd)$ quarks in this case), as it really occurs at $\mx\ll m$ in section 2.1. Clearly, {\it there are no any additional contributions to particle masses in the Lagrangian \eqref{(2.4.3)},
\eqref{(2.4.4)} at the scale $\mu=\mu_{\rm x}^{\rm pole}/({\rm several})$ from $\langle X^{\rm adj}_{SU(\nd)}\rangle_{\mu_{\rm x}^{\rm pole}/(\rm several)}=0$}, see \eqref{(2.4.19)}, because all $\nd^{\,2}-1$ fields $X^{\rm adj}_{SU(\nd)}$ with masses $\mx^{\rm pole}$ {\it decoupled already as heavy} at $\mu=\mx^{\rm pole}/(\rm several)$ where \eqref{(2.4.19)} is definitely valid, and do not influence the dynamics of the lower energy ${\cal N}=1$ theory at $\mu < \mx^{\rm pole}/(\rm several)$. This is the reason for a qualitative difference in patterns of color symmetry breaking and mass spectra at $\mx\ll m$ in section 2.1 and at $\mx\gg m$ in this section. Remind that the color breaking (and corresponding mass splitting) occurs at the scale $\mu\sim m\gg (\mx m)^{1/2}\gg\mx$ and looks as $SU(\nd)\ra SU(\no)\times U^{(2)}(1)\times SU(\nd-\no)$ in section 2.1, while it occurs at the scale $m\ll\mu\sim (\mx m)^{1/2}\ll\mx$ and looks as $SU(\nd)\ra SU(\nd-\no)$ in this section.

Besides, it is sufficient to say that if there were {\it additional contributions} to the quark masses in \eqref{(2.4.10)} from the formal $\langle X^{\rm adj}_{SU(\nd)}\rangle\sim m$ in \eqref{(2.4.20)}, then $2\no\nt$ massless Nambu-Goldstone fields $\Pi^1_2,\,\Pi^2_1$ from \eqref{(2.4.10)},\eqref{(2.4.11)} (which are in essence the quarks with $\nt$ flavors and $\no$ colors) will receive masses $\sim m$ and will be not massless. This will be clearly erroneous because the global flavor symmetry is broken spontaneously in these br2 vacua.

As it is seen from \eqref{(2.4.3)}-\eqref{(2.4.5)} and the whole content of this section, the whole contributions $\sim (\mx m)^{1/2}$ to the masses of $SU(\no)$ and hybrid gluons (see the footnote \ref{(f8)}) originate only at lower energies $\mu\sim (\mx m)^{1/2}\ll\mx$ in D-terms of the genuine ${\cal N}=1$ SQCD \eqref{(2.4.3)},\eqref{(2.4.4)} from higgsed quarks only. And the whole additional F-term contributions $\sim m$ to the quark masses originate really from the quark self-interaction term $\sim{\rm Tr\,}(\qq)^2/\mx$ with higgsed quarks $\langle\Qo\rangle_{\nd}=\langle{\ov Q}^1_1\rangle\langle Q^1_1\rangle$ in \eqref{(2.4.3)} or, what is the same, from $\langle\Phi_{1,2}\rangle\sim m$ in \eqref{(2.4.4)}, this last also originating finally from these higgsed quarks, forming the coherent condensate and giving masses corresponding gluons, see the last line of \eqref{(2.4.5)}. And the crucial difference between the section 2.1 with higgsed light $\langle X^{adj}_{SU(\nd)}\rangle\sim\langle a_2\rangle\sim m$ at $\mx\ll m$ in\eqref{(2.1.3)},
\eqref{(2.1.6)} contributing to quark masses, and replacing it $\langle\Phi^j_i\rangle\sim m$ with the mass $\sim\mx\gg m$ in this section, see \eqref{(2.4.4)},\eqref{(2.4.5)}, is that $ X^{adj}_{SU(\nd)}$ acts in the color space, while $\Phi^j_i$ acts in the flavor space.\\

\section{Unbroken flavor symmetry, \,\,S-vacua,\,\,$SU(N_c)$}
\numberwithin{equation}{section}

As in the br2 vacua in section 2.1, the non-trivial discrete  $Z_{\bb\geq 2}$ symmetry is also unbroken in these vacua, i.e. they also belong to the baryonic branch in the language \cite{APS}, this case corresponds to $\no=0$. From \eqref{(2.1.9)}, the quark condensates $\langle\qq\rangle_{N_c}$ in these vacua  with the multiplicity $\nd$ look as (neglecting smaller power corrections), see section 3 in \cite{ch4} or section 4 in \cite{ch5},
\bq
\langle\qq\rangle_{N_c}\approx -\,\frac{N_c}{\nd}\,\mx m\approx -\mx\tm\,,\quad \langle S\rangle_{N_c}=\Bigl ( \frac{\det\langle\qq\rangle_{N_c}}{\lm^{\bb}\mx^{N_c}}\Bigr )^{1/\nd}\approx\mx \tm^2\Bigl (\frac{-\tm}{\lm}\Bigr )^{(\bb)/\nd}\ll\mx m^2,\,\,\, \label{(3.1)}
\eq
\bbq
\langle{\rm Tr\,}(\sqrt{2}\,X^{adj}_{SU(N_c)})^2\rangle=\Bigl [(2N_c-N_F)\langle S\rangle_{N_c}+m\langle{\rm Tr\,}\qq\rangle_{N_c}\Bigr ]\approx m\,(-N_F\mx\tm)\,.
\eeq

As before, the scalar $X^{\rm adj}_{SU(N_c)}$ higgses the $SU(N_c)$ group at the scale $\mu\sim\lm$ as $SU(N_c)\ra SU(\nd)\times U^{(1)}(1)\times U^{\bt-1}(1)$,
\bq
\langle X \rangle=\langle\,X^{adj}_{SU(\nd)}+ X^{(1)}_{U(1)}+ X^{adj}_{\bt}\,\rangle\,,\quad \nd=N_F-N_c\,,\quad \bt=2N_c-N_F\,,\label{(3.2)}
\eq
\bbq
\langle\sqrt{2}\,X^{adj}_{\bt}\,\rangle\sim\lm\,{\rm diag}(\,\underbrace{\,0}_{\nd}\,; \underbrace
{\,\omega^0,\,\,\omega^1,\,...\,,\,\omega^{\bt-1}}_{\bt}\,)\,,\quad \omega=\exp\{\frac{2\pi i}{\,\bt}\,\}\,,
\eeq
\bbq
\sqrt{2}\,X^{(1)}_{U(1)}=a_1\,{\rm diag}(\,\underbrace{\,1}_{\nd}\,;\, \underbrace{\,c_1}_{\bt}\,),\quad c_1=-\,\frac{\nd}{\bt}\,,\quad \langle a_1\rangle=\frac{1}{c_1}m=\,-\frac{\bt}{\nd} m\,.
\eeq

I.e., as in br2 vacua, {\it the same} $\bt=2N_c-N_F\geq 2$ dyons $D_j$, massless in the limit $\mx\ra 0$ (this number is required by the unbroken $Z_{2N_c-N_F\geq 2}$ discrete symmetry) are formed at the scale $\mu\sim\lm$ and at sufficiently small $\mx$ the superpotential $\wh\w$ at $m\ll\mu\ll\lm$ is as in \eqref{(2.1.2)}.

The difference is in the behavior of the $SU(\nd)$ part. In these vacua with $\no=0$, there is no analog of the additional gauge symmetry breaking $SU(\nd)\ra SU(\no)\times U^{(2)}(1)\times SU(\nd-\no)$ in br2 vacua at the scale $\mu\sim m$.

{\bf A)}\,\, At $\mx\ll\Lambda^{SU(\nd)}_{{\cal N}=2\,\,SYM}\ll m$ and $N_c<N_F<2N_c-1$, {\it all quarks} in the $SU(\nd)$ part in these S vacua have now masses $\tm=\langle m-a_1\rangle=m N_c/\nd$ and decouple {\it in the weak coupling regime} at scales $\mu\,<\, \tm$, there remains ${\cal N}=2\,\,\,SU(\nd)$ SYM with the scale factor $\Lambda^{SU(\nd)}_{{\cal N}=2\,\, SYM}$ of its gauge coupling, see \eqref{(2.1.15)},
\bq
\langle\Lambda^{SU(\nd)}_{{\cal N}=2\,\, SYM}\rangle^{2\nd}=\Bigl (\Lambda_{SU(\nd)}= -\lm\Bigr )^{2\nd-N_F}\,\tm^{N_F}\,,\quad \tm=m\,\frac{N_c}{\nd}\,,\quad \langle\Lambda^{SU(\nd)}_{{\cal N}=2\,\, SYM}\rangle\ll \,m\,.\label{(3.3)}
\eq

And the field $X^{adj}_{SU(\nd)}$ higgses this ${\cal N}=2\,\,\, SU(\nd)$ SYM at the scale $\sim\langle\Lambda^{SU(\nd)}_{{\cal N}=2\,\, SYM}\rangle$ in a standard way, $SU(\nd)\ra U^{\nd-1}(1)$ \cite{DS}. This results in the right multiplicity $N_s=\nd=N_F-N_c$ of these S vacua. The two single roots with $(e^{+}-e^{-})\sim \Lambda^{SU(\nd)}_{{\cal N}=2\,\,SYM}$ of the curve \eqref{(1.2)} originate here from this ${\cal N}=2\,\,\, SU(\nd)$ SYM. Other $\nd-1$ unequal double roots originating from this $SU(\nd)$ sector correspond to massless at $\mx\ra 0\,\,\nd-1$ pure magnetic monopoles $M_{\rm n}$ with the $SU(\nd)$ adjoint charges.

As a result of all described above, the low energy superpotential at scales $\mu\ll\langle\Lambda^
{SU(\nd)}_{{\cal N}=2\,\, SYM}\rangle$ can be written in these S-vacua as (the coefficients $f_i=O(1)$ in \eqref{(3.4)} are known from \cite{DS})
\bq
\w^{\,\rm low}_{\rm tot}=\w^{\,(SYM)}_{SU(\nd)}+\w^{\,\rm low}_{\rm matter}+\dots\,,\quad
\w^{\,\rm low}_{\rm matter}=\w_{D}+\w_{a_1}\,,\label{(3.4)}
\eq
\bbq
\w^{\,(SYM)}_{SU(\nd)}=\nd \mx(1+\delta_2)\Bigl (\Lambda^{SU(\nd)}_{{\cal N}=2\,\, SYM}\Bigr )^2+\w^{\,(M)}_{SU(\nd)}\,,
\eeq
\bbq
\w^{\,(M)}_{SU(\nd)}= - \sum_{n=1}^{\nd-1} {\tilde a}_{M,\rm n}\Biggl [\, {\ov M}_{\rm n} M_{\rm n}+\mx(1+\delta_2)\langle\Lambda^{SU(\nd}_{{\cal N}=2\,\,SYM}\rangle\Biggl (1+O\Bigl (\frac{\langle\Lambda^{SU(\nd)}_{{\cal N}=2\,\,SYM}\rangle}{m}\Bigr )\Biggr )\,f_{\rm n}\,\Biggr ]\,,
\eeq
\bbq
\w_{D}=(m-c_1 a_1)\sum_{j=1}^{\bb}{\ov D}_j D_j-\sum_{j=1}^{\bb} a_{D,j}\,{\ov D}_j D_j\,-
\,\mx\lm\sum_{j=1}^{\bb}\omega^{j-1}\,a_{D,j}+\mx \, L_S \sum_{j=1}^{\bb} a_{D,j}\,,
\eeq
\bbq
\w_{a_1}=\frac{\mx}{2}(1+\delta_1)\frac{\nd N_c}{\bt} a_1^2+\mx N_c \delta_3 a_1(m-c_1 a_1)+\mx N_c\delta_4 (m-c_1 a_1)^2\,,
\eeq
where $\delta_{1,2}$ are given in \eqref{(2.1.12)}, while $\delta_3=0$, see \eqref{(B.4)}.

From \eqref{(3.4)}:
\bq
\langle a_1\rangle=\frac{1}{c_1} m= - \frac{2N_c-N_F}{\nd} m\,,\quad \langle a_{D,j}\rangle=\langle {\tilde a}_{M,\rm n}\rangle=0\,,\label{(3.5)}
\eq
\bbq
\langle{\ov M}_{\rm n} M_{\rm n}\rangle=\langle{\ov M}_{\rm n}\rangle\langle M_{\rm n}\rangle\approx \mx\langle\Lambda^{SU(\nd}_{{\cal N}=2\,\,SYM}\rangle f_n\,,\quad f_n=O(1)\,,\quad (2N_c-N_F)\mx\langle L_S\rangle=\langle\Sigma_D\rangle\,,
\eeq
\bbq
\langle{\ov D}_j D_j\rangle=\langle{\ov D}_j\rangle\langle D_j\rangle\approx -\mx\lm\omega^{j-1}+ \frac{1}{2N_c-N_F}\langle\Sigma_D\rangle\,,\quad
\langle\Sigma_D\rangle=\sum_{j=1}^{2N_c-N_F}\langle{\ov D}_j D_j\rangle\approx - \frac{N_c N_F}{\nd}\mx m\,.
\eeq

And the qualitative situation with the condensates $\sim \mx m$ of $N_F$ flavors of heavy non-higgsed quarks with $SU(2N_c-N_F)$ colors and masses $\sim\lm$ in these S-vacua with $\no=0$ is the same as those in br2 vacua described at the end of section 2.1, see \eqref{(2.1.16)}-\eqref{(2.1.26)}. ( The condensates of all lighter non-higgsed quarks with $N_F$ flavors, $\nd$ colors, and masses $\sim m$ in these S-vacua are power suppressed, see \eqref{(2.4.1)} with $\no=0,\, \nt=N_F$ or  (8.2.2) in \cite {ch6}, their small non-zero values originate in this case only from the Konishi anomaly for quarks in the $SU(\nd)$ sector, see \eqref{(3.3)}, $\,\langle\qq\rangle_{\nd}\sim\langle S\rangle_{\nd}/m\,\sim\mx\langle\Lambda^{SU(\nd)}_{{\cal N}=2\,\, SYM}\rangle^{2}/m\sim\mx m\Bigl (m/\lm\Bigr)^{(\bb)/\nd}\ll\mx m\,$).

Besides, from \eqref{(3.1)},\eqref{(3.5)} (the leading terms only for simplicity, compare with \eqref{(2.1.19)}\,)
\bq
\langle{\rm Tr\,}\qq\rangle_{N_c}\approx\langle{\rm Tr\,}\qq\rangle_{\bb}\approx\langle\Sigma_D\rangle\approx- \frac{N_c N_F}{\nd}\mx m\,. \label{(3.6)}
\eq

On the whole for the mass spectrum in these $\nd$ S-vacua with the unbroken flavor symmetry at $\mx\ll\Lambda^{SU(\nd)}_{{\cal N}=2\,\,SYM}\ll\, m\,$. - \\
1) All original electric quarks $Q, {\ov Q}$ are not higgsed but confined. The masses of all original electric particles charged under $SU(\bb)$ are the largest ones, $\sim\lm\,$, and they all are weakly confined because all $\bb$ dyons $D_j$ are higgsed, $\langle{\ov D}_j\rangle=\langle D_j\rangle\sim (\mx\lm)^{1/2}$, the string tension in this sector is $\sigma^{1/2}_{SU(\bb)}\sim \langle D_j\rangle\sim (\mx\lm)^{1/2}\ll\lm\,$. These confined particles form a large number of hadrons with masses $\sim\lm$.\\
2) The next mass scale is $\tm$, these are masses of original electric quarks with $SU(\nd)$ colors and $N_F$ flavors, they are also weakly confined due to higgsing of $\nd-1$ pure magnetic monopoles $M_{\rm n}$ of ${\cal N}=2\,\,SU(\nd)$\,\,SYM, the string tension in this sector is much smaller, $\sigma^{1/2}_{SU(\nd)}\sim \langle M_{\rm n}\rangle_{SU(\nd)}\sim (\mx\langle\Lambda^{SU(\nd)}_{{\cal N}=2\,\,SYM}\rangle)^{1/2}\ll (\mx\lm)^{1/2}$. They form hadrons with masses $\sim m$.\\
3) The next mass scale is $\langle\Lambda^{SU(\nd)}_{{\cal N}=2\,\,SYM}\rangle\ll m$, see \eqref{(3.3)}, these are masses of charged $SU(\nd)$ gluons and scalars. They are also weakly confined due to higgsing of $\nd-1$ magnetic monopoles $M_{\rm n}$, the tension of the confining string is the same, $\sigma^{1/2}_{SU
(\nd)}\sim \langle M_{\rm n}\rangle_{SU(\nd)}\sim (\mx\langle\Lambda^{SU(\nd)}_{{\cal N}=2\,\,SYM}\rangle)^{1/2}$. But the mass scale of these hadrons is $\sim\langle\Lambda^{SU(\nd)}_{{\cal N}=2\,\,SYM}\rangle\,$.\\
4) The other mass scale is $\sim (\mx\lm)^{1/2}$ due to higgsing of $\bt$ dyons $D_j$. As a result, $\bt=2N_c-N_F$ long ${\cal N}=2$ multiplets of massive photons with masses $\sim (\mx\lm)^{1/2}$ are formed (including $U^{(1)}(1)$ with its scalar $a_1$).\\
5) The lightest are $\nd-1$ long ${\cal N}=2$ multiplets of massive dual photons with masses $\sim (\mx\langle\Lambda^{SU(\nd)}_{{\cal N}=2\,\,SYM}\rangle)^{1/2}$.\\
6) Clearly, there are no massless Nambu-Goldstone particles because the global flavor symmetry remains unbroken. And there are no massless particles at all at $\mx\neq 0,\,\, m\neq 0$.

The corresponding scalar multiplets have additional small contributions $\sim\mx$ to their masses, these small corrections break ${\cal N}=2$ down to ${\cal N}=1$.\\

{\bf B)}\,\, At larger $\langle\Lambda^{SU(\nd)}_{{\cal N}=2\,\,SYM}\rangle\ll\mx\ll\, m$, the difference is that all $SU(\nd)$ adjoint scalars $X^{adj}_{SU(\nd)}$ are now {\it too heavy}, their $SU(\nd)$ phases fluctuate freely at all scales $\mu\gtrsim\langle\Lambda^{SU(\nd)}_{{\cal N}=1\,\,SYM}\rangle$ and they are not higgsed. Instead, they decouple as heavy at scales $\mu<\wmu^{\,\rm pole}=|g\mx|\ll m$ and there remains ${\cal N}=1 \,\, SU(\nd)$ SYM with the scale factor $\langle\Lambda^{SU(\nd)}_{{\cal N}=1\,\,SYM}\rangle=[\,\wmu \langle\Lambda^{SU(\nd)}_{{\cal N}=2\,\,SYM}\rangle^2\,]^{1/3}$ of its gauge coupling, $\,\,\langle\Lambda^{SU(\nd)}_{{\cal N}=2\,\,SYM}\rangle\ll\langle\Lambda^{SU(\nd)}_{{\cal N}=1\,\,SYM}\rangle\ll\mx\ll\, m$. (The small non-zero value ${\rm Tr\,}\langle (X^{adj}_{SU(\nd)})^2\rangle=\nd\langle S\rangle_{\nd}/\wmu=\nd\langle\Lambda^{SU(\nd)}_{{\cal N}=2\,\, SYM}\rangle^2\ll\langle\Lambda^{SU(\nd)}
_{{\cal N}=1\,\, SYM}\rangle^2$, see \eqref{(3.3)}, arises here only due to the Konishi anomaly, i.e. from one-loop Feynman diagrams with heavy scalars $X^{adj}_{SU(\nd)}$ and their fermionic superpartners with masses $\sim\mx$ inside, not because $X^{adj}_{SU(\nd)}$ are higgsed).

The multiplicity of vacua of this ${\cal N}=1 \,\, SU(\nd)$ SYM is also $\nd$ as it should be. There appears now a large number of strongly coupled gluonia with the mass scale $\sim\langle\Lambda^{SU(\nd)}_{{\cal N}=1\,\,SYM}\rangle$. All $SU(\nd)$ electrically  charged particles are still confined, the tension of the confining string is larger now, $\sigma^{1/2}_{SU(\nd)}\sim\langle\Lambda^{SU(\nd)}_{{\cal N}=1\,\,SYM}\rangle\ll\mx\ll\, m$.

The case with $m\ll\mx\ll\lm$ is described in section 8.2 of  \cite{ch6}.

\addcontentsline{toc}{section}
{\hspace*{4cm} \bf {Part II.\,\,Large quark masses}, $\Large\mathbf{m\gg\lm}$}
\vspace*{3mm}

\begin{center}{\hspace*{1cm}\bf\Large {Part II.\,\,Large quark masses}, $\Large\mathbf{m\gg\lm}$} \end{center}

\section{Broken flavor symmetry}
\numberwithin{equation}{section}

\hspace*{4mm} We present below in this section the mass spectra at $m\gg\lm$ of the direct (electric) $SU(N_c)$ theory \eqref{(1.1)} in vacua  with spontaneously broken flavor symmetry, $U(N_F)\ra U(\no)\times U(\nt)$,\, $N_c+1<N_F<2N_c-1$. There are the br1-vacua (br=breaking) with $1\leq\no< N_F/2\,,\,\, N_F=\no+\nt\,,$ in which $\langle\Qo\rangle_{N_c}\gg\langle\Qt\rangle_{N_c}$, see section 3 in \cite{ch4} and \eqref{(1.3)},\eqref{(1.4)}. The quark and gluino condensates look in these br1 vacua of $SU(N_c)$ as (the leading terms only, see the footnote \ref{(f4)})
\bq
\langle\Qo\rangle_{N_c}\approx\mx m_3\,,\quad \langle\Qt\rangle_{N_c}\approx\mx m_3\Bigl (\frac{\lm}{m_3}\Bigr )^{\frac{2N_c-N_F}{N_c-\no}} \,,\quad m_3=\frac{N_c}{N_c-\no}\, m\,, \label{(4.1)}
\eq
\bbq
\langle S\rangle_{N_c}=\frac{\langle\Qo\rangle_{N_c}\langle\Qt\rangle_{N_c}}{\mx}\approx\mx m_3^2\Bigl (\frac{\lm}{m_3}\Bigr )^{\frac{2N_c-N_F}{N_c-\no}}\,,\quad\frac{\langle\Qt\rangle_{N_c}}
{\langle\Qo\rangle_{N_c}}\approx\Bigl (\frac{\lm}{m_3}\Bigr)^{\frac{2N_c-N_F}{N_c-\no}}\ll 1\,,
\eeq
while $\no\leftrightarrow\nt$ in \eqref{(4.1)} in br2-vacua, but with $N_F/2<\nt<N_c$ in this case. The multiplicity of these br1 and br2 vacua is correspondingly $N_{\rm br1}=(N_c-\no)C_{N_F}^{\,\no}$ and $N_{\rm br2}=(N_c-\nt)C_{N_F}^{\,\nt},\,\, C_{N_F}^{\,\no}=C_{N_F}^{\,\nt}=[\,N_{F} !/\no!\,\nt!\,],\,\,\, 1\leq \no< N_F/2,\,\, \nt> N_F/2$. It is seen from \eqref{(4.1)} that the discrete $Z_{\bb}$ symmetry is unbroken at $m\gg\lm$ in all these br1 and br2 vacua with $\no\neq\nd$. In contrast, $Z_{\bb}$ symmetry is (formally) broken spontaneously in special vacua with $\no=\nd,\,\,\nt=N_c$ and the multiplicity $N_{spec}=(\bb)  C_{N_F}^{\,\nd}$ (as it is really broken spontaneously in these special vacua at $m\ll\lm$).

Out of them, see section 3 in \cite{ch4} and/or section 4 in ~\cite{ch5}, $(\nd-\no)C_{N_F}^{\,\no}$ part of br1-vacua with $1\leq\no <\nd$ evolves at $m\ll\lm$ into the br2-vacua of section 2.1 with $\langle\Qt\rangle^{(\rm br2)}_{N_c}\sim\mx m\gg\langle\Qo\rangle^{(\rm br2)}_{N_c}$, while the other part of br1-vacua and br2-vacua evolve at $m\ll\lm$ into $Lt$ vacua with $\langle\Qo\rangle^{(\rm Lt)}_{N_c}\sim\langle\Qt\rangle^{(\rm Lt)}_{N_c}\sim\mx\lm,\, ,\, \langle S\rangle_{Lt}\sim\mx\lm^2$.

\subsection{$SU(N_c),$  br1 vacua, smaller $\mx\,,\,\,\,\mx\ll\Lambda^{SU(N_{c}-\no)}_{{\cal N}=2\,\,SYM}$}
\numberwithin{equation}{subsection}

In the case considered, $X^{\rm adj}_{SU(N_c)}$ breaks the $SU(N_c)$ group {\it in the weak coupling regime} at the largest scale $\mu\sim m\gg\lm$ as\,: $\,SU(N_c)\ra SU(\no)\times U(1)\times SU(N_c-\no)$, see \eqref{(4.1.4)} below,
\bq
\langle X^{\rm adj}_{SU(N_c)}\rangle= \langle\, X^{adj}_{SU(\no)}+ X_{U(1)}+ X^{adj}_{SU(N_c-\no)}\,\rangle\,,\label{(4.1.1)}
\eq
\bbq
\sqrt{2}\, X_{U(1)}=a \,{\rm diag}\Bigl(\underbrace{\,1}_{\no}\,;\,\underbrace{\,c}_{N_c-\no}\, \Bigr )\,,\quad c=-\frac{\no}{N_c-\no}\,,\quad a\equiv \langle a\rangle+{\hat a}\,,\quad \langle a \rangle=m\,,
\quad \langle{\hat a}\rangle\equiv 0\,.
\eeq
As a result, all quarks $Q^i_a,\,{\ov Q}_j^{\,a}$ with flavors $i, j=1...N_F$ and colors $a=(\no+1)...N_c$ have large masses $m_3=m-c\langle a\rangle=m N_c/(N_c-\no)$, the hybrid gluons and hybrid $X$ also have masses $m_3\gg\lm$, and they all decouple at scales $\mu\lesssim\, m_3$. The lower energy theory at $\mu<\, m_3$ consists of ${\cal N}=2\,\, SU(N_c-\no)$\,\, SYM, then ${\cal N}=2\,\, SU(\no)$ SQCD with $N_F$ flavors of massless at $\mx\ra 0$ quarks $Q^{i}_{\rm b},\, {\ov Q}^{\,\rm b}_j$ with $\no$ colors, ${\rm b}=1\,...\,\no$, and finally one ${\cal N}=2\,\, U(1)$ photon multiplet with its scalar superpartner $X_{U(1)}$. The scale factor of the ${\cal N}=2\,\,\, SU(N_{c}-\no)$ SYM gauge coupling is $\langle\Lambda^{SU(N_{c}-{\rm n}_1)}_{{\cal N}=2\,\,SYM}\rangle\ll m$. The field $X^{\rm adj}_{SU(N_c-\no)}$ in \eqref{(4.1.1)}) breaks this ${\cal N}=2\,\, SU(N_{c}-\no)$ SYM at the scale $\mu\sim\langle\Lambda^{SU(N_{c}-{\rm n}_1)}_{{\cal N}=2\,\,SYM}\rangle$ in a standard way, $SU(N_c-\no)\ra U(1)^{N_c-\no-1}$ \cite{DS}, see also \eqref{(4.1)},
\bq
\langle\sqrt{2}\, X^{adj}_{SU(N_c-\no)}\rangle\sim\langle\Lambda^{SU(N_{c}-\no)}_{{\cal N}=2\,\,SYM}\rangle
\,{\rm diag}\Bigl (\,\underbrace{\,0}_{\no}\,; k_1,k_2,\,...\,, k_{N_c-\no}\Bigr )\,,\quad k_i=O(1)\,,\label{(4.1.2)}
\eq
\bbq
\langle\Lambda^{SU(N_{c}-\no)}_{{\cal N}=2\,\,SYM}\rangle^2=\Bigl (\frac{\lm^{\bb}{(m_3)}^{N_F}}{{(m_3)}^{2\no}}\Bigr )^{\frac{1}{N_c-\no}}=
m^2_3\Bigl (\frac{\lm}{m_3}\Bigr )^{\frac{2N_c-N_F}{N_c-\no}}\,,
\eeq
\bbq
\langle S\rangle_{N_c-\no}=\mx\langle\Lambda^{SU(N_{c}-\no)}_{{\cal N}=2\,\,SYM}\rangle^2\approx\langle S\rangle_{N_c}\,,
\eeq
where $k_i$ are known numbers \cite{DS}. We note that $\Lambda^{SU(N_{c}-\no)}_{{\cal N}=2\,\,SYM}$ in \eqref{(4.1.2)} and $\langle a\rangle=m$ in \eqref{(4.1.1)} have the same charge 2 under the discrete $Z_{2N_c-N_F}$ transformations as $X$ itself, this is due to unbroken $Z_{\bb}$ symmetry. There are $N_c-\no$ physically equivalent vacua and $N_c-\no-1$ light pure magnetic monopoles ${\ov M}_n,\, M_n,\,\, n=1... N_c-\no-1$ with the $SU(N_{c}-\no)$ adjoint charges (massless at $\mx\ra 0$) in each of these ${\cal N}=2$ SYM vacua. The low energy superpotential of this SYM part looks as
\bq
{\cal W}^{\,\rm low}_{\rm tot}=(N_c-\no)\mx\Bigl (\Lambda^{SU(N_{c}-{\rm n}_1)}_{{\cal N}=2\,\,SYM}\Bigr )^2+{\cal W}^{(M)}_{SYM}+{\cal W}_{\no}+\dots\,,\label{(4.1.3)}
\eq
\bbq
{\cal W}^{(M)}_{SYM}= - \sum_{n=1}^{N_c-\no-1} {\tilde a}_{M,\,\rm n}\Biggl [\, {\ov M}_{\rm n} M_{\rm n}+\mx\langle\Lambda^{SU(N_{c}-{\rm n}_1)}_{{\cal N}=2\,\,SYM}\rangle\Biggl (1+O\Bigl (\frac{\langle\Lambda^{SU(N_{c}-{\rm n}_1)}_{{\cal N}=2\,\,SYM}\rangle}{m}\Bigr )\Biggr )\,z_{\rm n}\, \,\Biggr ]\,,\quad z_{\rm n}=O(1),
\eeq
where $z_{\rm n}$ are known numbers \cite{DS} and dots denote as always smaller power corrections.

All monopoles in \eqref{(4.1.3)} are higgsed at $\mx\ne 0$ with $\langle M_{\rm n}\rangle=\langle{\ov M}_{\rm n} \rangle\sim [\,\mx\langle\Lambda^{SU(N_{c}-{\rm n}_1)}_{{\cal N}=2\,\,SYM}\rangle\,]^{1/2}$, so that all light particles of this ${\cal N}=2\,\, U(1)^{N_c-\no-1}$ Abelian magnetic part form $N_c-\no-1$ long ${\cal N}=2$ multiplets of massive dual photons with masses $\sim (\mx\langle\Lambda^{SU(N_{c}-\no}_{{\cal N}=2\,\,SYM})^{1/2}\rangle$ .

Besides, for this reason, all heaviest original $SU(N_{c}-\no)$ electrically charged particles  with masses $\sim m$ are weakly confined (the confinement is weak in the sense that the tension of the confining string is much smaller than their masses, $\sigma^{1/2}_{SYM}\sim (\mx\langle\Lambda^{SU(N_{c}-{\rm n}_1)}_{{\cal N}=2\,\,SYM}\rangle)^{1/2}\ll\langle\Lambda^{SU(N_{c}-{\rm n}_1)}_{{\cal N}=2\,\,SYM}\rangle\ll m$).\\

And finally, as for the (independent) ${\cal N}=2\,\,SU(\no)\times U(1)$ part with $N_F$ flavors of original electric quarks ${\ov Q}^b_j,  Q^{\,i}_b,\,\, i,j=1...N_F,\,\, b=1...\no$. Its superpotential at the scale $\mu=m$ looks as
\bq
{\cal W}_{\no}=(\, m-a\,)\,{\rm Tr}\,({\ov Q} Q)_{\no} - {\rm Tr}\,({\ov Q}\sqrt{2} X^{adj}_{SU(\no)} Q) +\mx {\rm Tr}\,( X^{adj}_{SU(\no)})^2+\frac{\mx}{2}\,\frac{\no N_c}{N_c-\no}\,a^2\,.\label{(4.1.4)}
\eq

From \eqref{(4.1.4)}, compare with \eqref{(4.1)},
\bq
\langle a\rangle=m\,,\quad \langle X^{adj}_{SU(\no)}\rangle=0\,,\quad\langle{\rm Tr}\,({\ov Q} Q)_{\no}\rangle=\no\langle\Qo\rangle_{\no}+\nt\langle\Qt\rangle_{\no}\approx\no \mx m_3\,,\label{(4.1.5)}
\eq
\bbq
\langle\Qo\rangle^{SU(N_c)}_{\no}\approx\mx m_3\approx\langle\Qo\rangle^{SU(N_c)}_{N_c}\,,\quad \langle\Qt\rangle_{\no}=\sum_{a=1}^{\no}\langle{\ov Q}^{\,a}_2\rangle\langle Q^2_a\rangle=0\,,\quad \langle S\rangle_{\no}=0\,,
\eeq
\bbq
\langle Q^i_b\rangle=\langle {\ov Q}_i^{\,b}\rangle\approx\delta^i_b\,(\mx m_3)^{1/2},\,\, b=1... \no,\,\,i=1...N_F\,,\,\, m_3=\frac{N_c}{N_c-\no}\,.
\eeq

On the whole, all this results in the right multiplicity of these br1 vacua, $N_{\rm br1}=(N_c-\no) C_{N_F}^{\,\no}$, the factor $N_c-\no$ originates from ${\cal N}=2\,\,\, SU(N_{c}-\no)$ SYM, while $C_{N_F}^{\,\no}$ is due to the spontaneous breaking $U(N_F)\ra U(\no)\times U(\nt)$ by $\no$ higgsed quarks in the $SU(\no)\times U(1)$ sector.

It is worth emphasizing that the value $\langle a\rangle=m$ in \eqref{(4.1.5)} is {\it exact} (i.e. there is no any small correction). This is seen from the following. At very small $\mx\ra 0$, the possible correction $\delta m\neq 0$ is independent of $\mx$, and so $\delta m\gg\mx$. Then quarks $Q^i_b, {\ov Q}_i^{\,b},\,\, i=1...N_F,\,\, b=1...\no$ will acquire masses $\delta m$, much larger than the scale of their potentially possible coherent condensate, $\delta m\gg (\mx m)^{1/2},\,\, \mx\ra 0$. In this case these quarks will be not higgsed but will decouple as heavy ones at scales $\mu<\delta m$ and the flavor symmetry will remain unbroken. There will remain ${\cal N}=2\,\, SU(\no)\times U(1)$ SYM  at scales $\mu<\delta m$. This lowest energy ${\cal N}=2\,\, SU(\no)$ SYM will give then its own multiplicity factor $\no$, so that the overall multiplicity of vacua will be $\no (N_c-\no)$, while the right multiplicity is $(N_c-
\no)C^{\,\no}_{N_F}$, see \eqref{(4.1)} and section 3 in \cite{ch4}. To have this right multiplicity and spontaneous flavor symmetry breaking all $\no$ quarks have to be higgsed at arbitrary small $\mx$, and this is only possible when they are {\it exactly massless} at $\mx=0$, i.e. at $\delta m=0$.~
~\footnote{\, These considerations clearly concern also all other similar cases.
}

As a result of higgsing of $\no$ out of $N_F$ quark flavors in $SU(\no)\times U(1)$ in \eqref{(4.1.4)}, the flavor symmetry is broken spontaneously as $U(N_F)\ra U(\no)\times U(\nt)$, the quarks $Q^{\,l}_b, {\ov Q}^{\,b}_l$ with flavors $l=(\no+1)...N_F$ and colors $b=1...\no$ will be the massless Nambu-Goldstone particles ($\,2\no\nt$ complex degrees of freedom), while all other particles in \eqref{(4.1.4)} will acquire masses $\sim (\mx m)^{1/2}$ and will form $\no^2$ long ${\cal N}=2$ multiplets of massive gluons.

On the whole, the hierarchies of non-zero masses look in this case as
\bq
\langle M_{\rm n}\rangle\sim (\mx\langle\Lambda^{SU(N_{c}-{\rm n}_1)}_{{\cal N}=2\,\,SYM}\rangle)^{1/2}\ll\langle Q^i_{b=i}\rangle\sim (\mx m)^{1/2}\,,\quad \lm\ll\langle\Lambda^{SU(N_{c}-\no)}_{{\cal N}=2\,\,SYM}\rangle\ll m\,,\label{(4.1.6)}
\eq
where ${\rm n}=1...N_{c}-\no-1\,,\,\,\, i=1...\no$\,. There are no massless particles, except for the $\,2\no\nt$ Nambu-Goldstone multiplets.

The curve \eqref{(1.2)} has $N_c-1$ double roots in these vacua. $N_c-\no-1$ unequal roots correspond to pure magnetic monopoles $M_{\rm n}$ in \eqref{(4.1.3)} and $\no$ equal ones correspond to higgsed quarks from $SU(\no)$. The two single roots with $(e^{+}-e^{-})\sim\langle\Lambda^{SU(N_{c}-\no)}_{{\cal N}=2\,\,SYM}\rangle$ originate from SYM.

The calculation of leading power corrections to $\langle{\cal W}^{\,\rm low}_{\rm tot}\rangle$ and $\langle\Qo\rangle^{SU(N_c)}_{\no}$ is presented in Appendix~ A.\\

Similarly as it has been done in section 2.1, let us emphasize also the following point. The numerically very large {\it total mean value} $\langle a\rangle\equiv\langle a\rangle_{\mu^{\rm lowest}_{\rm cut}=0}=
m\gg\langle\Lambda^{SU(N_{c}-\no)}_{{\cal N}=2\,\,SYM}\rangle\gg\lm$ in \eqref{(4.1.1)}, {\it giving the largest additional contributions $\sim m$ to  masses of quarks and adjoint hybrids}, does not mean really that this large {\it number} originates from the high energy region $\mu\sim m$. Because at least all $SU(\no)$ gluons are effectively massless at scales $\mu > \mu^{\rm low}_{\rm cut}=(\rm several)(\mx m)^{1/2}$, the physical $SU(\no)$ phases of all $SU(\no)$ quarks $Q^i_a, {\ov Q}^{\, a}_i,\, a=1...\no,\, i=1,,,N_F$ fluctuate freely and independently in this higher energy region, so that
\bq
\langle{\ov Q}^{\,b}_j Q^i_a\rangle_{\mu^{\rm low}_{\rm cut}}=\langle{\ov Q}^{\,b}_j\rangle_{\mu^{\rm low}_{\rm cut}}\langle Q^i_a\rangle_{\mu^{\rm low}_{\rm cut}}=0\,,\quad a,b=1...\no\,,\quad i,j=1...N_F\,.\label{(4.1.7)}
\eq

Therefore, the non-zero values of the quark mean values in \eqref{(4.1.5)} following from $\langle\partial
\w_{\no}/\partial a\rangle=0$, i.e. $\langle\Qo\rangle_{\no}/\mx\approx [\langle a\rangle=m]N_c/(N_c-\no)
\approx m_3$, originate only from the region $\mu\sim g(\mx m)^{1/2}$, after $\no$ out of $N_F$ quarks are higgsed (i.e. form the coherent condensate), {\it giving masses} $\sim g(\mx m)^{1/2}$ {\it to all} $SU(\no)\times U(1)$ {\it gluons}. And so, the large total mean value $\langle a\rangle=m$ originates also from this low energy region, see \eqref{(4.1.7)}.\\

On the whole, the total decomposition of quark condensates $\langle (\qq)_{1,2}\rangle_{N_c}$ over their separate color parts look in these br1 vacua as follows.\\
I) The condensate $\langle\Qo\rangle_{N_c}$.\\
a) From \eqref{(A.5)}, the (factorizable) condensate of higgsed quarks in the $SU(\no)$ part
\bq
\langle\Qo\rangle_{\no}=\mx m_3\Biggl [1-\frac{2N_c-N_F}{N_c-\no}\Bigl (\frac{\lm}{m_3}\Bigr )^{\frac{2N_c-N_F}{N_c-\no}}\Biggr ]\,,\quad m_3=\frac{N_c}{N_c-\no} m\gg
\langle\Lambda^{SU(N_{c}-\no)}_{{\cal N}=2\,\,SYM}\rangle\gg\lm\,. \label{(4.1.8)}
\eq
b) The (non-factorizable) condensate $\langle\Qo\rangle_{N_c-\no}$ is determined by the one-loop Konishi anomaly for the heavy non-higgsed quarks with the mass $m_3$ in the $SU(N_c-\no)$ SYM sector, see \eqref{(4.1.2)},
\bq
\langle\Qo\rangle_{N_c-\no}=\frac{\langle S\rangle_{N_c-\no}}{m_3}\approx\mx m_3\Bigl (\frac{\lm}{m_3}\Bigr )^{\frac{\bb}{N_c-\no}}\,.\label{(4.1.9)}
\eq

Therefore, on the whole
\bq
\langle\Qo\rangle_{N_c}=\langle\Qo\rangle_{\no}+\langle\Qo\rangle_{N_c-\no}\approx
\mx m_3\Biggl [ 1+\frac{\nt-N_c}{N_c-\no}\Bigl (\frac{\lm}{m_3}\Bigr )^{\frac{\bb}{\nt-N_c}}\Biggr ]\,,\,\,\quad \label{(4.1.10)}
\eq
as it should be, see \eqref{(A.1)}.\\
II) The condensate $\langle\Qt\rangle_{N_c}$.\\
a) From \eqref{(4.1.5)} the (factorizable) condensate of non-higgsed massless Nambu-Goldstone particles in the $SU(\no)$ part
\bq
\langle\Qt\rangle_{\no}=\sum_{a=1}^{\no}\langle{\ov Q}^{\,a}_2\rangle\langle
Q^2_{a}\rangle=0\,.\label{(4.1.11)}
\eq
b) The (non-factorizable) condensate $\langle\Qt\rangle_{N_c-\no}$ is determined by the same one-loop Konishi anomaly for the heavy non-higgsed quarks with the mass $m_3$ in the $SU(\nd-\no)$ SYM sector,
\bq
\langle\Qt\rangle_{N_c-\no}=\frac{\langle S\rangle_{N_c-\no}}{m_3}\approx\mx m_3\Bigl (\frac{\lm}{m_3}\Bigr )^{\frac{\bb}{N_c-\no}}\,.\label{(4.1.12)}
\eq

Therefore, on the whole
\bq
\langle\Qt\rangle_{N_c}=\langle\Qt\rangle_{\no}+\langle\Qt\rangle_{N_c-\no}\approx
\mx m_3\Bigl (\frac{\lm}{m_3}\Bigr )^{\frac{\bb}{N_c-\no}}\,, \label{(4.1.13)}
\eq
as it should be, see \eqref{(A.1)}.\\

\subsection{$U(N_c)$, br1 vacua, smaller $\mx\,,\,\, \mx\ll\Lambda^{SU(N_{c}-\no)}_{{\cal N}=2\,\,SYM}$}

When the additional $U^{(0)}(1)$ is introduced, the Konishi anomalies look as in \eqref{(2.2.2)} but $\langle\Qo\rangle_{N_c}$ is dominant in br1 vacua, i.e. (see the footnote \ref{(f4)}, the leading terms only)
\bq
\langle\Qo+\Qt\rangle_{N_c}=\mx m\,,\quad \langle\Qo\rangle_{N_c}=\mx m-\langle\Qt\rangle_{N_c}\approx\mx m\,,\label{(4.2.1)}
\eq
\bbq
\langle\Qt\rangle_{N_c}\approx\mx m\Bigl (\frac{\lm}{m}\Bigr )^{\frac{2N_c-N_F}{N_c-\no}}\approx
\mx m\,\frac{\langle\Lambda^{SU(N_{c}-\no)}_{{\cal N}=2\,\,SYM}\rangle^2}{m^2}\,,
\eeq
while
\bq
\frac{\langle a_0\rangle}{m}=\frac{1}{\mx m\, N_c}\langle{\rm Tr}\,\qq\rangle_{N_c}\approx\frac{\no}{N_c} +\frac{\nt-\no}{N_c}\Bigl (\frac{\lm}{m}\Bigr )^{\frac{2N_c-N_F}{N_c-\no}}
\approx\frac{\no}{N_c}\,.\label{(4.2.2)}
\eq
Instead of \eqref{(4.1.4)} the superpotential of the $SU(\no)\times U^{(0)}(1)\times U(1)$ part looks now as
\bbq
{\cal W}^{\,\rm low}_{\rm matter}=\w_{\no}+\w_{a_0,a}+\dots\,,
\eeq
\bq
{\cal W}_{\no}=( m-a_0-a){\rm Tr}({\ov Q} Q)_{\no}-{\rm Tr}({\ov Q}\sqrt{2} X^{adj}_{SU(\no)} Q)+\mx {\rm Tr}( X^{adj}_{SU(\no)})^2,\,\,\,\label{(4.2.3)}
\eq
\bbq
\w_{a_0,a}=\frac{\mx}{2}N_c a^2_0+\frac{\mx}{2}\frac{\no N_c}{N_c-\no} a^2\,,
\eeq
where dots denote smaller power corrections.

From \eqref{(4.2.3)} (for the leading terms)
\bbq
\langle a_0\rangle\approx\frac{\no}{N_c} m\,,\,\,\,\langle a\rangle=\langle m-a_0\rangle\approx\frac{N_c-
\no}{N_c} m\,,\,\,\, \langle{\rm Tr}\,({\ov Q} Q)\rangle_{\no}\approx\mx N_c\langle a_0\rangle
\approx\mx\frac{\no N_c}{N_c-\no}\langle a\rangle\approx\no \mx m\,,
\eeq
\bq
\langle X^{adj}_{SU(\no)}\rangle=0,\,\,\langle\Qo\rangle^{U(N_c)}_{\no}=\sum_{a=1}^{\no}\langle{\ov Q}^{\,a}_1 Q^1_a\rangle=\langle{\ov Q}^{\,a}_1\rangle\langle Q^1_a\rangle\approx\mx m\approx\langle\Qo\rangle^{U(N_c)}_{N_c},\,\, \langle\Qt\rangle_{\no}=0.\quad\label{(4.2.4)}
\eq

The qualitative difference with section 4.1 is that one extra ${\cal N}=1$ photon multiplet remains massless now in this $U(N_c)$ theory while corresponding scalar multiplet has the smallest mass $\sim\mx$ because ${\cal N}=2$ is broken down to ${\cal N}=1$ at the level $O(\mx)$\,.

The calculation of the leading power corrections to $\langle{\cal W}^{\,\rm low}_{\rm tot}\rangle$ and $\langle\Qo\rangle^{U(N_c)}_{\no}$ in these $U(N_c)$ vacua is presented in Appendix A.

\subsection{$SU(N_c)$, br1 vacua, larger $\mx\,,\, \Lambda^{SU(N_{c}-\no)}_{{\cal N}=2\,\,SYM}\ll\mx\ll m$}

Nothing changes significantly in this case with the $SU(\no)\times U(1)$ part in \eqref{(4.1.4)}. The mean value $\langle a\rangle=m\gg\mx$ of the field $\langle\sqrt{2}\,X_{U(1)}\rangle$ in \eqref{(4.1.4)} stays intact as far as $\mx\ll m$, and so the contributions $\sim (\mx m)^{1/2}$ to the masses of $\no^2-1$ long ${\cal N}=2\,\,\,SU(\no)$ multiplets of massive gluons and to the mass of the long ${\cal N}=2\,\,\,U(1)$ multiplet of the massive photon due to higgsing of $\no$ quarks remain dominant. At this level the ${\cal N}=2$ SUSY remains unbroken in this sector. But the ${\cal N}=2$ SUSY breaks down to ${\cal N}=1$ at the level $\sim\mx$ by the additional smaller contributions $\sim\mx\ll (\mx m)^{1/2}$ to the masses of corresponding scalar multiplets.

The situation with $SU(N_c-\no)$ SYM in \eqref{(4.1.2)},\eqref{(4.1.3)} is different. At $\langle\Lambda^{SU(N_c-\no)}_{{\cal N}=2\,\,SYM}\rangle\ll\mx\ll m$, the adjoint scalars $X^{adj}_{SU(N_c-\no)}$ of the $SU(N_c-\no)$ subgroup in \eqref{(4.1.2)} become too heavy, their $SU(N_c-\no)$ phases fluctuate now freely at all scales $\mu\gtrsim\langle\Lambda^{SU(N_c-\no)}_{{\cal N}=1\,\,SYM}\rangle$ and they are not higgsed. Instead, they decouple as heavy at scales $\mu<\mx^{\rm pole}=g\mx\ll m$ in the weak coupling region and can all be integrated out. There remains ${\cal N}=1 \,\, SU(N_c-\no)$ SYM with the scale factor $\langle\Lambda^{SU(N_c-\no)}_{{\cal N}=1\,\,SYM}\rangle=[\,\mx \langle\Lambda^{SU(N_c-\no)}_{{\cal N}=2\,\,SYM}\rangle^2\,]^{1/3}$ of its gauge coupling, $\,\,\langle\Lambda^{SU(N_c-\no)}_{{\cal N}=2\,\,SYM}\rangle\ll\langle\Lambda^{SU(N_c-\no)}_{{\cal N}=1\,\,SYM}\rangle\ll\mx\ll\, m$,
\bq
\langle S\rangle_{N_c-\no}=\langle\Lambda^{SU(N_c-\no)}_{{\cal N}= 1\,\,SYM}\rangle^3=\mx m_3^2\Bigl (\frac{\lm}{m_3}\Bigr )^{\frac{2N_c-N_F}{N_c-\no}},\quad \frac{\langle\Lambda^{SU(N_c-\no)}_{{\cal N}=1\,\,SYM}\rangle}{\langle\Lambda^{SU(N_{c}-{\rm n}_1)}_{{\cal N}=2\,\,SYM}\rangle}\sim\Biggl (\frac{\mx}{\langle\Lambda^{SU(N_{c}-{\rm n}_1)}_{{\cal N}=2\,\,SYM}\rangle}\Biggr )^{1/3}\gg 1\,. \label{(4.3.1)}
\eq
(The small non-zero value ${\rm Tr\,}\langle (X^{adj}_{SU(N_c-\no)})^2\rangle=(N_c-\no)\langle S\rangle_{N_c-\no}/\mx=(N_c-\no)\langle\Lambda^{SU(N_c-\no)}_{{\cal N}=2\,\, SYM}\rangle^2\ll\langle\Lambda^{SU(N_c-\no)}_{{\cal N}=1\,\, SYM}\rangle^2$, see \eqref{(3.3)}, arises here only due to the Konishi anomaly, i.e. from one-loop diagrams with heavy scalars $X^{adj}_{SU(N_c-\no)}$ and their fermionic superpartners with masses $\sim\mx$ inside, not because $X^{adj}_{SU(N_c-\no)}$ are higgsed).

There is now a large number of strongly coupled gluonia with the mass scale $\sim\langle\Lambda^{SU(N_c-\no)}_{{\cal N}=1\,\,SYM}\rangle$ in this ${\cal N}=1$ SYM. The multiplicity of vacua in this sector remains equal $N_c-\no$ as it should be. All heavier original electric particles with masses $\sim m$ charged with respect to $SU(N_c-\no)$ remain weakly confined, but the string tension is larger now, $\sigma^{1/2}_{{\cal N}=2}\ll\sigma^{1/2}_{{\cal N}=1}\sim\langle\Lambda^{SU(N_c-\no)}_{{\cal N}=1\,\,SYM}\rangle\ll m$.

\subsection{$SU(N_c)$\,, \, special vacua}

The values of quark condensates in these special vacua with $\no=\nd,\, \nt=N_c$, see \cite{ch4,ch6} and \eqref{(4.1.2)},
\bq
\langle\Qo+\Qt-\frac{1}{N_c}{\rm Tr}\,(\qq)\rangle_{N_c}=(1-\frac{\no=\nd}{N_c})\langle\Qo\rangle_{N_c}=\mx m\,,\label{(4.4.1)}
\eq
\bbq
\langle\Qo\rangle_{N_c}=\frac{N_c}{\bb}\mx m=\mx m_3\,,\quad \langle\Qt\rangle_{N_c}=\mx\lm\,,\quad
\langle S\rangle_{N_c}=\frac{\langle\Qo\rangle_{N_c}\langle\Qt\rangle_{N_c}}{\mx}=
\eeq
\bbq
=\Bigl (\frac{\det \langle\qq\rangle_{N_c}}{\lm^{2N_c-N_F}\mx^{N_c}} \Bigr )^{1/\nd}=\langle S\rangle_{2N_c-N_F}=\mx\langle\Lambda^{SU(\bb)}_{{\cal N}=2\,\,SYM}\rangle^2=\mx m_3\lm\,,
\eeq
are exact and valid at any values $m\gtrless\lm$. Therefore, the discrete $Z_{\bb}$ symmetry is formally broken spontaneously therein at large $m\gg\lm$ also, as it is really broken at $m\ll\lm$. But practically they behave at $m\gg\lm$ as the br1 vacua described above in section (4.1) with $\no=\nd\,$. I.e., the factor $\bb$ in their multiplicity $N_{\rm spec}=(\bb) C^{\,\nd}_{N_F}$ originates at $m\gg\lm$ from the multiplicity $\bb$ of the $SU(\bb)\,\,{\cal N}=2$ SYM at $\mx\ll\langle\Lambda^{SU(\bb)}_{{\cal N}=2\,\,SYM}\rangle$ or ${\cal N}=1$ SYM at $\langle\Lambda^{SU(\bb)}_{{\cal N}=1\,\,SYM}\rangle\ll\mx\ll m$, after the breaking $SU(N_c)\ra SU(\nd)\times U(1)\times SU(\bb)$ by $\langle X_{U(1)}\rangle$ at the scale $\mu= m_3\gg\lm$, see \eqref{(4.1.1)}. The factor $C^{\,\nd}_{N_F}$ originates due to higgsing of $\no=\nd$ original electric quarks in the $SU(\nd)$ sector, $\langle\Qo\rangle_{\nd}=\langle{\ov Q}^{\,1}_1
\rangle\langle Q^1_1\rangle=\mx m_3$, see \eqref{(4.1.4)}, and spontaneous breaking $U(N_F)\ra U(\nd)\times U(N_c)$. The two single roots with $( e^{+}-e^{-})\sim\langle\Lambda^{SU(\bb)}_{{\cal N}=2\,\,SYM}\rangle$ of the curve \eqref{(1.2)} at small $\mx$ originate from the ${\cal N}=2\,\,SU(\bb)$ SYM. There are no massless particles at $\mx\neq 0$, except for $2\no\nt$ Nambu-Goldstone multiplets.

\section{Unbroken flavor symmetry,\,\, SYM vacua,\,\, $SU(N_c)$}

These SYM vacua of the $SU(N_c)$ theory have the multiplicity $N_c\,$. $\langle X^{\rm adj}\rangle_{N_c}\sim
\langle\Lambda^{SU(N_c)}_{{\cal N}=2\, SYM}\rangle\ll m$ in these vacua, so that all quarks have here large masses $m\gg\langle\Lambda^{SU(N_c)}_{{\cal N}=2\, SYM}\rangle$ and decouple in the weak coupling regime
at $\mu<m^{\rm pole}$. There remains at the scale $\mu<m^{\rm pole}$ the ${\cal N}=2\,\, SU(N_c)$ SYM with the scale factor of its gauge coupling: $\langle\Lambda^{SU(N_c)}_{{\cal N}=2\, SYM}\rangle^2=(\lm^{\bb} m^
{N_F})^{1/N_c}=m^2\, (\lm/m)^{(\bb)/N_c},\,\,\,\lm\ll\langle\Lambda^{SU(N_c)}_{{\cal N}=2\, SYM}\rangle\ll m$.

If $\mx\ll\langle\Lambda^{SU(N_c)}_{{\cal N}=2\, SYM}\rangle$, then the field $X^{\rm adj}_{SU(N_c)}$ is higgsed at the scale $\mu\sim\langle\Lambda^{SU(N_c)}_{{\cal N}=2\, SYM}\rangle$ in a standard way for the ${\cal N}=2$ SYM, $\langle\sqrt{2} X^{\rm adj}_{SU(N_c)}\rangle\sim\langle\Lambda^{SU(N_c)}_{{\cal N}=2\, SYM}\rangle\,{\rm diag}\,(k_1,k_2,\,...\,k_{N_c}),\,\, k_i=O(1)$ \cite{DS}, $SU(N_c)\ra U^{N_c-1}(1)$. Note that the value $\sim\langle\Lambda^{SU(N_c)}_{{\cal N}=2\, SYM}\rangle$ of $\langle X^{\rm adj}_{SU(N_c)}\rangle$ is consistent with the unbroken $Z_{\bb}$ discrete symmetry.

All original electrically charged gluons and scalars acquire masses $\sim\langle\Lambda^{SU(N_c)}_{{\cal N}=2\, SYM}\rangle$ and $N_c-1$ lighter Abelian pure magnetic monopoles $M_k$ are formed. These are higgsed at $\mx\neq 0$ and $N_c-1$ long ${\cal N}=2$ multiplets of massive dual photons are formed, with masses $\sim (\mx\langle\Lambda^{SU(N_c)}_{{\cal N}=2\, SYM}\rangle)^{1/2}\ll\langle\Lambda^{SU(N_c)}_{{\cal N}=2\, SYM}\rangle$ (there are non-leading contributions $\sim\mx\ll (\mx\langle\Lambda^{SU(N_c)}_{{\cal N}=2\, SYM}\rangle)^{1/2}$ to the masses of corresponding scalars, these break slightly ${\cal N}=2$ down to ${\cal N}=1$ ). As a result, all original electrically charged quarks, gluons and scalars $X$ are weakly confined (i.e. the tension of the confining string, $\sigma^{1/2}_2\sim (\mx\langle\Lambda^{SU(N_c)}_{{\cal N}=2\, SYM}\rangle)^{1/2}$, is much smaller than their masses $\sim m$ or $\sim\langle\Lambda^{SU(N_c)}_{{\cal N}=2\, SYM}\rangle\,$).

All $N_c-1$ double roots of the curve \eqref{(1.2)} correspond in this case to $N_c-1$ pure magnetic monopoles (massless at $\mx\ra 0$), while two single roots with $(e^+ - e^-)\sim\langle
\Lambda^{SU(N_c)}_{{\cal N}=2\, SYM}\rangle$ also originate from this ${\cal N}=2 \,\,SU(N_c)$ SYM.\\

The mass spectrum is different if $\langle\Lambda^{SU(N_c)}_{{\cal N}=2\, SYM}\ll\mx\ll m$. All scalar fields $X^{\rm adj}_{SU(N_c)}$ are then too heavy and not higgsed. Instead, they decouple as heavy at scales $\mu<\mx^{\rm pole}$, still in the weak coupling regime, and there remains at lower energies ${\cal N}=1\,\, SU(N_c)$ SYM with its $N_c$ vacua and with the scale factor $\langle\Lambda^{SU(N_c)}_{{\cal N}=1\, SYM}\rangle$ of its gauge coupling, $\langle\Lambda^{SU(N_c)}_{{\cal N}=1\, SYM}\rangle=[\,\mx(\langle
\Lambda^{SU(N_c)}_{{\cal N}=2\, SYM}\rangle)^{\,2}\,]^{1/3},\,\, \langle\Lambda^{SU(N_c)}_{{\cal N}=2\, SYM}\rangle\ll\langle\Lambda^{SU(N_c)}_{{\cal N}=1\, SYM}\rangle\ll\mx\ll m$. A large number of strongly coupled gluonia with the mass scale $\sim\langle\Lambda^{SU(N_c)}_{{\cal N}=1\, SYM}\rangle$ is formed in this ${\cal N}=1$ SYM theory, while all heavier charged particles are weakly confined (the tension of the confining string, $\sigma^{1/2}_1\sim\langle\Lambda^{SU(N_c)}_{{\cal N}=1\, SYM}\rangle$, is much smaller than the quark masses $\sim m$ or the scalar masses $\sim\mx$).

\section{Very special vacua with $\langle S\rangle_{N_c}=0$ in $U(N_c)$ gauge theory}
\numberwithin{equation}{section}

We consider in this section the vs (very special) vacua with $\no=\nd,\,\nt=N_c,\, N_c\leq N_F<2 N_c-1$\,, $\langle S\rangle_{N_c}=0$ (\,as usual, $\langle S\rangle_{N_c}$ is the gluino condensate summed over all its colors) and the multiplicity $N_{vs}=C^{\nd}_{N_F}=C^{N_c}_{N_F}$ in the $U(N_c)=SU(N_c)\times U^{(0)}(1)$ theory, when the additional Abelian $U^{(0)}(1)$ with $\mu_0=\mx$ is added to the $SU(N_c)$ theory (see e.g. \cite{SY2} and references therein, these are called $r=N_c$ vacua in \cite{SY2}). Note that vacua with $\langle S\rangle_{N_c}=0$ are absent in the $SU(N_c)$ theory with $m\neq 0,\,\, \mx\neq 0$, see Appendix in \cite{ch4}.

The superpotential of this $U(N_c)$ theory looks at high energies as
\bq
{\cal W}_{\rm matter}=\mx\Bigl [{\rm Tr}\,(X_0)^2=\frac{1}{2}N_c a_0^2\Bigr ]+\mx{\rm Tr}\,(X^{\rm adj}_{SU(N_c)})^2+{\rm Tr}\,\Biggl [\,(m-a_0)\,{\ov Q} Q-{\ov Q}\sqrt{2} X^{adj}_{SU(N_c)} Q \Biggr ]_{N_c}\,,\label{(6.1)}
\eq
\bbq
\sqrt{2} X_{0}=a_0\, {\rm diag}(\,\underbrace{\,1}_{N_c}\,)\,,\quad
X^{\rm adj}_{SU(N_c)}=X^A T^A\,,\,\, {\rm Tr}\,(T^A T^B)=\frac{1}{2}\,\delta^{AB}\,, \quad A, B=1,\,...,\,N_c^2-1\,.
\eeq

Taking $\mx$ sufficiently large
\footnote{\,
The scale factor $\Lambda_0$ of the Abelian coupling $g_0(\Lambda_0/\mu)$ is taken sufficiently large, the theory (6.1) is considered only at scales $\mu\ll\Lambda_0$ where $g_0$ is small, the large $\mx$ means here  $\lm\ll\mx\ll\Lambda_0$.
}
and integrating out all $N^2_c$ scalars $X$ as heavy, the superpotential looks then as
\bq
{\cal W}_{\rm matter}=m\,{\rm Tr}\, (\,{\ov Q} Q)_{N_c}-\frac{1}{2\mx}\,\sum_{i,j=1}^{N_F}\,({\ov Q}_j Q^i)_{N_c}({\ov Q}_i Q^j)_{N_c}\,,\quad ({\ov Q}_j Q^i)_{N_c}=\sum_{a=1}^{N_c}({\ov Q}_j^{\,a} Q^i_a)\,.\label{(6.2)}
\eq

From \eqref{(6.2)}, the Konishi anomalies look here as
\bq
\langle\Qo+\Qt\rangle_{N_c}=\mx m\,,\quad\langle S\rangle_{N_c}=\frac{\langle\Qo\rangle_{N_c}\langle\Qt\rangle_{N_c}}{\mx}=0\,,\label{(6.3)}
\eq
\bbq
\langle\Qt\rangle_{N_c}=\mx m\,,\quad \langle\Qo\rangle_{N_c}=0\,,
\eeq
while from \eqref{(6.1)}
\bq
\langle a_0\rangle=\frac{\langle {\rm Tr}({\ov Q}\, Q)\rangle_{N_c}
=\nd\langle\Qo\rangle_{N_c}+N_c\langle\Qt\rangle_{N_c}}{\mx N_c}=\frac{\langle\Qt\rangle_{N_c}}{\mx} = m\,,\label{(6.4)}
\eq
\bbq
\mx\langle{\rm Tr}\,(\sqrt{2} X^{\rm adj}_{SU(N_c)})^2\rangle=(2N_c-N_F)\langle S\rangle_{N_c}+\langle m-a_0\rangle\langle{\rm Tr}\,(\qq)\rangle_{N_c}=0\,.
\eeq
Note that, as a consequence of a unique property $\langle S\rangle_{N_c}=0$ of these $U(N_c)$ vs -vacua, the curve \eqref{(1.2)} at small $\mx\ra 0$ for theory \eqref{(6.1)} with equal mass quarks has not $N_c-1$ but $N_c$ double roots, $\nd$ equal double roots $e_k= - m$ corresponding to $SU(\nd)$, and $\bt=2N_c-N_F$ unequal double roots $e_j= -m+\omega^{j-1}\lm$, see e.g. \cite{Bo}. There are no single roots of the curve \eqref{(1.2)} in these vacua.

\subsection {$m\gg\lm\,,\,\, \mx\gg\lm^2/m$}
\numberwithin{equation}{subsection}

\hspace*{4mm} Consider first this simplest case with $0<\mx\ll\lm,\,\,m\gg\lm$ and $\mx m\gg\lm^2$. Note that $\langle a_0\rangle=m$ "eats" all quark masses $m$ in \eqref{(6.1)}, see \eqref{(6.4)}. In this region of parameters, $\nt=N_c$ quarks are higgsed {\it in the weak coupling regime} \eqref{(6.3)}, e.g. (compare with \eqref{(2.4.19)},\eqref{(2.4.20)}),
\bq
\langle{\ov Q}_i^{\, b}\rangle=\langle Q^i_b\rangle= (\mx m)^{1/2}\gg\lm\,,\quad  b=1...N_c\,,\quad i=\nd+1...N_F\,,\label{(6.1.1)}
\eq
\bbq
\langle X^{adj,A}_{SU(N_c)}\rangle={\sqrt 2}\,{\rm Tr\,}\Bigl [\langle{\ov Q}\rangle T^{A}\langle Q\rangle\Bigr ]/\mx =0\,,\quad A=1...N_c^2-1\,,\quad \langle a_0\rangle=m\,,\quad \langle S\rangle_{N_c}=0\,.
\eeq

Higgsed quarks with $N_c$ colors and $\nt=N_c$ flavors give in this case large masses $\sim (\mx m)^{1/2}\gg\lm$ to all $N_c^2$ gluons, $a_0$, and $X^{\rm adj}_{SU(N_c)}$, and simultaneously {\it prevent} these adjoint scalars (which are light by itself, in the sense $\mx\ll\lm$) from higgsing.

From \eqref{(6.3)},\eqref{(6.1.1)}, the multiplicity of these vs-vacua is $N_{vs}=C^{N_c}_{N_F}$, the factor $C^{N_c}_{N_F}$ originates in this phase from the spontaneous flavor symmetry breaking, $U(N_F)\ra U(N_c)\times U(\nd)$, due to higgsing of $\nt=N_c$ out of $N_F$ quarks. This multiplicity shows that the non-trivial at $2N_c-N_F\geq 2$ discrete $Z_{\bb}$ symmetry is {\it unbroken}.

As far as $\mx$ remains sufficiently larger than $\lm^2/m$, the mass spectrum includes in this phase $N_c^2$ long ${\cal N}=2$ multiplets of equal mass gluons and their ${\cal N}=2$ superpartners (up to small corrections $\sim\mx\ll\lm\ll (\mx m)^{1/2}$ to the masses of corresponding scalar multiplets),
all with masses $\mu_{\rm gl}\sim g(\mx m)^{1/2}\gg\lm$, and $2\no\nt=2\nd N_c$ massless Nambu-Goldstone multiplets (these are remained original electric quarks with $\nd$ flavors and $N_c$ colors). There are no heavier particles with masses~ $~\sim m$.

\subsection{$m\gg\lm\,,\,\, \mx\ll\lm^2/m\,,\,\, N_c\leq N_F < 2 N_c-1$}
\numberwithin{equation}{subsection}

\hspace*{4mm} Consider now the case of smaller $\mx$, such that $(\mx m)^{1/2}\ll\lm$, while $m$ remains large, $m\gg\lm$. The quark condensates still look as in \eqref{(6.3)}, and $\,\langle a_0\rangle=m$ {\it stays intact} and still "eats"  all quark masses $"m"$ in \eqref{(6.1)}. Therefore, because the $SU(N_c)$ theory is UV free, its gauge coupling grows logarithmically with diminished energy and, if nothing prevents, it will become $g^2(\mu)<0$ at $\mu<\lm$. Clearly, even if quarks were higgsed as above in section 6.1, i.e. $\langle Q^2\rangle=\langle{\ov Q}_2\rangle\sim (\mx m)^{1/2}$ (while $\langle X^{adj}_{SU(N_c)}\rangle
\ll\lm$), see \eqref{(6.1.1)}, this will result only in appearance of small particle masses $\sim (\mx m)^{1/2}\ll\lm$, so that {\it all particles will remain effectively massless at the scale $\mu\sim\lm\gg (\mx m)^{1/2}$} (this is especially clear in the unbroken ${\cal N}=2$ theory at $\mx= 0$). Therefore, this will not help and the problem with $g^2(\mu<\lm)<0$ cannot be solved in this way (these quarks ${\ov Q}_2, Q^2$ are really not higgsed now at all, i.e. $\langle Q^i_b\rangle=\langle{\ov Q}_i^{\,b}\rangle=0,\, i=\nd+1...N_F,\, b=1...N_c$, see below).

As explained in Introduction, to really avoid the unphysical $g^2(\mu<\lm)<0$, the field $X^{adj}_{SU(N_c)}$ is higgsed necessarily in this case at $\mu\sim\lm$. Because the non-trivial $Z_{\bb\geq 2}$ discrete symmetry is {\it unbroken} in these vs -vacua, this results, as in section 2.1, in $SU(N_c)\ra SU(\nd)\times U^{(1)}(1)\times U^{\bb-1}(1)$, and main contributions $\sim\lm$ to particle masses originate now from this higgsing of $X^{adj}_{SU(N_c)}$,
\bq
\langle\, X^{adj}_{SU(N_c)}\rangle=\langle  X^{adj}_{SU(\nd)}+X^{(1)}_{U(1)}+ X^{adj}_{SU(\bb)}\rangle,\label{(6.2.1)}
\eq
\bbq
\langle \sqrt{2} X^{adj}_{SU(\bb)}\rangle\sim\lm\,{\rm diag}\,\Bigl (\,\underbrace{0}_{\nd}\,;\,
\underbrace{\omega^0,\,\omega^1,\,...\,\omega^{\bb-1}}_{\bb} \,\Bigr )\,,\quad \omega=\exp\{\frac{2\pi i}{\bb}\}\,,
\eeq
\bbq
\sqrt{2}\, X^{(1)}_{U(1)}=a_{1}\,{\rm diag}(\,\underbrace{\,1}_{\nd}\,;\,\underbrace{\,c_1}
_{\bb}),\, c_1=-\,\frac{\nd}{\bb},\,\, \langle Q^i_k\rangle=\langle{\ov Q}_i^{\,k}\rangle=0,\,\,k=\nd+1...\,N_c,\, i=1...\,N_F.
\eeq

Now, vice versa, higgsed $\langle X^{A}_{SU(\bb)}\rangle\sim\lm$ {\it prevent} quarks $Q^i_k,\, {\ov Q}^{\,k}_i,\, k=\nd+1...\,N_c,\, i=1...\,N_F$ from higgsing, i.e. $\langle Q^i_k\rangle=\langle{\ov Q}_i^{\,k}\rangle=0$. The reason is that these quarks acquire now masses $\sim\lm$, much larger than a potentially possible scale $\sim (\mx m)^{1/2}$ of their coherent condensate, see \eqref{(6.1.1)}, while their light physical $U^{2N_c-N_F}(1)$ phases fluctuate independently and freely at least at all scales $\mu > \mu^{\rm low}_{\rm cut}=(\rm several){\rm max}\{(\mx m)^{1/2},\, (\mx\lm)^{1/2} \}$. I.e., in any case, $\langle Q^i_k\rangle_{\mu^{\rm low}_{\rm cut}}=\langle{\ov Q}_i^{\,k}\rangle_{\mu^{\rm low}_{\rm cut}}=0$. Clearly, this concerns also all other heavy charged particles with masses $\sim\lm$. Therefore, after integrating out the theory \eqref{(6.1)} over the interval from the high energy down to the scale $\mu^{\rm low}_{\rm cut}$, all particles with masses $\sim\lm$ {\it decouple} as heavy already at the scale $\mu<\lm/(\rm several)$, i.e. much above $\mu^{\rm low}_{\rm cut}$, and definitely {\it the heavy quarks by itself do not give any contributions to masses of remaining lighter degrees of freedom in the Lagrangian \eqref{(6.2.2)}} because $\langle Q^i_k\rangle_{\mu^{\rm low}_{\rm cut}}=\langle{\ov Q}_i^{\,k}\rangle_{\mu^{\rm low}_{\rm cut}}=0$. This is qualitatively the same situation as with heavy quarks with masses $\sim\lm$ discussed in detail in section 2.1, and with $X^{adj}_{SU(\nd)}$ with masses $\sim\mx\gg (\mx m)^{1/2}$ in section 2.4 (and only $m\gg\lm$ in this section while $m\ll\lm$ in sections 2.1 and 2.4, while always $\mx\ll\lm$).

The lower energy original electric ${\cal N}=2\,\,SU(\nd)$ theory at the scale $\mu=\mu^{\rm low}_{\rm cut}\ll\lm$, with $2\nd<N_F<2N_c-1$ flavors of remained light quarks $Q^i_a, {\ov Q}_j^{\,a},\,\, a=1...\nd,\,\, i,j=1...N_F$, is IR free and weakly coupled. The whole matter superpotential has the same form as \eqref{(2.1.5)}, but with  omitted $a_2$ and $SU(\nd-\no)$ SYM part, $\no=\nd$, $m\ra (m-a_0)$, and with the addition of $\sim\mx a_0^2\,$,
\bq
\w_{N_c}=\w_{\,\nd}+\w_{D}+\w_{a_0,\,a_1}\,,\label{(6.2.2)}
\eq
\bbq
\w_{\,\nd} =(m-a_0-a_1){\rm Tr}\,({\ov Q} Q)_{\nd}-{\rm Tr}\,\Bigl ({\ov Q}\sqrt{2}X_{SU(\nd)}^{\rm adj} Q\Bigr )_{\nd}+\mx (1+\delta_2){\rm Tr}\,(X^{\rm adj}_{SU(\nd)})^2\,,
\eeq
\bbq
\w_{D}=\Bigl ( m-a_0-c_1 a_1 \Bigr )\sum_{j=1}^{\bb}{\ov D}_j D_j\,-\sum_{j=1}^{\bb} a_{D,j}{\ov D}_j D_j
-\,\mx\lm \sum_{j=1}^{\bb}\omega^{j-1} a_{D,j}+\mx  L\,\Bigl (\,\sum_{j=1}^{\bb} a_{D,j}\Bigr )\,,
\eeq
\bbq
\w_{a_0,\,a_1}=\frac{\mx}{2}N_c\,a^2_0+\frac{\mx}{2}\frac{\nd N_c}{\bt}\,(1+\delta_1)\,a^2_1+
\mx N_c\delta_3\, a_1(m-a_0-c_1 a_1)+\mx N_c\delta_4\, (m-a_0-c_1 a_1)^2\,,
\eeq
where $\delta_1$ and $\delta_2$ are the same as in \eqref{(2.1.12)} (and $\delta_3=0$, see \eqref{(B.4)}).

Proceeding now similarly to section 2.1, we obtain from \eqref{(6.2.2)}
\bq
\langle a_0\rangle=m\,,\quad \langle a_1\rangle=0\,,\quad \langle a_{D,j}\rangle=0\,,
\quad \langle X_{SU(\nd)}^{\rm adj}\rangle=0\,,\label{(6.2.3)}
\eq
\bbq
\langle{\ov D}_j D_j\rangle=\langle{\ov D}_j\rangle\langle D_j \rangle=\,-\mx\lm\,\omega^{j-1}+\mx \langle L\rangle,\,\, \langle L\rangle=m,\,\, \langle\Sigma_D\rangle=\sum_{j=1}^{\bb}\langle{\ov D}_j\rangle\langle D_j \rangle=(\bb)\mx m,
\eeq
\bbq
\langle\Qo\rangle_{\nd}=\langle{\ov Q}^1_1\rangle\langle Q^1_1\rangle=\mx m=\langle\Qt\rangle_{N_c},\,\, \langle\Qt\rangle_{\nd}=\sum_{a=1}^{\nd}\langle{\ov Q}^{\,a}_2\rangle\langle Q^2_a\rangle=0=\langle\Qo\rangle_{N_c}\,,
\eeq
\bbq
\langle{\rm Tr} ({\ov Q} Q)\rangle_{\nd}=\nd\langle\Qo\rangle_{\nd}+N_c\langle\Qt\rangle_{\nd}=\nd\mx m\,, \quad \wmu\langle S\rangle_{\nd}=\langle\Qo\rangle_{\nd}\langle\Qt\rangle_{\nd}=0.\,\,\,
\eeq

From \eqref{(6.2.3)}, the multiplicity of these vs -vacua in this phase is $N_{vs}=C^{\nd}_{N_F}=C^{N_c}_{N_F}$ as it should be, the factor $C^{\nd}_{N_F}$ originates now from the spontaneous flavor symmetry breaking, $U(N_F)\ra U(\nd)\times U(N_c)$, due to higgsing of $\nd$ quarks in the $SU(\nd)$ color sector.

On the whole, the mass spectrum looks now as follows.\\
1) Due to higgsing of $X^{adj}_{SU(\bb)}$ at the scale $\sim\lm$, $\,\,SU(N_c)\ra SU(\nd)\times U^{(1)}(1)\times U^{\bb-1}(1)$, there is a large number of original pure electrically $SU(\bb)$ charged particles with the largest masses $\sim\lm$. They all are weakly confined due to higgsing of mutually non-local with them $\bb$ BPS dyons $D_j, \ov D_j$ with the non-zero $SU(\bb)$ adjoint magnetic charges, the string tension is $\sigma^{1/2}_D\sim \langle{\ov D}_j D_j\rangle^{1/2}\sim (\mx m)^{1/2}\ll\lm$. These original electrically charged particles form hadrons with the mass scale $\sim\lm\,$. {\it The global flavor symmetry $U(N_F)$ is unbroken in this heaviest sector}. All heavy quarks have equal masses $|\omega^{j-1}
\lm|=|\lm|$ and are in the (anti)fundamental representation of $U(N_F)$, while all heavy gluons and scalars are flavorless. There is no color-flavor locking in this heaviest sector. \\
2) Due to higgsing of these dyons $D_j, {\ov D}_j$, there are $\bb$ long ${\cal N}=2\,\, U(1)$ multiplets of massive photons, all with masses $\sim \langle{\ov D}_j D_j\rangle^{1/2}\sim (\mx m)^{1/2}$.\\
3) There are $\nd^{\,2}$ long ${\cal N}=2$ multiplets of massive gluons with masses $\sim \langle\Qo\rangle^{1/2}_{\nd}\sim (\mx m)^{1/2}\ll\lm$ due to higgsing of $\nd$ flavors of original electric quarks with $SU(\nd)$ colors.\\
4) $2 \no\nt=2\nd N_c$ Nambu-Goldstone multiplets are massless (in essence, these are remained original electric quarks with $\nt=N_c=N_F-\nd$ flavors and $SU(\nd)$ colors). There are no other massless particles at $\mx\neq 0$.\\
5) There are no particles with masses $\sim m\gg\lm$.

The $U(N_c)$ curve \eqref{(1.2)} has in this phase at $\mx\ra 0$\,\, $\nd$ equal double roots $e_k= - m$ corresponding to $\nd$ higgsed original electric quarks with $SU(\nd)$ colors, and $\bb$ unequal double roots $e_j= -m+\omega^{j-1}\lm$ corresponding to dyons $D_j$.\\

In essence, the situation here with the heavy non-higgsed quarks $Q^i_k, {\ov Q}^{\,k}_i,\, i=1...N_F,\, k=\nd+1...N_c$ with masses $\sim\lm$, and with their chiral bilinear condensates (really, more precisely is to speak about mean values, using the word "condensate" only for "genuine", i.e. coherent, condensates of higgsed fields) {\it is the same} as in br2 vacua of section 2.1, see \eqref{(2.1.16)}-\eqref{(2.1.26)} and accompanying text. I.e., in the lower energy phase of this section at $\mx\ll\lm^2/m$, the non-zero mean values of {\it non-factorizable} bilinears of these heavy quarks with masses $\sim\lm$ originate from: a) the quantum loop effects at the scale $\sim\lm$ transforming bilinears of heavy quark fields into bilinears of light dyon and quark fields, b) and finally, from formed at much lower scales $\mu\sim (\mx m)^{1/2}\ll\lm$ the "genuine" (i.e. coherent) condensates of light higgsed quarks and dyons, $\langle{\ov Q}^1_1\rangle=\langle Q^1_1\rangle\neq 0$, and $\langle{\ov D}_j\rangle=\langle D_j\rangle\neq 0$  see \eqref{(6.2.3)},\eqref{(6.2.4)}.

In this section, the corresponding values of condensates of heavy quarks with $SU(\bb)$ colors, $N_F$ flavors, and masses $\sim\lm$ in these vs-vacua with $\nt=N_c,\, \no=\nd$, and $\langle S\rangle_{N_c}=0$ differ from those in the br2 vacua with $\nt>N_c,\, \no<\nd$ and $\langle S\rangle_{N_c}\neq 0$ in section 2.1 only by: 1) $m_1\ra m$ due to $SU(N_c)\ra U(N_c)$,\,\, 2) the absence of non-leading power corrections originating in br2 vacua from the SYM part, see \eqref{(6.3)},\eqref{(6.2.3)}, compare with \eqref{(6.1.1)}, \eqref{(2.1.18)},\eqref{(2.1.19)},
\bq
\langle\Qt\rangle_{N_c}=\langle\Qt\rangle_{\bb}+[\langle\Qt\rangle_{\nd}=\sum_{a=1}^{\nd}\langle{\ov Q}^{\,a}_2\rangle\langle Q^2_a\rangle=0\,]=\mx m\,, \label{(6.2.4)}
\eq
\bbq
\langle\Qo\rangle_{N_c}=[\langle\Qo\rangle_{\nd}=\langle{\ov Q}^{\,1}_1\rangle
\langle Q^1_1\rangle=\mx m\,]+[\langle\Qo\rangle_{\bb}= - \mx m\,]=0\,,
\eeq
\bbq
\langle{\rm Tr\,}\qq\rangle_{\bb}=N_c\langle\Qt\rangle_{\bb}+\nd\langle\Qo\rangle_{\bb}=\langle
\Sigma_D\rangle=(2N_c-N_F)\mx m\,,
\eeq
while $\langle \sqrt{2} X^{adj}_{SU(\bb)}\rangle$ in \eqref{(6.2.1)} is the same as \eqref{(2.1.17)} and differs from \eqref{(6.1.1)}.\\

But there is one important qualitative difference between the vs vacua and br2 vacua. Because \eqref{(6.3)},
\eqref{(6.4)} remain the same at $m\lessgtr\lm$, and $\langle a_0\rangle=m$ still "eats" all quark masses $"m"$ in \eqref{(6.1)} at all $m\lessgtr\lm$, the whole physics in vs vacua is qualitatively independent separately of the ratio $m/\lm$ and {\it depends only on a competition between $(\mx m)^{1/2}$ and $\lm$}. For this reason, the lower energy phase with $m\gg\lm,\, \mx\ll\lm,\, (\mx m)^{1/2}\ll\lm$ in vs vacua of this section continues smoothly into the region $\mx\ll m\ll\lm$. This is in contrast with the br2 vacua with $\mx\ll m\ll\lm$ in section 2.1. In these last, all properties depend essentially on the ratio $m/\lm$ and they evolve into the qualitatively different br1 vacua at $m\gg\lm$.

Comparing the mass spectra at $(\mx m)^{1/2}\gg\lm$ in section 6.1 with those in this section at $(\mx m)^{1/2}\ll\lm$ one can see the qualitative difference between these two cases. {\it The lower energy phase with $\mx\ll\lm^2/m$ in this section is, in essence, the phase of the unbroken ${\cal N}=2$ theory because it continues down to $\mx\ra 0$. While the higher energy phase of section 6.1 with $\mx\gg\lm^2/m$, because it cannot be continued as it is down to $\mx\ra 0$, is, in essence, the phase of the strongly broken ${\cal N}=2$ theory (i.e. already the genuine ${\cal N}=1$ theory in this sense)}. Remind that the applicability of the whole machinery with the use of roots of the curve \eqref{(1.2)} is guaranteed only at $\mx\ra 0$. I.e., it is applicable only to the lower energy phase of this section, and clearly not applicable to the higher energy phase of section 6.1, in spite of that $\mx\ll\lm$ therein also.

And these two phases at $\mx\gtrless\lm^2/m$ are qualitatively different, compare e.g. \eqref{(6.1.1)} and \eqref{(6.2.1)}. In the higher energy phase in section 6.1 all charged gluons from $SU(\bb)$ have {\it exactly equal masses $| g (\mx m)^{1/2}|$ in the whole region from $(\mx m)^{1/2}\gg\lm$ down to $(\mx m)^{1/2} >(\rm several)\lm$}, see  \eqref{(6.1.1)}. (Otherwise, if $\langle X^{adj}_{SU(\bb)}\rangle$ were higgsed as in \eqref{(6.2.1)}, it will give then in particular the non-zero masses $\sim\lm\ll (\mx m)^{1/2}$ to $2\nd (\bb)$ out of previously massless $2\nd N_c$ Nambu-Goldstone multiplets in section 6.1. And this will be clearly wrong because the number of massless Nambu-Goldstone multiplets has to be exactly $2\nd N_c$).

On the other hand, in the lower energy phase of this section, the masses of these $SU(\bb)$ gluons differ greatly at $\mx\ra 0:\, \mu_{\rm gl}^{ij}\sim |(\omega^{i}-\omega^{j})\lm |\,,\,\, i\neq j,\,\,i,j=0...\bb-1$. This shows that the dependence of masses of these gluons on $\mx m$ is non-analytic. Besides, the mean values of $\langle X^{adj}_{SU(N_c)}\rangle$ and $\langle Q^i_k\rangle=\langle{\ov Q}_i^{\,k}\rangle,\, k=\nd+1...\,N_c,\, i=1...\,N_F$ behave non-analytically, compare \eqref{(6.1.1)} and \eqref{(6.2.1)}. In this sense, we can say that there is a phase transition somewhere at $(\mx m)^{1/2}\sim\lm$. This is in contradiction with a widespread opinion that there are no phase transitions in the supersymmetric theories because the gauge invariant condensates of chiral fields ( e.g. $\langle (\qq)_{1,2}\rangle_{N_c},\,\langle S\rangle_{N_c},\, \,\langle{\rm Tr\,}(X^{adj}_{SU(N_c)})^2\rangle$ in \eqref{(6.3)},\eqref{(6.4)}), depend holomorphically on the superpotential parameters.

This phenomenon with the phase transition of the type described above occurring at very small $\mx\ll\lm$ to the strongly broken ${\cal N}=2$, i.e. ${\cal N}=2\ra {\cal N}=1$, where the use of roots of the curve \eqref{(1.2)} becomes not legitimate, appears clearly {\it only} at $m\gg\lm$. Let us start e.g. with the nearly unbroken ${\cal N}=2$ theory at {\it sufficiently small} $\mx$ where, in particular, the whole machinery with the use of roots of the curve \eqref{(1.2)} is still legitimate. In this section, when $\mx$ begins to increase, the phase transition occurs at $\mx\sim (\lm^2/m)\ll\lm$. But in other cases, see sections 7 and 8 below, it occurs even at parametrically smaller values of $\mx$ than in this section.
In br2 vacua of section 7 it occurs at $\mx\sim \langle\Lambda_{SU(\nt)}\rangle^2/m,\, \langle\Lambda_
{SU(\nt)}\rangle\ll\lm$, see \eqref{(7.1.6)}, and in S- vacua of section 8 it occurs at $\mx\sim \langle\Lambda_{SU(N_F)}\rangle^2/m,\, \langle\Lambda_{SU(\nt)}\rangle\ll\lm$, see \eqref{(8.4)}. \\

Let us compare now the corresponding parts of global symmetries of two phases, those in section 6.1 and here. It is $SU(N_c)_{C+F}\times SU(\nd)\times U(1)$ in section 6.1, where "C+F" denotes the color-flavor locking. The higgsed massive electric quarks $Q^i_b, {\ov Q}^{\,b}_i,\,\, b=1...N_c,\, i=\nd+1...N_F$, with masses $m_Q\sim g(\mx m)^{1/2}\gg\lm$, together with massive electric $SU(N_c)$ gluons and $X^{adj}_{SU(N_c)}$ scalars form the long ${\cal N}=2$ multiplet in the adjoint representation of  $SU(N_c)_{C+F}$ (and one $SU(N_c)$ singlet multiplet). The $2\nd N_c$ Nambu-Goldstone multiplets are bifundamental.

The global symmetry in this section is $SU(\nd)_{C+F}\times SU(N_c)\times U(1)$ in the lighter sector, and higgsed massive electric quarks $Q^i_a, {\ov Q}^{\,a}_i,\,\, a,i=1...\nd$, with masses $m_Q\sim g(\mx m)^{1/2}\ll\lm$, together with $SU(\nd)$ gluons and $X^{adj}_{SU(\nd)}$ scalars form the long ${\cal N}=2$ multiplet in the adjoint representation of $SU(\nd)_{C+F}$  (and one $SU(\nd)$ singlet multiplet), see the point "3" just above. The $2\nd N_c$ Nambu-Goldstone multiplets are also bifundamental. But, in addition, all original heavy quarks ${\ov Q}^{\,k}_i, Q^i_k,\,i=1...N_F,\, k=\nd+1...N_c$, have equal masses $\sim |\omega^j\lm|=|\lm |$. At $\mx=0$ they are not confined and form (anti)fundamental representation of $U(N_F)$. All original heavy unconfined at $\mx=0$ $\,SU(\bb)$ adjoint and hybrid gluons and scalars with masses $\sim\lm$ are clearly flavor singlets. At small $\mx\neq 0$ all these heavy charged particles are weakly confined and form hadrons with masses $\sim\lm$. The heavy quarks form e.g. a number of the $SU(N_F)$ adjoint hadrons.

The difference of adjoint representations $SU(N_c)_{C+F}\leftrightarrow SU(\nd)_{C+F}$ at $\mx m\gtrless
\lm^2$ was considered in \cite{SY4},\cite{SY2} as an evidence that all $SU(\nd)$ particles cannot be the same original pure electric particles and, by analogy with \cite{SY3}, it was claimed that they are all dyons with non-zero magnetic charges. As it is seen from the above, this line of reasonings is not right. The reason for different adjoint representations is that the phase and mass spectrum are determined by $N_c$ out of $N_F$ higgsed quarks of $SU(N_c)$ at $\mx m\gg\lm^2$ (while $X^{adj}_{SU(N_c)}$ is not higgsed), and they are determined by higgsed both $X^{\rm adj}_{SU(2N_c-N_F)}$ and $\nd$ out of $N_F$ quarks of $SU(\nd)$ at $\mx m\ll\lm^2$ (both phases in section 6.1 and here at $m\gg\lm$, and the case $\mx\ll m\ll\lm$ clearly corresponds to the regime $\mx m <\lm^2$, see section 6.4 below).

Besides, it is worth emphasizing that it is erroneous to imagine that, because $\langle\Qt\rangle_{N_c}=\mx m$ still in the $\mx m\ll\lm^2$ phase of this section, then all $\nt=N_c$ quarks are still higgsed as previously in section 6.1 at $\mx m\gg\lm^2$, i.e. $\langle{\ov Q}^{\,b}_i\rangle=\langle Q^i_b\rangle= (\mx m)^{1/2},\, b=1...N_c,\, i=\nd+1...N_F$, and so give corresponding contributions to particle masses, but because $(\mx m)^{1/2}\ll\lm$ now, this will be of little importance. This is not so, because such higgsed quarks will give not only small corrections $\sim (\mx m)^{1/2}\ll\lm$ to masses $\sim\lm$ of heavy particles, but all $2\nd N_c$ massless Nambu-Goldstone multiplets in \eqref{(6.2.3)} will also acquire additional masses $\sim (\mx m)^{1/2}$ in this case, and this is clearly unacceptable. (See also a more detailed discussion of similar situation with the heavy quark condensates in br2 vacua, \eqref{(2.1.16)}
-\eqref{(2.1.34)}).

And finally, we would like to especially emphasize (and this concerns the whole content of this paper, not only those of this section) that, for {\it higgsed fields giving non-zero contributions to particle masses}, {\it the mean values of fields itself should be non-zero}, i.e. $\langle Q\rangle=\langle{\ov Q}\rangle\neq 0$ or $\langle X\rangle\neq 0$ (but even this may be sometimes insufficient as explained e.g. at the end of section 2.4), {\it not only} something like $\langle Q^{\dagger} Q\rangle^{1/2}\neq 0$ or $\langle X^{\dagger} X\rangle^{1/2}\neq 0$ for the D-terms, or $\langle{\ov Q} Q\rangle^{1/2}\neq 0,\, \langle X^2\rangle^{1/2}\neq 0$ for the F-terms (which all are, as a rule, non-zero even for those fields which are really not higgsed).

This is especially clearly seen  e.g. from the mass terms of fermionic superpartners. For the D-terms of {\it higgsed} quarks or $X$ these look as:
\bq
{\rm Tr\,}[\,(\,\langle Q^{\,\dagger}\rangle\, \chi+\langle{\ov Q}^{\,\dagger}\rangle\,{\ov\chi}\,)\lambda\,]\quad {\rm or}\quad {\rm Tr\,}\Bigl [\,\langle X^{\dagger}\rangle\,\psi\lambda\,\Bigr ]\,,\label{(6.2.5)}
\eq
and for the F-terms they look as
\bq
{\rm Tr\,}\Bigl [\,(\,\langle Q\rangle\,\chi+\langle{\ov Q}\rangle\,{\ov\chi\,)}\psi\,\Bigr ]\quad {\rm or}\quad {\rm Tr\,}\Bigl [\langle X\rangle\,{\ov \chi} \chi\,\Bigr ]\,.\label{(6.2.6)}
\eq

\subsection{$m\ll\lm$}

\hspace*{4mm} Comparing with the case $\mx\ll\lm,\,\, m\gg\lm$ in sections 6.1-6.2, the case $\mx\ll m\ll\lm,$ corresponds clearly the regime $\mx\ll\lm^2/m$.
The quark condensates remain the same as in \eqref{(6.3)}, and $\langle a_0\rangle=m$ remains the same as in \eqref{(6.4)} and still "eats" all quark masses $"m"$. {\it The phase remains the same as in section} 6.2, and at $m\ll\lm$ the difference with the case $m\gg\lm,\,\, \mx\ll\lm^2/m$ in section 6.2 is only quantitative, not qualitative, see \eqref{(6.2.3)}. The main difference is that $\langle D_j\rangle\sim (\mx m)^{1/2}\gg (\mx\lm)^{1/2}$ at $m\gg\lm$, while $\langle D_j\rangle\sim (\mx\lm)^{1/2}\gg (\mx m)^{1/2}$ at $m\ll\lm$. Still, there are no particles with masses $\sim m$, the only particle masses independent of $\mx$ are $\sim\lm$.

\subsection{$N_F=N_c$} 

And finally, in short about the vs vacuum with $N=N_F=N_c=\nt,\, \nd=0=\no$ at $m\lessgtr\lm$ and $\mx\ll\lm$. There is only one such vacuum. And really, there is nothing extraordinary with this case.

\subsubsection{Higher energy phase at $(\mx m)^{1/2}\gg\lm$}

As always in all vs vacua, see \eqref{(6.3)},\eqref{(6.4)}, $\langle\qq\rangle_{N_c}=\mx m,\, \langle S\rangle_{N_c}=0$ in this vacuum, and $\langle a_0\rangle=m$ "eats" all quark masses. The whole color group $U(N_c)$ is broken in this higher energy phase as before in br2 vacua of section 6.1 by higgsed quarks at the scale $\mu\sim g(\mx m)^{1/2}\gg\lm$ in the weak coupling regime, while $\langle X^{adj}_{SU(N_c)}\rangle=0$ is not higgsed, see \eqref{(6.1.1)}. {\it All $N^2$ gluons have equal masses $\sim g(\mx m)^{1/2}$ in this phase in the whole region from $(\mx m)^{1/2}\gg\lm$ down to $(\mx m)^{1/2} >(\rm several)\lm$}. There is the residual global $U(N)_{C+F}$ symmetry (i.e. color-flavor locking). $N^2$ long ${\cal N}=2$ multiplets of massive gluons are formed in the adjoint plus singlet representations of this remained unbroken global symmetry $U(N)_{C+F}$.

There are no massless Nambu-Goldstone particles in this vacuum. And there are no massless particles at all in this phase.

\subsubsection{Lower energy phase at $(\mx m)^{1/2}\ll\lm$}

In this phase, at all $m\gtrless\lm$, to avoid $g^2(\mu<\lm)<0$ in this UV free theory, the non-Abelian $SU(N_c)$ is also broken, but now at the scale $\mu\sim \lm$ by higgsed $\langle X^{adj}_{SU(N_c)}
\rangle\sim\lm,\, SU(N_c)\ra U^{N_c-1}(1)$,
\bq
\langle\sqrt{2} X^{adj}_{SU(N_c)}\rangle\sim\langle\lm\rangle\,{\rm diag}\,\Bigl(\,
\rho^0,\,\rho^1,\,...,\,\rho^{N_c-1}\,\Bigr ),\quad \rho=\exp\{\,\frac{2\pi i}{N_c}\,\}\,. \label{(6.4.1)}
\eq

{\it There is no flavor-color locking, the global flavor symmetry $U(N_F)=U(N)_F$ remains unbroken, while the non-Abelian color $SU(N_c)=SU(N)_C$ is broken down to Abelian one}. The unbroken non-trivial discrete symmetry is $Z_{2N_c-N_F}=Z_{N\geq 2}$ in this vacuum, and it determines the standard pattern of this color symmetry breaking by higgsed $\langle X^{adj}_{SU(N_c)}\rangle$ giving the largest contributions $\sim\lm$ to particle masses. All original charged particles acquire masses $\sim\lm$, while $N_c=N$ light BPS dyons $D_j$ (massless at $\mx=0$) are formed. As before in section 6.2 for heavy quarks with $SU(\bb)$ colors, now {\it all quarks have equal masses} $\sim |\rho^{j}\lm|=|\lm|$, they are too heavy and {\it not higgsed} in this phase, $\langle Q^i_a\rangle=\langle{\ov Q}^{\,b}_k\rangle=0,\, i,k,a,b=1...N$.  All equal mass quarks are clearly {\it in the (anti)fundamental representation} of this unbroken global $U(N)_F$, while {\it the non-Abelian color $SU(N)_C$ is broken at the scale $\mu\sim\lm$ down to Abelian $U^{N-1}(1)$, and masses of all charged flavorless gluons (and scalars) differ greatly at $\mx\ra 0:\, \mu_{\rm gl}^{ij}\sim |(\rho^{i}-\rho^{j})\lm |\,,\,\, i\neq j,\,\,i,j=1...N$}, see \eqref{(6.4.1)}. There is the phase transition at $(\mx m)^{1/2}\sim\lm$.

After all heavy particles with masses $\sim\lm$ decoupled at $\mu=\lm/(\rm several)$, the lower energy superpotential at the scale $\mu=\lm/(\rm several)$ looks as
\bq
\w_{N_F}=\w_{D}+\w_{a}\,, \label{(6.4.2)}
\eq
\bbq
\w_{D}= (m-a_0)\sum_{j=1}^{N}{\ov D}_j D_j -\sum_{j=1}^{N} a_{D,\,j}{\ov D}_j D_j -\mx\lm\sum_{j=1}^{N}\rho^{\,{j-1}} a_{D,\,j}+\mx L\,\Bigl (\sum_{j=1}^{N} a_{D,\,j}\Bigr )\,
\eeq
\bbq
\w_{a}=\frac{\mx}{2} N \,a_0^2+\mx N\,{\hat\delta}_4\, (m-a_0)^2\,.
\eeq

From \eqref{(6.4.2)}
\bq
\langle a_0\rangle=m\,,\quad \langle a_{D,\,j}\rangle=0\,,\quad \langle{\ov D}_j\rangle\langle D_j \rangle= -\mx\lm\rho^{\,{j-1}}+\mx m\,, \quad \langle\Sigma_D\rangle=\sum_{j=1}^{N}\langle{\ov D}_j\rangle\langle D_j\rangle=N_c\mx m\,.\,\,\,\label{(6.4.3)}
\eq

All $N$ dyons are higgsed at small $\mx\neq 0$ in the weak coupling regime and $N$ long ${\cal N}=2$ multiplets of massive photons are formed with masses $\sim\langle{\ov D}_j D_j\rangle^{1/2}\ll\lm$, i.e. $\sim (\mx m)^{1/2}$ at $m\gg\lm$ or $\sim (\mx\lm)^{1/2}$ at $\mx\ll m\ll\lm$. All original heavy charged particles with masses $\sim\lm$ are weakly confined, the string tension is $\sigma^{1/2}\sim\langle{\ov D}_j D_j\rangle^{1/2}\ll\lm$. In particular, the heavy flavored quarks form a number of $SU(N_F)$ adjoint hadrons with masses $\sim\lm$. There are no massless Nambu-Goldstone particles because the global flavor symmetry $U(N_F)$ is unbroken.

Moreover, at $\mx=0$, because $\langle a_0\rangle=m$ "eats" all quark masses, all charged solitons will have masses either $\sim\lm$ or zero. But the $U(N_c)$ curve \eqref{(1.2)} has only $N_c$ unequal double roots $e^{(D)}_j= - m +\rho^{j-1}\lm,\, j=1...N_c$ in this vacuum corresponding to $N_c=N$ massless BPS dyons $D_j$, and this shows that {\it there are no additional charged ${\cal N}=2$ BPS solitons massless at $\mx=0$}. In particular, there are no massless pure magnetic monopoles, they all have masses $\sim\lm$. And there no massless particles at all at small $\mx\neq 0$. Besides, there are no particles with masses $\sim m\gg\lm$, the largest masses are $\sim\lm$.

The global unbroken symmetry in this lower energy phase is $U(N)_F\times U^{N}(1)_C$, i.e. the non-Abelian flavor symmetry is unbroken while the non-Abelian color symmetry is broken down to Abelian, {\it there is no color-flavor locking}. There is no confinement or charge screening at $\mx=0$ due to higgsed quarks or dyons. Therefore, because pure magnetic monopoles with $SU(N_c)$ adjoint charges are flavorless, the flavored quarks {\it can not decay into these flavorless monopoles}. And this remains true in this phase at small $\mx\ll\lm^2/m$ (and not only at $m\gg\lm$, but at $\mx\ll\lm,\, m\ll\lm$ as well), when these heavy non-higgsed quarks in the (anti)fundamental representation of unbroken $U(N)_F$, and with masses $\sim\lm$, are weakly confined. (Remind that the global flavor symmetry $U(N_F)$ also remains unbroken in the sector of heaviest particles with masses $\sim\lm$ in this lower energy phase in br2 vacua of sections 6.2 and 6.3). This is in contradiction with the proposed by M. Shifman and A. Yung "instead-of-confinement" phase, see e.g. the most recent paper \cite{SY6} and refs therein to their previous papers with this "instead-of-confinement" phase in other similar cases.

\subsection{$m\gg\lm\,,\,\, 2N_c-N_F=1$}
\numberwithin{equation}{subsection}

At $\mx m\gg\lm^2$ the behavior of all vs-vacua is the same and is described in section 6.1. Therefore, we consider below only the case $\mx m\ll\lm^2$ when the vs-vacua with $\bb=1$ and $\bb\geq 2$ behave differently.

The pattern of flavor symmetry breaking remains fixed, $\no=\nd=N_c-1\,,\,\, \nt=N_c\,,\,\, U(N_F)\ra U(\no)\times U(\nt)$, as well as the multiplicity $N_{vs}=C^{\,\nt=N_c}_{N_F}=C^{\,\no=\nd}_{N_F}$. And $\langle a_0\rangle=m$ also remains the same in all vs -vacua (really, at all $m\gtrless\lm$) and still "eats" all quark masses $"m"$. But the discrete symmetry $Z_{2N_c-N_F}$ becomes trivial at $\bb=1$ and gives no restrictions on the form of $\langle X^{\rm adj}_{SU(N_c)}\rangle\sim\lm$ (remind that $\langle X^{\rm adj}_{SU(N_c)}\rangle\sim\lm$ is higgsed necessarily at $\mx m\ll\lm^2$ to avoid $g^2(\mu<\lm)<0$ ). As above in section 6.2, the right flavor symmetry breaking implies the unbroken $SU(\nd)$ lower energy group and a presence of $N_F$ flavors of quark-like particles massless at $\mx\ra 0$, then $\no=\nd$ of them will be higgsed at small $\mx\neq 0$ and there will remain $2\nd N_c$ Nambu-Goldstone multiplets.

However, it is not difficult to see that a literal picture with light BPS original electric particles
of remained unbroken at the scale $\mu\sim\lm$ electric subgroup $SU(\nd)$ cannot be right at $\bb=1$. Indeed, if this were the case, then $\langle X^{\rm adj}_{SU(N_c)}\rangle\sim\lm\, {\rm diag}(\,1... 1\,;\, -\nd\,),\,\,\nd=N_c-1$. But all quarks $Q^i,\,{\ov Q}_i$ with $SU(N_c)$ colors will acquire then masses $\sim\lm$ and decouple at $\mu\lesssim\lm$. The remained ${\cal N}=2\,\, SU(\nd)$ SYM with the scale factor $\sim\lm$ of its gauge coupling is not IR free and, to avoid unphysical $g^2(\mu\lesssim\lm)<0$, it will be necessarily higgsed as $\langle X^{\rm adj,A}_{SU(\nd)}\rangle\sim\lm,\,\, SU(\nd)\ra U^{\nd-1}(1)$, in contradiction with the assumption that it remains unbroken at $\mu\lesssim\lm$.
\footnote{\,
Clearly, at $m\ll\lm,\, \mx\ra 0$ and $\bb=1$, this argument concerns also $SU(N_c)$ br2 vacua with $0<\no<\nd$ and the multiplicity $(\nd-\no)C^{\no}_{N_F}$, and S vacua with the multiplicity $\nd$, both with $\langle S\rangle_{N_c}\neq 0$ and with the $SU(\nd=N_c-1)$ gauge group at lower energy $\mu<\lm$.
}

Therefore, the vs-vacua with $2N_c-N_F=1$ are not typical vacua, they are really exceptional. The absence of the non-trivial unbroken $Z_{2N_c-N_F}$ symmetry is crucial.
\footnote{\,
Considered in \cite{SY3} example of $U(N_c)$ vs-vacua ($r=N_c$ vacua in the language of \cite{SY3}) with $N_c=3,\, N_F=5,\, \no=\nd=N_c-1=2$ and $\langle S\rangle_{N_c}=0$ belongs just to this exceptional type of vs-vacua with $2N_c-N_F=1$.
}
The possible exceptional properties of vacua with $\bb=1$ at $\mx\ra 0$ are indicated also by the curve \eqref{(1.2)}, its form is changed at $\bb=1$\,: $m\ra m+(\lm/N_c)$ \cite{APS}.

For this reason we do not deal with such vacua in this paper.

\section{$\mathbf{SU(N_c),\,\,m\gg\lm,\,\, U(N_F)\ra U(\no)\times U(\nt)},$\,\,  br2 vacua}

\subsection{Larger $\mx$}

\hspace*{4mm} The quark condensates in these br2 vacua with $\nt<N_c$ are obtained from \eqref{(6.1)} by the replacement $\no\leftrightarrow\nt$, i.e.\,: $\langle\Qt\rangle_{N_c}\approx\mx m_1\gg\langle\Qo\rangle_{N_c},\, m_1=m N_c/(N_c-\nt)$, see sections 3 and 11 in \cite{ch4}. The discrete $Z_{\bb\geq 2}$ symmetry is also unbroken in these vacua and the multiplicity is $N_{\rm br2}=(N_c-\nt) C^{\,\nt}_{N_F}$. They evolve at $m\ll\lm$ to Lt-vacua with $\langle\Qo\rangle_{N_c}\sim\langle\Qt\rangle_{N_c}\sim\mx\lm$, see section 3 in \cite{ch4}.

Similarly to br1 vacua, the scalars $X^{adj}$ are also higgsed at the largest scale $\mu\sim m\gg\lm$ in the weak coupling region, now as $SU(N_c)\ra SU(\nt)\times U(1)\times SU(N_c-\nt),\,\, N_F/2<\nt<N_c$, so that (the leading terms only)
\bq
\langle\Qt\rangle_{N_c}\approx\mx m_1\,,\quad \langle\Qo\rangle_{N_c}\approx\mx m_1\Bigl (\frac{\lm}{m_1}\Bigr )^{\frac{2N_c-N_F}{N_c-\nt}} \,,\quad m_1=\frac{N_c}{N_c-\nt}\, m\,, \label{(7.1.1)}
\eq
\bbq
\langle S\rangle_{N_c}=\frac{\langle\Qo\rangle_{N_c}\langle\Qt\rangle_{N_c}}{\mx}\approx\mx m_1^2\Bigl (\frac{\lm}{m_1}\Bigr )^{\frac{2N_c-N_F}{N_c-\nt}}\ll\mx m^2\,,
\quad\frac{\langle\Qo\rangle_{N_c}}{\langle\Qt\rangle_{N_c}}\approx\Bigl (\frac{\lm}{m_1}\Bigr)^{\frac{2N_c-N_F}{N_c-\nt}}\ll 1\,,
\eeq
\bq
\langle X^{\rm adj}_{SU(N_c)}\rangle\equiv \langle\, X^{adj}_{SU(\nt)}+ X_{U(1)}+ X^{adj}_{SU(N_c-\nt)}\,\rangle\,,\label{(7.1.2)}
\eq
\bbq
\sqrt{2}\, X_{U(1)}=a \,{\rm diag}\Bigl (\underbrace{\,1}_{\nt}\,;\,\underbrace{\,\wh{c}}_{N_c-\nt}\, \Bigr )\,,\quad \wh{c}=-\frac{\nt}{N_c-\nt}\,,\quad \langle a \rangle= m\,,
\eeq
\bbq
\mx\langle{\rm Tr}\,(\sqrt{2} X^{\rm adj}_{SU(N_c)})^2\rangle=(2N_c-N_F)\langle S\rangle_{N_c}+ m \langle{\rm Tr}\,(\qq)\rangle_{N_c}\approx m (\nt\mx m_1)\,.
\eeq
As a result, all quarks charged under $SU(N_c-\nt)$ and hybrids $SU(N_c)/[\,SU(\nt)\times SU(N_c-\nt)\times U(1)\,]$ acquire large masses $m_1=m-{\wh c}\,\langle a\rangle=m N_c/(N_c-\nt)$ and decouple at scales $\mu<\,m_1$, there remains ${\cal N}=2\,\, SU(N_c-\nt)$\, SYM with the scale factor $\Lambda^{SU(N_c-\nt)}_
{{\cal N}=2\, SYM}$ of its gauge coupling, see also \eqref{(7.1.1)},
\bq
\langle\Lambda^{SU(N_c-\nt)}_{{\cal N}=2\, SYM}\rangle^2=\Bigl (\frac{\lm^{\bb}{(m_1)}^{N_F}}{{(m_1)}^{2\nt}
}\Bigr )^{\frac{1}{(N_c-\nt)}}={m^2_1}\Bigl (\frac{\lm}{{m_1}}\Bigr )^{\frac{\bb}{(N_c-\nt)}}\ll\lm^2\,,\quad m_1=\frac{N_c}{N_c-\nt} m\,,\label{(7.1.3)}
\eq
\bbq
\langle S\rangle_{N_c-\nt}=\mx\langle\Lambda^{SU(N_c-\nt)}_{{\cal N}=2\, SYM}\rangle^2\approx\langle S\rangle_{N_c}\,.
\eeq

At $\mx\ll\langle\Lambda^{SU(N_c-\nt)}_{{\cal N}=2\, SYM}\rangle$ it is higgsed in a standard way \cite{DS}, $SU(N_c-\nt)\ra U^{N_c-\nt-1}(1)$, due to higgsing of $X^{adj}_{SU(N_c-\nt)}$
\bq
\langle {\sqrt 2} X^{adj}_{SU(N_c-\nt)}\rangle\sim\langle\Lambda^{SU(N_c-\nt)}_{{\cal N}=2\, SYM}\rangle\, {\rm diag} \Bigl (\underbrace{0}_{\nt}\,;\,\underbrace{k_1,\,...\,k_{N_c-\nt}}_{N_c-\nt}\Bigr )\,,\quad k_i=O(1)\,.\label{(7.1.4)}
\eq
Note that the value $\langle\Lambda^{SU(N_c)}_{{\cal N}=2\, SYM}\rangle$ of $\langle X^{adj}_{SU(N_c-\nt)}\rangle$ in \eqref{(7.1.3)},\eqref{(7.1.4)} is consistent with the unbroken $Z_{\bb}$ discrete symmetry.

As a result, $\,N_c-\nt-1$ pure magnetic monopoles $M_{\rm n}$ (massless at $\mx\ra 0$) with the $SU(N_c-\nt)$ adjoint charges are formed at the scale $\mu\sim\langle\Lambda^{SU(N_c-\nt)}_{{\cal N}=2\, SYM}\rangle$ in this SYM sector. These correspond to $N_c-\nt-1$ unequal double roots of the curve \eqref{(1.2)} connected with this SYM sector. Note also that two single roots with $(e^{+} -e^{-})\sim\langle\Lambda^{SU(N_c)}_{{\cal N}=2\, SYM}\rangle$ of the curve \eqref{(1.2)} also originate from this SYM sector. Other $\nt$ double roots of the curve at $\mx\ra 0$ originate in these vacua from the $SU(\nt)$ sector, see below.

These $N_c-\nt-1$ monopoles are all higgsed with $\langle M_{\rm n}\rangle=\langle {\ov M}_{\rm n}\rangle\sim (\mx\langle\Lambda^{SU(N_c-\nt)}_{{\cal N}=2\, SYM}\rangle)^{1/2}$, so that $N_c-\nt-1$ long ${\cal N}=2$ multiplets of massive photons appear, with masses $\sim (\mx\langle\Lambda^{SU(N_c-\nt)}_{{\cal N}=2\, SYM}\rangle)^{1/2}$. Besides, this leads to a weak confinement of all heavier original $SU(N_c-\nt)$ electrically charged particles, i.e. quarks and all hybrids with masses $\sim m$, and all ${\cal N}=2\,\, SU(N_c-\nt)$ SYM adjoint charged particles with masses $\sim\langle\Lambda^{SU(N_c-\nt)}_{{\cal N}=2\, SYM}\rangle$, the tension of the confining string is $\sigma^{1/2}_2\sim (\mx\langle\Lambda^{SU(N_c-\nt)}_{{\cal N}=2\, SYM}\rangle)^{1/2}\ll\langle\Lambda^
{SU(N_c-\nt)}_{{\cal N}=2\, SYM}\rangle\ll\lm\ll m$. The factor $N_c-\nt$ in the multiplicity of these br2 vacua arises just from this ${\cal N}=2\,\, SU(N_c-\nt)$\, SYM part.

At larger $\mx$ in the range $\langle\Lambda^{SU(N_c-\nt)}_{{\cal N}=2\, SYM}\rangle\ll\mx\ll m$, the scalars $X^{adj}_{SU(N_c-\nt)}$ become too heavy, their lighter  $SU(N_c-\nt)$ phases fluctuate then freely at the scale $\sim\mx^{\rm pole}=g(\mu=\mx^{\rm pole})\mx$ and they are not higgsed. Instead, they decouple as heavy in the weak coupling region at scales $\mu<\mx^{\rm pole}$. There remains then ${\cal N}=1\,\, SU(N_c-\nt)$ SYM with the scale factor $\Lambda^{SU(N_c-\nt)}_{{\cal N}=1\, SYM}$ of its gauge coupling,
\bq
\langle S\rangle_{{\cal N}=1\, SYM}=\langle\Lambda^{SU(N_c-\nt)}_{{\cal N}=1\, SYM}\rangle^3
=\mx\langle\Lambda^{SU(N_c-\nt)}_{{\cal N}=2\, SYM}\rangle ^2,\,\,
\,\, \langle\Lambda^{SU(N_c-\nt)}_{{\cal N}=2\, SYM}\rangle\ll\langle\Lambda^{SU(N_c-\nt)}_{{\cal N}=1\, SYM}\rangle\ll\mx\ll m\,.\quad\label{(7.1.5)}
\eq
Therefore, there will be in this case a large number of strongly coupled $SU(N_c-\nt)$ gluonia with the mass scale $\sim\langle\Lambda^{SU(N_c-\nt)}_{{\cal N}=1\, SYM}\rangle$, while all original heavier charged particles with $SU(N_c-\nt)$ electric charges and masses either $\sim m\gg\langle\Lambda^{SU(N_c-\nt)}_{{\cal N}=1\, SYM}\rangle$ or $\sim g\mx\gg\langle\Lambda^{SU(N_c-\nt)}_{{\cal N}=1\, SYM}\rangle$ still will be weakly confined, the tension of the confining string is larger in this case, $\sigma^{1/2}_1\sim\langle\Lambda^{SU(N_c-\nt)}_{{\cal N}=1\, SYM}\rangle$. This ${\cal N}=1\,\,SU(N_c-\nt)\,$ SYM gives the same factor $N_c-\nt$ in the multiplicity of these vacua.\\

Now, about the electric $SU(\nt)$ part decoupled from $SU(N_c-\nt)$ SYM. The scale factor of its gauge coupling is
\bq
\langle\Lambda_{SU(\nt)}\rangle=\Bigl (\,\frac{\lm^{\bb}}{(m_1)^{2(N_c-\nt)}}\,\Bigr )^{\frac{1}{2\nt-N_F}}=m_1\Bigl (\frac{\lm}{m_1} \Bigr )^{\frac{\bb}{2\nt-N_F}}\ll\lm\,,\quad m_1=\frac{N_c}{N_c-\nt} m\,.\label{(7.1.6)}
\eq
Note that the value of $\langle\Lambda_{SU(\nt)}\rangle$ in \eqref{(7.1.6)} is also consistent with the unbroken $Z_{\bb}$ discrete symmetry.

What {\it qualitatively differs} this $SU(\nt)$ part in br2 vacua at scales $\mu<\, m$ from its analog $SU(\no)$ in br1 vacua of section 4.1 is that $SU(\no)$ with $N_F>2\no$ is IR free, while $SU(\nt)$ with $N_F<2\nt$ is UV free, and its small ${\cal N}=2$ gauge coupling at the scale $\mu\sim m\gg\lm$ begins to grow logarithmically with diminished energy at $\mu<m$. Therefore, {\it if nothing prevents}, $\,\,X^{adj}_{SU(\nt)}$ will be higgsed necessarily at the scale $\sim\langle\Lambda_{SU(\nt)}\rangle$, with $\langle X^{adj}_{SU(\nt)}\rangle\sim\langle\Lambda_{SU(\nt)}\rangle$, to avoid unphysical $g^2(\mu<\langle\Lambda_{SU(\nt)}\rangle)<0$ (see section 7.2 below).

But if $(\mx m)^{1/2}\gg\langle\Lambda_{SU(\nt)}\rangle$, {\it the phase will be different}. The leading effect in this case will be the breaking of the whole $SU(\nt)$ group due to higgsing of $\nt<N_c$ out of $N_F$ quarks $Q^i,\,{\ov Q}_i$ in the weak coupling region, with $\langle Q^i_b\rangle\sim (\mx m)^{1/2},\,\, b=1...\nt,\,\, i=\no+1... N_F$. This will give the additional factor $C^{\,\nt}_{N_F}$ in the multiplicity of these br2 vacua due to spontaneous flavor symmetry breaking $U(N_F)\ra U(\no)\times U(\nt)$, so that the overall multiplicity will be $(N_c-\nt)C^{\,\nt}_{N_F}$, as it should be. The corresponding parts of the superpotential will be as in \eqref{(4.1.3)},\eqref{(4.1.4)}, with a replacement $\no\ra\nt$,
\bq
{\cal W}_{\nt}=(\, m-a\,)\,{\rm Tr}\,({\ov Q} Q)_{\nt} - {\rm Tr}\,({\ov Q}\sqrt{2} X^{\rm adj}_{SU(\nt)} Q)_{\nt} +\mx {\rm Tr}\,( X^{\rm adj}_{SU(\nt)})^2+\frac{\mx}{2}\,\frac{\nt N_c}{N_c-\nt}\,a^2\,. \label{(7.1.7)}
\eq

From \eqref{(7.1.7)} in the case considered (no summation over $i$ or $j$ in \eqref{(7.1.8)}\,)
\bbq
\langle a\rangle=m,\,\, \langle  X^{adj}_{SU(\nt)}\rangle=0,\,\,\langle \Qt\rangle_{\nt}=
\sum_{b=1}^{\nt}\langle{\ov Q}_2^{\,b} Q^2_b\rangle=\langle{\ov Q}_2^{\,2}\rangle\langle Q^2_2\rangle\approx\frac{N_c}{N_c-\nt}\mx m=\mx m_1\,,
\eeq
\bq
\langle \Qo\rangle_{\nt}=\sum_{b=1}^{\nt}\langle{\ov Q}_j^{\,b}\rangle\langle Q^j_b\rangle=0\,,\,\, j=1...\no,\,\,\langle { S}\rangle_{\nt}=\frac{1}{\mx}\langle \Qt\rangle_{\nt}\langle \Qo\rangle_{\nt}=0\,.\quad\label{(7.1.8)}
\eq

$2\no\nt$ massless Nambu-Goldstone multiplets are formed as a result of the spontaneous flavor symmetry breaking, $\Pi_j^i= ({\ov Q}_{j} Q^i)_{\nt}\,\ra\,({\ov Q}_{j}\langle Q^i\rangle)_{\nt},\,\,\Pi_i^j= ({\ov Q}_{i} Q^j)_{\nt}\,\ra\,(\langle{\ov Q}_{i}\rangle Q^j)_{\nt},\,\, i=\no+1...N_F,\,\, j=1...\no,\,\,
\langle\Pi^i_j\rangle=\langle \Pi^j_i\rangle=0$ (in essence, these are non-higgsed quarks with $\no$ flavors and $SU(\nt)$ colors). Besides, there are $\nt^2$ long ${\cal N}=2$ multiplets of massive gluons with masses $\sim (\mx m)^{1/2}$.

We emphasize that higgsing of $\nt$ quarks at the higher scale $(\mx m)^{1/2}\gg\langle\Lambda_{SU(\nt)}
\rangle$, see \eqref{(7.1.7)}, \eqref{(7.1.8)}, {\it prevents $X^{adj}_{SU(\nt)}$ from higgsing}, i.e. $\langle  X^{adj}_{SU(\nt)}\rangle_{\nt}$ is not simply smaller, but exactly zero (see also section 6.1 for a similar regime).

On the whole for these br2 vacua in the case considered, i.e. at $(\mx m)^{1/2}\gg\langle\Lambda_{SU(\nt)}
\rangle$, all qualitative properties are similar to those of br1 vacua in section 4.1.

\subsection{Smaller $\mx$}

\hspace*{4mm} Consider now the behavior of the $SU(\nt)\times U(1)$ part in the opposite case of smaller $\mx,\, \mx\ll\langle\Lambda_{SU(\nt)}\rangle^2/m,\, \langle\Lambda_{SU(\nt)}\rangle\ll\lm\ll m$. As will be seen below, in this case even all qualitative properties of mass spectra in this part will be quite different (the behavior of the $SU(N_c-\nt)$ SYM part decoupled from $SU(\nt)\times U(1)$ was described above in section 7.1).

Now, at smaller $\mx$, instead of quarks, the scalar field $X^{adj}_{SU(\nt)}$ is higgsed first at the scale $\sim\langle\Lambda_{SU(\nt)}\rangle$ to avoid unphysical $g^2(\mu<\langle\Lambda_{SU(\nt)}\rangle)<0$ of UV free ${\cal N}=2\,\, SU(\nt)$. Although $\langle\Lambda_{SU(\nt)}\rangle$ respects the original $Z_{\bb}$ discrete symmetry of the UV free $SU(N_c)$ theory, see \eqref{(7.1.6)}, because the $SU(N_c-\nt)$ SYM part is now  decoupled, there is the analog of $Z_{\bb}$ in the $SU(\nt)$ group with $N_F$ flavors of quarks, this is the unbroken non-trivial $Z_{2\nt-N_F}=Z_{\nt-\no}$ discrete symmetry with $\nt-\no\geq 2$. Therefore, the field $X^{adj}_{SU(\nt)}$ is higgsed at $\nt-\no\geq 2$ qualitatively similarly to \eqref{(2.1.1)},\eqref{(2.1.2)} in section 2.1\,: $SU(\nt)\ra SU(\no)\times U^{(1)}(1)\times U^{\nt-\no-1}(1)$. There will be similar dyons $D_j$ (massless at $\mx\ra 0$) etc., and a whole qualitative picture will be similar to those in section 2.1, with evident replacements of parameters. The only qualitative difference is that the additional SYM part is absent here because $\no$ quark flavors are higgsed now in $SU(\no)$,
\bq
\langle X^{adj}_{SU(\nt)}\rangle=\langle  X^{adj}_{SU(\no)}+X^{(1)}_{U(1)}+ X^{adj}_{SU(\nt-\no)}\rangle,\label{(7.2.1)}
\eq
\bbq
\langle\sqrt{2} X^{adj}_{SU(\nt-\no)}\rangle\sim\langle\Lambda_{SU(\nt)}\rangle\,{\rm diag}\,\Bigl(\,\underbrace{0}_{\no}\,;\,\underbrace{\tau^0,\,\tau^1,\,...,\,\tau^{\nt-\no-1}}_{\nt-\no}\,;
\underbrace{0}_{N_c-\nt}\,\Bigr )\,,\quad \tau=\exp\{\frac{2\pi i}{\nt-\no}\}\,,
\eeq
\bbq
\sqrt{2}\, X^{(1)}_{U(1)}=a_{1}\,{\rm diag}\,(\,\underbrace{\,1}_{\no}\,;\, \underbrace{\,\wh{c}_1}
_{\nt-\no}\,;\underbrace{0}_{N_c-\nt}\,),\quad \wh{c}_1=-\,\frac{\no}{\nt-\no}\,.\quad
\eeq
Therefore, similarly to \eqref{(2.1.2)},\eqref{(6.2.2)}, in this case the low energy superpotential of this $SU(\nt)\times U(1)$ sector can be written as
\footnote{\,
In other words, in this case the $SU(N^\prime_c=\nt)\times U(1)$ part of $SU(N_c)$ with $m\gg\lm$ and with decoupled $SU(N_c-\nt)$ SYM part, {\it is in its own vs-vacuum} of section 6.2, with $\langle\,{S}\,\rangle_
{N^\prime_c}=0$ and with $a$ of $U(1)$ in \eqref{(7.2.2)} playing a role of $a_0$ in \eqref{(6.2.2)}, etc.
}
\bq
\w_{\nt}=\w_{\no}+\w_{D}+\w_{a,\,a_1}+\dots\,,\label{(7.2.2)}
\eq
\bbq
\w_{\no} =(m-a-a_1){\rm Tr}\,({\ov Q} Q)_{\no}-{\rm Tr}\,\Bigl ({\ov Q}\sqrt{2}X_{SU(\no)}^{\rm adj} Q\Bigr )_{\no}+\mx (1+\wh{\delta}_2){\rm Tr}\,(X^{\rm adj}_{SU(\no)})^2\,,
\eeq
\bbq
\w_{D}=\Bigl ( m-a-\wh{c}_1 a_1 \Bigr )\sum_{j=1}^{\nt-\no}{\ov D}_j D_j -\sum_{j=1}^{\nt-\no} a_{D,j}{\ov D}_j D_j -\mx\Lambda_{SU(\nt)}\sum_{j=1}^{\nt-\no}\tau^{j-1} a_{D,j}+\mx{\wh L}\,\Bigl (\sum_{j=1}^{\nt-\no} a_{D,j}\Bigr ),
\eeq
\bbq
\w_{a,\,a_1}=\frac{\mx}{2}\frac{\nt N_c}{N_c-\nt}\,a^2+\frac{\mx}{2}
\frac{\no\nt}{\nt-\no}(1+\wh{\delta}_1)\,a^2_1+\mx\nt\wh{\delta}_3\,a_1(m-a-\wh{c}_1 a_1)+\mx\nt\wh{\delta}_4\, (m-a-\wh{c}_1 a_1)^2\,,
\eeq
where $\wh{L}$ is the Lagrange multiplier field, $\langle\wh{L}\rangle=O(m)$ and dots denote small power corrections. The additional terms with $\wh{\delta}_{1,2,3,4}$, as previously in br2 vacua in section 2.1 and in vs -vacua in section 6.2, originate from integrating out all heavier fields with masses $\sim\Lambda_{SU(\nt)}$ in the soft background. Here these soft background fields are $(m-a-\wh{c}_1 a_1)$ and $[(1-\wh{c}_1) a_1+\sqrt{2}X^{\rm adj}_{SU(\no)}]$ (\,while $a_1$ plays here a role of $a_1$ in \eqref{(2.1.1)}\,). These constants $\wh\delta_i$ differ from \eqref{(2.1.12)} only because the number of colors is different, $N_c\ra N^\prime_c=\nt$, i.e. $\wh\delta_1=[\,- 2N^\prime_c/(2N^\prime_c-N_F)]
=[\,-2\nt/(\nt-\no)\,],\,\, \wh\delta_2=-2$ (while $\wh\delta_3=0$ as in br2 vacua, see \eqref{(B.4)}).

The charges of fields and parameters entering \eqref{(7.2.2)} under $Z_{\nt-\no}=\exp\{i\pi/(\nt-\no)\}$ transformation are\,: $q_{\lambda}=q_{\rm\theta}=1,\,\, q_{X_{SU(\no)}^{\rm adj}}=q_{a}=q_{a_1}=q_{a_{D,j}}=q_{\rm m}=q_{\wh{L}}=2,\,\,q_{Q}=q_{\ov Q}=q_{D_j}=q_{{\ov D}_j}=q_{\Lambda_{SU(\nt)}}
=0,\,\, q_{\mx}=-2$. The non-trivial $Z_{\nt-\no\geq 2}$ transformations change only numerations of dual scalars $a_{D,j}$ and dyons in \eqref{(7.2.2)}, so that $\int d^2\theta\,\w_{\nt}$ is $Z_{\nt-\no}$-invariant.

From \eqref{(7.2.2)}:
\bq
\langle a\rangle=m\,,\quad\langle a_1\rangle=0\,,\quad \langle a_{D,j}\rangle=0\,,
\quad \langle X_{SU(\no)}^{\rm adj}\rangle=0\,,\quad \langle S\rangle_{\no}=0\,,\label{(7.2.3)}
\eq
\bbq
\langle{\ov D}_j D_j\rangle=\langle{\ov D}_j\rangle\langle D_j \rangle\approx -\mx\langle\Lambda_{SU(\nt)}\rangle\tau^{j-1}+\mx \langle{\wh L}\rangle\,,\quad \langle\Sigma_D\rangle=\sum_{j=1}^{\nt-\no}\langle{\ov D}_j\rangle\langle D_j \rangle=(\nt-\no)\mx \langle{\wh L}\rangle,
\eeq
\bbq
\langle{\wh L}\rangle\approx\frac{N_c}{N_c-\nt}\, m=m_1\,, \quad \langle{\rm Tr}\,({\ov Q} Q)\rangle_{\no}=\no\langle\Qo\rangle_{\no}+\nt\langle\Qt\rangle_{\no}\approx\no\mx m_1\,,
\eeq
\bbq
\langle\Qo\rangle_{\no}=\langle{\ov Q}^{\,1}_1\rangle\langle Q^1_1\rangle\approx\frac{N_c}{N_c-\nt}\,\mx m\approx\mx m_1\approx\langle\Qt\rangle_{N_c}\,,\quad \langle\Qt\rangle_{\no}=\sum_{a=1}^{\no}\langle{\ov Q}^{\,a}_2\rangle\langle Q^2_a\rangle=0\,.
\eeq

The multiplicity of these br2 vacua in the case considered is $N_{\rm br2}=(N_c-\nt)C^{\,\no}_{N_F}=(N_c-\nt)C^{\,\nt}_{N_F}$, as it should be. The factor $(N_c-\nt)$ originates from the $SU(N_c-\nt)$ SYM. The factor $C^{\,\no}_{N_F}$ is finally a consequence of the color breaking $SU(\nt)\ra SU(\no)\times U(1)^{\nt-\no}$ with the unbroken $Z_{2\nt-N_F}=Z_{\nt-\no}$ discrete symmetry, and spontaneous breaking of flavor symmetry due to higgsing of $\no$ quarks flavors (massless at $\mx\ra 0$) in the $SU(\no)$ color subgroup of $SU(\nt)$, $\langle Q^i_a\rangle=\langle{\ov Q}^{\,a}_i\rangle\approx (\mx m_1)^{1/2}$,  $a=1...\no,\,\,i=1...\no$.

Note also that if the non-trivial discrete symmetry $Z_{\nt-\no\geq 2}$ were broken spontaneously in the $SU(\nt)$ subgroup at the scale $\mu\sim\langle\Lambda_{SU(\nt)}\rangle\gg(\mx m)^{1/2}$, this will lead then
to the factor $\nt-\no$ in the multiplicity of these vacua, and this is a wrong factor.

Due to higgsing of $\nt-\no$ dyons with the non-zero $SU(\nt-\no)$ magnetic charges, $\langle{\ov D_j}\rangle\langle D_j\rangle\sim \mx m\gg\mx\lm\gg\mx\langle\Lambda_{SU(\nt)}\rangle$, all original pure electrically charged particles with $SU(\nt-\no)$ charges and masses either $\sim m$ or $\sim\langle\Lambda_{SU(\nt)}\rangle$ are weakly confined, the string tension is $\sigma^{1/2}\sim (\mx m)^{1/2}\ll\langle\Lambda_{SU(\nt)}\rangle\ll\lm\ll m$. Besides, $\nt-\no$ long ${\cal N}=2$ multiplets of massive photons with masses $\sim (\mx m)^{1/2}$ are formed. Remind that, due to higgsing of $N_c-\nt-1$ magnetic monopoles $M_{\rm n}$ from $SU(N_c-\nt)$ SYM at $\mx < \langle\Lambda^{SU(N_c-\nt)}_{{\cal N}=2\, SYM}\rangle$, all original pure electrically charged particles with $SU(N_c-\nt)$ charges and masses either $\sim m$ or $\langle\Lambda^{SU(N_c-\nt)}_{{\cal N}=2\, SYM}\rangle$ are also weakly confined, but this string tension $\sigma^{1/2}_2\sim (\mx\langle\Lambda^{SU(N_c-\nt)}_{{\cal N}=2\, SYM}\rangle)^{1/2}\ll (\mx m)^{1/2}$ is much smaller.

Due to higgsing of $\no$ flavors of original electric quarks with $SU(\no)$ colors there are $\no^2$ long ${\cal N}=2$ multiplets of massive gluons with masses $\sim (\mx m)^{1/2}\ll\langle\Lambda_{SU(\nt)}\rangle$, while the original non-higgsed quarks with $\nt$ flavors and $SU(\no)$ colors form $2\no\nt$ massless Nambu-Goldstone multiplets. There are no other massless particles at $\mx\ne 0$. And there are no particles with masses $\sim m\gg\lm$ in this $SU(\nt)\times U(1)$ sector.

Remind that the curve \eqref{(1.2)} has $N_c-1$ double roots in these br2 vacua at $\mx\ra 0$. Of them: $\nt-\no$ unequal roots correspond to $\nt-\no$ dyons $D_j,\,\, \no$ equal roots correspond to $\no$ higgsed quarks $Q^i$ of $SU(\no)$, the remaining unequal $N_c-\nt-1$ double roots correspond to $N_c-\nt-1$ pure magnetic monopoles $M_{\rm n}$ from $SU(N_c-\nt)\,\, {\cal N}=2$ SYM. The two single roots with $(e^{+}-e^
{-})\sim\langle\Lambda^{SU(N_c-\nt)}_{{\cal N}=2\, SYM}\rangle$ originate from this ${\cal N}=2$ SYM.

Besides, similarly to vs vacua in section 6, there is a phase transition in these br2 vacua in the region $\mx\sim\langle\Lambda_{SU(\nt)}\rangle^2/m$. E.g., all $SU(\nt-\no)$ charged gluons have equal masses in the whole range of the higher energy phase of section 7.1 at $\mx >(\rm a\, few)\langle\Lambda_{SU(\nt)}
\rangle^2/m$, while masses of these gluons differ greatly at $\mx\ra 0$ in the lower energy phase of this section, $\mu^{ij}_{gl}\sim |(\tau^i-\tau^j)\langle\Lambda_{SU(\nt)}\rangle|,\, i,j=0...\nt-\no-1,\, i\neq j$. And $\langle X^{adj}_{SU(\nt-\no)}\rangle$, $\,\langle\Qt\rangle_{\nt}$ also behave non-analytically at $\mx\gtrless\langle\Lambda_{SU(\nt)}\rangle^2/m$. But, unlike the phase transition at $\mx\sim\lm^2/m$ in vs vacua of section 6, the phase transition in these br2 vacua is at much smaller $\mx\sim\langle\Lambda_
{SU(\nt)}\rangle^2/m\,,\,\,\langle\Lambda_{SU(\nt)}\rangle\ll\lm$, see \eqref{(7.1.6)}.\\

And finally, remind that the mass spectra in these br2 vacua with $\nt<N_c$ and $m\gg\lm$ depend essentially on the value of $m/\lm$ and all these vacua  evolve at $m\ll\lm$ to Lt vacua with the spontaneously broken discrete $Z_{\bb}$ symmetry, see section 3 in \cite{ch4} or section 4 in \cite{ch5}.

\section{\bf Large quark masses $\mathbf {m\gg\lm}$,\quad $SU(N_c),\,\,\, N_F<N_c$}
\numberwithin{equation}{section}

\hspace*{4mm} The vacua with the unbroken $U(N_F)$ flavor symmetry (here and in what follows at not too small $N_c$ and $N_F$) are in this case the SYM-vacua with the multiplicity $N_c$ and S-vacua with the multiplicity $N_c-N_F$, see section 2 in \cite{ch4} and section 3 in \cite{ch5}. The vacua with the broken $U(N_F)\ra U(\no)\times U(\nt)$ symmetry are the br1 and br2 vacua with multiplicities $(N_c-{\rm n}_i)C^{\,{\rm n}_i}_{N_F},\, i=1,\,2$.

Except for S-vacua, all formulae for this case $N_F<N_c,\,\, m\gg\lm$ are the same as described above for br1 vacua in section 4, for SYM vacua in section 5 and for br2 vacua in section 7 (and only $N_F<N_c$ now). Therefore, we describe below only the mass spectra in S-vacua with $m\gg\lm,\,\mx\ll\lm\,$.

The quark and gluino condensates look in these $SU(N_c)$ S-vacua as, see \eqref{(1.3)},\eqref{(1.4)} and the footnote \ref{(f4)}, the leading terms only,
\bq
\langle\qq\rangle_{N_c}\approx\frac{N_c}{N_c-N_F}\,\mx m\approx\mx m_1\,,\quad  m_1=\frac{N_c}{N_c-(\nt=N_F)}\,m\,,\label{(8.1)}
\eq
\bbq
\langle S\rangle_{N_c}=\Bigl (\frac{\lm^{2N_c-N_F}\mx^{N_c}}{\det\langle\qq\rangle}\Bigr )^{\frac{1}{N_c-N_F}}\approx\mx m_1^2\Bigl (\frac{\lm} {m_1} \Bigr )^{\frac{2N_c-N_F}{N_c-N_F}}\ll\mx m_1^2\,.
\eeq
It is seen from \eqref{(8.1)} that the non-trivial discrete symmetry $Z_{2N_c-N_F\geq 2}$ is not broken.

The scalars $ X^{adj}_{SU(N_c)}$ are higgsed at the highest scale $\sim m\gg\lm$ {\it in the weak coupling regime}, $SU(N_c)\ra SU(N_F)\times U(1)\times SU(N_c-N_f)$, see \eqref{(8.7)} below,
\bq
\langle X^{adj}_{SU(N_c)}\rangle=\langle  X^{adj}_{SU(N_F)}+X_{U(1)}+ X^{adj}_{SU(N_c-N_F)}\rangle,\label{(8.2)}
\eq
\bbq
\sqrt{2}\, X_{U(1)}=a\,{\rm diag}\,(\,\underbrace{\,1}_{N_F}\,;\, \underbrace{\,\wh{c}}
_{N_c-N_F}\,),\quad \wh{c}=-\,\frac{N_F}{N_c-N_F}\,,\quad \langle a\rangle= m\,,
\eeq

The quarks in the $SU(N_c-N_F)$ sector and $SU(N_c)/[SU(N_F)\times SU(N_c-N_F)\times U(1)]$ hybrids
have large masses $m_Q=\langle m-{\wh c}\, a\rangle=N_c m/(N_c-N_F)=m_1$. After they are integrated out, the scale factor of the $SU(N_c-N_F)$ SYM gauge coupling is, see also \eqref{(8.1)},
\bq
\langle\Lambda^{SU(N_c-N_F)}_{{\cal N}=2\, SYM}\rangle^2=\Bigl (\frac{\lm^{\bb}{(m_1)}^{N_F}}{(m_1)^{2 N_F}}\Bigr )^{\frac{1}{N_c-N_F}}={m^2_1}\Bigl (\frac{\lm}{{m_1}}\Bigr )^{\frac{\bb}{N_c-N_F}}\ll\lm^2\,,\quad m_1=\frac{N_c}{N_c-N_F} m\,,\label{(8.3)}
\eq
\bbq
\langle S\rangle_{N_c-N_F}=\mx\langle\Lambda^{SU(N_c-N_F)}_{{\cal N}=2\, SYM}\rangle^2\approx\langle S\rangle_{N_c}\,,
\eeq
while those of $SU(N_F)$ SQCD is
\bq
\langle\Lambda_{SU(N_F)}\rangle=\Bigl (\,\frac{\lm^{\bb}}{(m_1)^{2(N_c-N_F)}}\,\Bigr )^{\frac{1}{N_F}}=m_1\Bigl (\frac{\lm}{m_1} \Bigr )^{\frac{\bb}{N_F}}\ll\lm\,.\label{(8.4)}
\eq

If $\mx\ll\langle\Lambda^{SU(N_c-N_F)}_{{\cal N}=2\, SYM}\rangle$, the scalars $X^{adj}_{SU(N_c-N_F)}$ of ${\cal N}=2$ SYM are higgsed at $\mu\sim\langle\Lambda^{SU(N_c-N_F)}_{{\cal N}=2\, SYM}\rangle$ \eqref{(8.3)} in a known way, $SU(N_c-N_F)\ra U^{N_c-N_F-1}(1)$ \cite{DS}, giving $N_c-N_F$ vacua and $N_c-N_F-1$ massless at $\mx\ra 0$ magnetic monopoles ${\ov M}_{\rm n}, M_{\rm n}$. These all are higgsed at $\mx\neq 0$ giving $N_c-N_F-1$ $\,{\cal N}=2$ long multiplets of dual massive photons with masses $\mu_{a_{M,{\rm n}}}\sim\langle M_{\rm n}\rangle\sim (\mx\langle\Lambda^{SU(N_c-N_F)}_{{\cal N}=2\, SYM}\rangle)^{1/2}$. All heavy original particles with masses $\sim m$ and all charged $SU(N_c-N_F)$ adjoints with masses $\sim\langle\Lambda^{SU(N_c-N_F)}_{{\cal N}=2\, SYM}\rangle$ are weakly confined, the string tension in this sector is $\sigma^{1/2}_2\sim (\mx\langle\Lambda^{SU(N_c-N_F)}_{{\cal N}=2\, SYM}\rangle)^{1/2}\ll\langle\Lambda^{SU(N_c-N_F)}_{{\cal N}=2\, SYM}\rangle$.

On the other hand, if e.g. $\langle\Lambda^{SU(N_c-N_F)}_{{\cal N}=2\, SYM}\rangle\ll\mx\ll m$, see \eqref{(8.3)}, all scalars  $X^{adj}_{SU(N_c-N_F)}$ are too heavy and not higgsed, they all decouple and can be integrated out at $\mu<g\mx$, there remains then ${\cal N}=1\,\, SU(N_c-N_F)$ SYM with $\langle\Lambda^{SU(N_c-N_F)}_{{\cal N}=1\, SYM}\rangle=[\,\mx\langle\Lambda^{SU(N_c-N_F)}_{{\cal N}=2\, SYM}\rangle^2\,]^{1/3}
\gg\langle\Lambda^{SU(N_c-N_F)}_{{\cal N}=2\, SYM}\rangle$. There will be a large number of strongly coupled $SU(N_c-N_F)$ gluonia with the mass scale $\sim\langle\Lambda^{SU(N_c-N_F)}_{{\cal N}=1\, SYM}\rangle$. All original heavier $SU(N_c-N_F)$ charged particles with masses $\sim m$ or $g\mx$ are still weakly confined, but the string tension in this sector is larger now, $\langle\Lambda^{SU(N_c-N_F)}_{{\cal N}=2\, SYM}\rangle\ll\sigma^{1/2}_1
\sim\langle\Lambda^{SU(N_c-N_F)}_{{\cal N}=1\, SYM}\rangle\ll\mx$.\\

Now, as for the $SU(N_F)$ sector. At scales $\langle\Lambda_{SU(N_F)}\rangle\ll\mu\ll m$ this is the UV free ${\cal N}=2$ SQCD with $N_F$ flavors of massless quarks ${\ov Q}^{\,a}_j, Q^i_a,\,\,i,a=1...N_F,\,\, \langle m-a\rangle=0$. When $\langle\Lambda_{SU(N_F)}\rangle^2/m\ll\mx\ll m$, see \eqref{(8.4)}, then the low energy superpotential in this sector will be as in \eqref{(7.1.7)} with $\nt=N_F$. I.e., all quarks in this $SU(N_F)$ sector will be higgsed (while $\langle X^{adj}_{SU(N_F)}\rangle=0$) and $N_F^2$ long ${\cal N}=2$ multiplets of massive gluons (including $U(1)$ with its scalar "a") with masses $\mu_{gl}\sim g\langle\qq\rangle^{1/2}_{N_F}\sim g(\mx m)^{1/2}$ will be formed, see \eqref{(7.1.8)}. The global symmetry $SU(N_F)_{C+F}$ is unbroken. The multiplicity of vacua in this phase is determined by the multiplicity $N_c-N_F$ of $SU(N_c-N_F)$ SYM.

But at $(\mx m)^{1/2}\ll\langle\Lambda_{SU(N_F)}\rangle$ the phase is different. To avoid $g^2(\mu<\langle\Lambda_{SU(N_F)}\rangle)<0$, the main contributions to masses originate now in this UV free $SU(N_F)$ sector from higgsed $\langle X^{adj}_{SU(N_F)}\rangle\sim\langle\Lambda_{SU(N_F)}\rangle$. The situation in this phase is qualitatively similar to those in section 7.2, the difference is that now the whole $SU(N_F)$ color group will be broken at the scale $\sim\langle\Lambda_{SU(N_F)}\rangle$, see below. Because $SU(N_c-N_F)$ sector is decoupled, there is the residual unbroken discrete symmetry $Z_{2N_F-N_F\geq\, 2}=Z_{N_F\geq\, 2}$. Therefore, $X^{adj}_{SU(N_F)}$ are higgsed as
\bq
\langle\sqrt{2} X^{adj}_{SU(N_F)}\rangle\sim\langle\Lambda_{SU(N_F)}\rangle\,{\rm diag}\,\Bigl(\,
\underbrace{\rho^0,\,\rho^1,\,...,\,\rho^{N_F-1}}_{N_F}\,;\underbrace{0}_{N_c-N_F}\,\Bigr ),\quad \rho=\exp\{\,\frac{2\pi i}{N_F}\,\}\,,\label{(8.5)}
\eq
and the whole $SU(N_F)$ color group is broken, $SU(N_F)\ra U^{N_F-1}(1)$. All quarks with $SU(N_F)$ colors acquire now masses $m_Q\sim\langle\Lambda_{SU(N_F)}
\rangle$ which are large in comparison with the potentially possible scale of their coherent condensate, $m_Q\gg (\mx m)^{1/2}$, so that {\it they are not higgsed but decouple as heavy at $\mu<\langle\Lambda_{SU(N_F)}\rangle$}, as well as all charged $SU(N_F)$ adjoints. Instead, $N_F$ massless at $\mx\ra 0$ dyons ${\ov D}_j, D_j$ are formed at the scale $\sim\langle\Lambda_{SU(N_F)}\rangle$. The relevant part of the low energy superpotential of this sector looks in this case as
\bq
\w_{N_F}=\w_{D}+\w_{a}+\w_{SYM}\,, \label{(8.6)}
\eq
\bbq
\w_{D}= (m-a)\sum_{j=1}^{N_F}{\ov D}_j D_j -\sum_{j=1}^{N_F} a_{D,\,j}{\ov D}_j D_j -\mx\langle\Lambda_{SU(N_F)}\rangle\sum_{j=1}^{N_F}\rho^{\,{j-1}} a_{D,\,j}+\mx{L^{\,\prime}}\,\Bigl (\sum_{j=1}^{N_F} a_{D,\,j}\Bigr )\,,
\eeq
\bbq
\w_{a}=\frac{\mx}{2}\frac{N_c N_F}{N_c-N_F}\,a^2+\mx N_F\,{\delta}^{\,\prime}_4\, (m-a)^2\,,
\eeq
\bbq
\w_{SYM}= (N_c-N_F)\mx \Bigl (\Lambda^{SU(N_{c}-N_F)}_{{\cal N}=2\,\,SYM}\Bigr )^2= (N_c-N_F)\mx\Biggl (\frac{\lm^{2N_c-N_F}(m_1-{\hat c}\,{\hat a})^{N_F}}{[\,m_1+(1-{\hat c}){\hat a}\,]^{2 N_F}}\Biggr )^{\frac{1}{N_c-N_F}}\approx
\eeq
\bbq
\approx \,\mx\langle\Lambda^{SU(N_c-N_F)}_{{\cal N}=2\,\,SYM}\rangle^2\Biggl [(N_c-N_F) -N_F \frac{(2N_c-N_F)}{N_c-N_F}\,\frac{\hat a}{m_1}\,\Biggr ]\,,\quad m_1=\frac{N_c}{N_c-N_F} m\,,\quad a=\langle a\rangle+{\hat a}\,.
\eeq

From \eqref{(8.6)}
\bbq
\langle a\rangle=m\,,\quad \langle a_{D,\,j}\rangle=0\,,\quad \langle{\ov D}_j\rangle\langle D_j \rangle= -\mx\langle\Lambda_{SU(N_F)}\rangle\rho^{\,{j-1}}+\mx \langle{L^{\,\prime}}\rangle\,,\quad \langle\Sigma_D\rangle=\sum_{j=1}^{N_F}\langle{\ov D}_j\rangle\langle D_j\rangle=N_F\mx \langle{L^{\,\prime}}\rangle,
\eeq
\bq
\langle\Sigma_D\rangle\approx N_F\mx m_1\Bigl [\,1-\frac{2N_c-N_F}{N_c-N_F}\frac{\langle
\Lambda^{SU(N_c-N_F)}_{{\cal N}=2\,\,SYM}\rangle^2}{m^2_1}\,\Bigr ]\approx N_F\mx m_1\Bigl [\,1-\frac{2N_c-N_F}{N_c-N_F}\Bigl (\frac{\lm}{{m_1}}\Bigr )^{\frac{\bb}{N_c-N_F}}\,\Bigr ]\,,\label{(8.7)}
\eq
\bbq
\langle S\rangle_{N_F}=0,\quad\langle{\ov D}_j\rangle\langle D_j\rangle\approx \mx m_1\Bigl [\,1-\frac{2N_c-N_F}{N_c-N_F}\frac{\langle
\Lambda^{SU(N_c-N_F)}_{{\cal N}=2\,\,SYM}\rangle^2}{m^2_1}\,\Bigr ]-\mx\langle\Lambda_{SU(N_F)}
\rangle\rho^{\,{j-1}}\,,\quad j=1...N_F\,.
\eeq

The curve \eqref{(1.2)} has $N_c-1$ double roots in these $N_c-N_F$ S-vacua. From these, $N_c-N_F-1$ unequal double roots correspond to massless at $\mx\ra 0$ magnetic monopoles $M_{\rm n}$ from $SU(N_c-N_F)\,\, {\cal N}=2$ SYM, and remaining $N_F$ unequal double roots correspond to massless at $\mx\ra 0$ dyons $D_j$. The two single roots with $(e^{+}-e^{-})\sim \langle\Lambda^{SU(N_c-N_F)}_{{\cal N}=2\, SYM}\rangle$ \eqref{(8.3)} originate at $\mx\ra 0$ from the $SU(N_c-N_F)\,\,{\cal N}=2$ SYM sector. The global flavor symmetry is unbroken and there are no massless Nambu-Goldstone particles. And there are no massless particles at all at $\mx\neq 0$. All original charged electric particles are confined in the lower energy phase at $\mx<\langle\Lambda_{SU(N_F)}\rangle^2/m_1$, see \eqref{(8.4)}. But the phase changes in the $SU(N_F)$ sector already at very small $\mx :\,\langle\Lambda_{SU(N_F)}\rangle^2/m_1<\mx\ll\langle\Lambda_{SU(N_F)}
\rangle\ll\lm\ll m$. Instead of higgsed $\langle X^{adj}_{SU(N_F)}\rangle\sim\langle\Lambda_{SU(N_F)}
\rangle,\,\, SU(N_F)\ra U^{N_F-1}(1)$ and subsequent condensation of $N_F$ dyons $D_j$ \eqref{(8.7)}, all $N_F$ flavors of original electric quarks ${\ov Q}^{\,a}_i,\,Q^i_a,\,\,a,i=1...N_F$ are then higgsed as $\langle{\ov Q}^{\,a}_i\rangle=\langle  Q_a^i\rangle\approx\delta^i_a (\mx m_1)^{1/2}>\langle\Lambda
_{SU(N_F)}\rangle$ (and clearly not confined), forming $N_F^2$ long ${\cal N}=2$ multiplets of massive gluons.

Remind that, at $N_F<N_c$ and $m\gg\lm$, the mass spectra in these S vacua with the multiplicity $N_c-N_F$, as well as in the SYM vacua with the multiplicity $N_c$, depend essentially on the value of $m/\lm$, and all these vacua with the unbroken global flavor symmetry $U(N_F)$ evolve at $\mx\ll m\ll\lm$ to L- vacua with the multiplicity $2N_c-N_F$ and with the spontaneously broken discrete $Z_{\bb}$ symmetry, see sections 3.1 and 12 in \cite{ch5}.

\section{Conclusions}

\hspace*{4mm} We presented above in this paper the calculations of mass spectra of softly broken ${\cal N}=2\,\, SU(N_c)$ or $U(N_c)$ gauge theories in a number of their various vacua with the unbroken non-trivial $Z_{\bb\geq 2}$ discrete symmetry. The charges of light particles in each vacuum were determined, as well as forms of corresponding low energy Lagrangians and mass spectra at different hierarchies between the Lagrangian parameters $m,\,\mx,\,\lm$. A crucial role in obtaining these results was played by the use\,: a) the unbroken $Z_{\bb\geq 2}$ symmetry,\, b) the pattern of spontaneous flavor symmetry breaking, $U(N_F)\ra U(\no)\times U(\nt),\, 0\leq \no<N_F/2,\,$ c) the knowledge of multiplicities of various vacua. Besides, the knowledge of the quark and gluino condensates $\langle\Qo\rangle_{N_c},\,\langle\Qt\rangle_{N_c}$, and $\langle S\rangle_{N_c}$ was also of importance. In addition, the two dynamical {\it assumptions} "A" and "B\," of general character formulated in Introduction were used. They concern the BPS properties of original particles and the absence of extra fields massless at $\mx\neq 0,\, m\neq 0$. All this appeared to be sufficient to calculate mass spectra in vacua considered at small but non-zero values of $\mx$, i.e. $0<\mx<M_{\rm min}$, where $M_{\rm min}$ is the independent of $\mx$ mass scale appropriate for each given vacuum, such that the phase and mass spectrum continue smoothly down to $\mx\ra 0$. In other words, the theory stays at $\mx<M_{\rm min}$ in the same regime and all hierarchies in the mass spectrum are the same as in the ${\cal N}=2$ theory at $\mx\ra 0$.

Within this framework we considered first in detail in section 2.1 the br2 vacua of $SU(N_c)$ theory with $N_c+1<N_F<2N_c-1$ (these are vacua of the baryonic branch in \cite{APS} or zero vacua in \cite{SY1,SY2}), at $m\ll\lm$ and $0<\mx\ll M^{(\rm br2)}_{\rm min}=\langle\Lambda^{SU(\nd-{\rm n}_1)}_{{\cal N}=2\,\,SYM}
\rangle\ll m\ll\lm$. The original color symmetry is broken spontaneously in these vacua by higgsed $\langle X^{adj}_{SU(N_c)}\rangle\neq 0$ in three stages. The first stage is at the highest scale $\mu\sim\lm$, $\,\langle X^{adj}_{SU(2N_c-N_F)}\rangle\sim\lm$, $\,SU(N_c)\ra SU(N_F-N_c)\times U^{(1)}(1)\times U^{2N_c-N_F-1}(1)$, this pattern is required by the unbroken discrete $Z_{2N_c-N_F\geq 2}$ symmetry and is necessary to avoid $g^2(\mu<\lm)$ in this effectively massless at the scale $\mu\sim\lm$ ${\cal N}=2$ UV free theory. The second stage is in the $SU(N_F-N_c)$ sector at the scale $\mu\sim m$, $\,\langle X^{adj}_{SU(N_F-N_c)}\rangle\sim m\ll\lm$, $\,SU(N_F-N_c)\ra SU(\no)\times U^{(2)}(1)\times SU(N_F-N_c-\no)$ . And the third stage is in the $SU(N_F-N_c-\no)$ sector at the scale $\mu\sim\langle\Lambda^{SU(\nd-{\rm n}_1)}_{{\cal N}=2\,\,SYM}\rangle\ll m$, $\,SU(N_F-N_c-\no)\ra U^{N_F-N_c-\no-1}(1)$.

The global flavor symmetry $U(N_F)$ is broken spontaneously at $\mx\neq 0$ at the scale $\mu\sim (\mx m)^{1/2}$ in these vacua in the $SU(\no)$ sector by higgsed original pure electric quarks, $\,\langle{\ov Q}^{\,a}_i\rangle=\langle Q^i_a\rangle\sim \delta ^a_i\,(\mx m)^{1/2}\,,\\ a=1...\no\,,\,\, i=1...N_F$, as\,: $U(N_F)\ra U(\no)\times U(\nt),\, 1\leq\no< N_F-N_c$.

It was shown that the lightest charged BPS particles (massless at $\mx\ra 0$) are the following.\,-\\
1) $\bb$ flavorless dyons $D_j, {\ov D}_j$ (this number $\bb$ is required by the unbroken $Z_{\bb\geq 2}$ discrete symmetry which operates interchanging them with each other), with the two non-zero pure electric charges, the $SU(N_c)$ baryon charge and $U^{(1)}(1)$ one and, on the whole, with non-zero all $\bb-1$ independent $SU(\bb)$ adjoint charges (with non-zero magnetic parts). These dyons are formed at the scale $\mu\sim\lm$, at the first stage of the color breaking, and are mutually local between themselves and with respect to all particles of the $SU(N_F-N_c)$ sector. 2) All original electric particles of remaining unbroken at $\mx\ra 0$ $\,SU(\no)$ subgroup of original $SU(N_c)$. 3) $\nd-\no-1$ pure magnetic monopoles with the $SU(\nd-\no)$ adjoint magnetic charges. These last are formed at the low scale $\mu\sim\langle\Lambda^{SU(\nd-{\rm n}_1)}_{{\cal N}=2\,\,SYM}\rangle\ll m\ll\lm$ in the ${\cal N}=2\,\,SU(\nd-\no)$ SYM sector, with the scale factor $\langle\Lambda^{SU(\nd-{\rm n}_1)}_{{\cal N}=2\,\,SYM}\rangle$ of its gauge coupling.  In addition, there are $N_c-1$ Abelian ${\cal N}=2$ photon multiplets massless at $\mx\ra 0$.

Massless at $\mx=0$ $\,2N_c-N_F$ dyons $D_j,\,\no$ original pure electric quarks $Q^i_a$ from $SU(\no)$, and $\nd-\no-1$ pure magnetic monopoles are all higgsed at small $\mx\neq 0$, and no massless particles remains (except for $2\no\nt$ Nambu-Goldstone multiplets due to spontaneous breaking of global flavor symmetry, $U(N_F)\ra U(\no)\times U(\nt)$). The mass spectrum was described in these br2 vacua at $0<\mx\ll\langle\Lambda^{SU(\nd-{\rm n}_1)}_{{\cal N}=2\,\,SYM}\rangle\ll m\ll\lm$.

The material of this section 2.1 served then as a basis for similar regimes in sections 3,\,6,\,7, and 8.

The mass parameter $\mx$ was increased then in two different stages\,:\, 1) $\langle\Lambda^{SU(\nd-{\rm n}_1)}_{{\cal N}=2\,\,SYM}\rangle\ll\mx\ll m$ in section 2.3\,;\,\,2) $m\ll\mx\ll\lm$ in section 2.4, and changes in the mass spectra were described. In all those cases when the corresponding ${\cal N}=1$ SQCD lower energy theories were weakly coupled, we needed no additional dynamical assumptions.  And only in those few cases when the corresponding ${\cal N}=1$ SQCD theories were in the strongly coupled conformal regime at $3N_c/2<N_F<2N_c$, we used additionally the assumption of the dynamical scenario introduced in \cite{ch3}. In essence, this assumption from \cite{ch3} looks here as follows\,: {\it no additional parametrically light composite solitons are formed in this ${\cal N}=1$ SQCD without decoupled corresponding colored adjoint scalars $X^{\rm adj}$ at those even lower scales where this ${\cal N}=1$ conformal regime is broken by non-zero particle masses} (see also the footnote \ref{(f3)}). As was shown in \cite{ch3,ch5,ch6}, this dynamical scenario does not contradict to any proven properties of ${\cal N}=1$ SQCD, satisfies all those checks of Seiberg's duality hypothesis for ${\cal N}=1$ SQCD which were used in \cite{S2}, and allows to calculate the corresponding mass spectra.

Many other vacua with the broken or unbroken flavor symmetry and at different hierarchies between the Lagrangian parameters $m,\, \mx,\, \lm$ were considered in sections 3-8 and corresponding mass spectra were calculated within the above described framework. Besides, calculations of power corrections to the leading terms of the low energy  quark and dyon condensates are presented in two important Appendices. These last calculations serve as {\it the independent numerous checks} of a self-consistency of the whole approach.

We consider that there is no need to repeat in detail in these conclusions a whole context of the paper (see the table of contents). But at least two points have to be emphasized. 1) As shown e.g. in section 6.2 (and similarly in sections 6.4, 7, and 8), the widespread opinion that the holomorphic dependence of gauge invariant chiral condensates on parameters of the superpotential implies the absence of phase transitions in supersymmetric theories is not right. 2) In contradiction with the appearance of "instead of confinement phase" proposed in a number of papers by M. Shifman and A. Yung, see e.g. the latest paper \cite{SY6} and references therein to their previous papers with this phase, we have found no such phase in all cases considered (see e.g. section 6.4 with the most clear case).

Clearly, this article does not pretend on strict proofs as the two dynamical assumptions of general character presented in Introduction were used (i.e., the assumption "A" about the BPS nature of original electric particles, and "B" about the absence of extra massless particles). But, to the best of our knowledge, the results obtained look self-consistent and do not contradict to any proven results.

In comparison with corresponding results from recent related papers \cite{SY1,SY2,SY6} of M. Shifman and A. Yung (and from a number of their previous numerous papers on this subject), our results are essentially different. In addition to critical remarks given in Introduction and in section 6, an extended criticism of a number of results from \cite{SY1,SY2} is given in section 8 of \cite{ch6}\,.\\
\vspace*{3mm}

\appendix{\Large \bf Power corrections to the quark and dyon condensates}\\

\hspace*{4mm} As examples, we present below in two Appendices the calculation of leading power corrections to the low energy quark condensates $\langle\Qo\rangle_{\no}$ in br1 vacua of $SU(N_c)$ and $U(N_c)$ in sections 4.1 and 4.2, and in br2 vacua of $SU(N_c)$ and $U(N_c)$ in sections 2.1 and 2.2, and to the dyon condensates in sections 2.1 and 2.2. These calculations confirm {\it independently} in a non-trivial way a self-consistency of the whole approach.

\section{Power corrections to quark condensates}

{\bf 1) br1 vacua of $SU(N_c),\,\, m\gg\lm$ in section 4.1}\\

From \eqref{(1.3)},\eqref{(1.4)}, in these vacua (up to even smaller power corrections)
\bq
\langle\Qo\rangle_{N_c}=\mx m_3-\frac{N_c-\nt}{N_c-\no}\langle\Qt\rangle_{N_c}\,,\quad
\langle\Qt\rangle_{N_c}\approx\mx m_3\Bigl (\frac{\lm}{m_3}\Bigr )^{\frac{2N_c-N_F}{N_c-\no}}\,,\label{(A.1)}
\eq
\bbq
\langle S\rangle_{N_c}=\frac{\langle\Qo\rangle_{N_c}\langle\Qt\rangle_{N_c}}{\mx}\approx m_3\langle\Qt\rangle_{N_c}\,,\quad m_3=\frac{N_c}{N_c-\no} m\,.
\eeq
From \eqref{(2.1.9)},\eqref{(A.1)}, the leading power correction to $\langle\w^{\,\rm eff}_{\rm tot}\rangle$ is
\bq
\langle\delta \w^{\,\rm eff}_{\rm tot}\rangle\approx (N_c-\no)\mx m^2_3\Bigl (\frac{\lm}{m_3}\Bigr )^{\frac{2N_c-N_F}{N_c-\no}}\approx (N_c-\no)\langle S\rangle_{N_c}\,.\label{(A.2)}
\eq
On the other hand, this leading power correction to $\langle\w^{\,\rm low}_{\rm tot}\rangle$ originates from the $SU(N_c-\no)$ SYM part, see \eqref{(4.1.3)},\eqref{(4.1.2)},
\bbq
\langle\delta \w^{\,\rm low}_{\rm tot}\rangle\approx\mx\langle\,{\rm Tr\,}\Bigl (X^{adj}_{SU(N_c-\no)}\Bigr )^2\,\rangle\approx(N_c-\no)\langle S\rangle_{N_{c}-\no}
\approx (N_c-\no)\mx\langle\Lambda^{SU(N_{c}-\no)}_{{\cal N}=2\,\,SYM}\rangle^2\approx
\eeq
\bq
\approx (N_c-\no)\mx  m_3^2\Bigl (\frac{\lm}{m_3}\Bigr )^{\frac{2N_c-N_F}{(N_c-\no)}}\,,\quad \ra\quad \langle\delta \w^{\,\rm low}_{\rm tot}\rangle=\langle\delta \w^{\,\rm eff}_{\rm tot}\rangle\,,\label{(A.3)}
\eq
as it should be.

To calculate the leading power correction $\delta\langle\Qo\rangle_{\no}\sim\mx\langle\Lambda^
{SU(N_{c}-\no)}_{{\cal N}=2\,\,SYM}\rangle^2/m$ to the low energy quark condensate $\langle\Qo\rangle_{\no}$, we have to account for the terms with the first power of quantum fluctuation ${\hat a}$ of the field $a=\langle a\rangle+{\hat a},\,\,\langle{\hat a}\rangle=0$, in the $SU(N_c-\no)$ SYM contribution to $\delta \w^{\,\rm low}_{\rm tot}$, see\eqref{(4.1.3)},\eqref{(4.1.2)},
\bbq
\delta \w^{\,\rm low}_{\rm tot}\approx (N_c-\no)\mx\, \delta\, \Bigl (\Lambda^{SU(N_{c}-\no)}_{{\cal N}=2\,\,SYM}\Bigr )^2\approx (N_c-\no)\mx\,\delta\,\Bigl (\frac{\lm^{2N_c-N_F}(m_3-c\,{\hat a})^{N_F}}{[\,m_3+(1-c){\hat a}\,]^{2\no}}\Bigr )^{\frac{1}{N_c-\no}}\approx
\eeq
\bq
\approx \,\mx\langle\Lambda^{SU(N_{c}-\no)}_{{\cal N}=2\,\,SYM}\rangle^2\,\Bigl [\,-\no \frac{(2N_c-N_F)}{N_c-\no}\,\Bigr ]\,\frac{{\hat a}}{m_3}\,,\quad m_3=\frac{N_c}{N_c-\no} m\,.\label{(A.4)}
\eq
As a result, from \eqref{(4.1.3)},\eqref{(4.1.4)},\eqref{(A.4)},\eqref{(A.1)},
\bbq
\delta\langle\Qo\rangle_{\no}\approx\frac{1}{\no}\langle\frac{\partial}{\partial {\hat a}}\delta \w^{\,\rm low}_{\rm tot}\rangle\approx - \frac{(2N_c-N_F)}{N_c-\no}\,\mx\frac{\langle\Lambda^{SU(N_{c}-\no)}_{{\cal N}=2\,\,SYM} \rangle^2}{m_3}\,,
\eeq
\bq
\langle\Qo\rangle^{SU(N_c)}_{\no}=\langle{\ov Q}^1_1\rangle\langle Q^1_1\rangle\approx \mx m_3\Biggl [1-\frac{2N_c-N_F}{N_c-\no} \,\frac{\langle\Lambda^{SU(N_{c}-\no)}_{{\cal N}=2\,\,SYM} \rangle^2}{m^2_3}\Biggr ]\approx \label{(A.5)}
\eq
\bbq
\approx\mx m_3\Biggl [1-\frac{2N_c-N_F}{N_c-\no}\Bigl (\frac{\lm}{m_3}\Bigr )^{\frac{2N_c-N_F}{N_c-\no}} \Biggr ]\,,\quad\langle\Qo\rangle^{SU(N_c)}
_{N_c}\approx\mx m_3\Biggl [1-\frac{N_c-\nt}{N_c-\no}\Bigl (\frac{\lm}{m_3}\Bigr )^{\frac{2N_c-N_F}{N_c-\no}} \Biggr ]\,.
\eeq
It is seen from \eqref{(A.5)} that, although the leading terms are the same in $\langle\Qo\rangle_{\no}$ and $\langle\Qo\rangle_{N_c}$, the leading power corrections are different.\\

{\bf 2) br1 vacua of $U(N_c),\,\, m\gg\lm$ in section 4.2}

From \eqref{(1.4)},\eqref{(4.2.1)},\eqref{(4.2.2)},\eqref{(4.2.3)} (keeping now leading power corrections)
\bq
\langle\Qo\rangle_{N_c}=\mx m-\langle\Qt\rangle_{N_c}\,,\label{(A.6)}
\eq
\bbq
\langle\Qt\rangle_{N_c}=\mx\lm^{\frac{\bb}{N_c-\no}}\Bigl (\frac{\langle\Qo\rangle_{N_c}}{\mx}\Bigr )^{\frac{\nd-\no}{N_c-\no}}\approx\mx m\Bigl (\frac{\lm}{m}\Bigr )^{\frac{2N_c-N_F}{N_c-\no}}
\Biggl (1+O\Bigl (\frac{\langle\Lambda^{SU(N_{c}-\no)}_{{\cal N}=2\,\,SYM}\rangle^2}{m^2} \Bigr )\Biggr )\,,
\eeq
\bbq
\langle S\rangle_{N_c}=\frac{\langle\Qo\rangle_{N_c}\langle\Qt\rangle_{N_c}}{\mx}\approx \mx m^2\Bigl (\frac{\lm}{m}\Bigr )^{\frac{2N_c-N_F}{N_c-\no}}\Biggl (1+O\Bigl (\frac{\langle\Lambda^{SU(N_{c}-\no)}_{{\cal N}=2\,\,SYM}\rangle^2}{m^2} \Bigr )\Biggr )\,,
\eeq
\bq
\frac{\langle a_0\rangle}{m}=\frac{1}{N_c\mx m}\langle{\rm Tr}\,\qq\rangle_{N_c}\approx\frac{\no}{N_c}  +\frac{\nt-\no}{N_c}\Bigl (\frac{\lm}{m}\Bigr )^{\frac{2N_c-N_F}{N_c-\no}}\,,\quad \langle a\rangle=\langle m-a_0\rangle\,, \label{(A.7)}
\eq
\bq
\langle\w^{\,\rm low}_{\rm tot}\rangle\approx \Bigl [\langle\w_{\no}\rangle=0\,\Bigr ]+\Biggl [
\langle\w_{a_0,a}\rangle=\frac{\mx}{2}\Bigl (\,N_c\langle a_0\rangle^2+\frac{\no N_c}{N_c-\no}\langle a\rangle^2\Bigr )\Biggr ]+\langle\w_{SU(N_c-\no)}^{\,SYM}\rangle\,,\label{(A.8)}
\eq
\bbq
\langle\w_{SU(N_c-\no)}^{\,SYM}\rangle=\mx\langle{\,\rm Tr\,}\Bigl ( X^{adj}_{SU(N_c-\no)}\Bigr )^2\rangle=
(N_c-\no)\langle S\rangle_{SU(N_c-\no)}=(N_c-\no)\mx\langle\Lambda^{SU(N_{c}-\no)}_{{\cal N}=2\,\,SYM}\rangle^2=
\eeq
\bbq
=(N_c-\no)\mx\langle\Biggl (\frac{\lm^{2N_c-N_F}(m-a_0-c \,a)^{N_F}}{[\,(1-c) a \,]^{2\no}} \Biggr )^{\frac{1}{N_c-\no}}\rangle\approx (N_c-\no)\mx m^2\Bigl (\frac{\lm}{m}\Bigr )^{\frac{2N_c-N_F}{N_c-\no}}\,.
\eeq
From \eqref{(A.6)}-\eqref{(A.8)}, the leading power correction $\langle\,\delta\,\w^{\,\rm low}_{\rm tot}\rangle\sim\mx\langle\Lambda^{SU(N_{c}-\no)}_{{\cal N}=2\,\,SYM}\rangle^2$ to $\langle\w^{\,\rm low}
_{\rm tot}\rangle$ looks as
\bbq
\langle\,\delta\, \w^{\,\rm low}_{\rm tot}\rangle=\Bigl (\langle\,\delta\,\w_{a_0,a}\rangle=0\Bigr )+
\langle\,\delta\,\w_{SU(N_c-\no)}^{\,SYM}\rangle\approx
\eeq
\bq
\approx (N_c-\no)\mx\langle\Lambda^{SU(N_{c}-\no)}_{{\cal N}=2\,\,SYM}\rangle^2\approx (N_c-\no)\,m\langle\Qt\rangle_{N_c}\approx (N_c-\no)\mx m^2\Bigl (\frac{\lm}{m}\Bigr )^{\frac{2N_c-N_F}{N_c-\no}}\,.\label{(A.9)}
\eq

Instead of \eqref{(2.1.9)},\eqref{(2.1.10)} we have now, see \eqref{(2.2.7)},
\bq
\w^{\,\rm eff}_{\rm tot}(\Pi)=m\,{\rm Tr}\,({\ov Q} Q)_{N_c}-\frac{1}{2\mx}\Biggl [ \,\sum_{i,j=1}^{N_F} ({\ov Q}_j Q^i)_{N_c}({\ov Q}_{\,i} Q^j)_{N_c}\Biggr ]-\nd \Biggl [\,S_{N_c}=\frac{\langle\Qo\rangle_{N_c}\langle\Qt\rangle_{N_c}}{\mx}\,\Biggr ],\,\,\,\,\,\label{(A.10)}
\eq
and using \eqref{(A.6)} (with the same accuracy)
\bbq
\langle\delta\w^{\,\rm eff}_{\rm tot}(\Pi)\rangle\approx (N_c-\no)m\langle\Qt\rangle_{N_c}\approx (N_c-\no)\mx m^2\Bigl (\frac{\lm}{m}\Bigr )^{\frac{2N_c-N_F}{N_c-\no}}\approx\langle\delta\w^{\,\rm low}
_{\rm tot}\rangle\,,
\eeq
as it should be.

As for the leading power correction $\delta\langle\Qo\rangle_{\no}\sim \mx m\Bigl (\lm/m\Bigr )^
{\frac{2N_c-N_F}{N_c-\no}}$ to the quark condensate, we have now instead of \eqref{(A.4)}
\bbq
\delta\w^{\,\rm low}_{\rm tot}=\delta \w_{a_0,a}+\delta \w_{SU(N_c-\no)}^{\,SYM}\,,\quad a_0=\langle a_0\rangle +{\hat a}_0\,,\quad a=\langle a\rangle +{\hat a}\,,
\eeq
\bbq
\delta \w_{SU(N_c-\no)}^{\,SYM}=(N_c-\no)\mx\, \delta\, \Bigl (\Lambda^{SU(N_{c}-\no)}_{{\cal N}=2\,\,SYM}\Bigr )^2= (N_c-\no)\mx\,\delta\,\Bigl (\frac{\lm^{2N_c-N_F}(m-{\hat a}_0-c\,{\hat a})^{N_F}}{[\,m+(1-c){\hat a}\,]^{2\no}}\Bigr )^{\frac{1}{N_c-\no}}\approx
\eeq
\bq
\approx \Bigl [\,-N_F{\hat a_0}-\frac{\no(2N_c-N_F)}{N_c-\no}\,{\hat a}\, \Bigr ) \mx\frac{\langle\Lambda^{SU(N_{c}-\no)}_{{\cal N}=2\,\,SYM}\rangle^2}{m}\,,
\label{(A.11)}
\eq
while from \eqref{(4.2.3)},\eqref{(A.7)}
\bq
\delta \w_{a_0,a}\approx (\nt-\no)\Bigl [\,{\hat a}_0-\frac{\no}{N_c-\no}\,{\hat a}\,\Bigr ]\mx\frac{\langle\Lambda^{SU(N_{c}-\no)}_{{\cal N}=2\,\,SYM}\rangle^2}{m}\,.\label{(A.12)}
\eq

Instead of \eqref{(A.5)} (with the same accuracy), on the one hand, see \eqref{(4.2.3)},\eqref{(4.2.4)},\eqref{(A.6)},
\bbq
\langle{\,\rm Tr\,}\qq\rangle_{\no}=\no\mx m+\delta\langle{\,\rm Tr\,}\qq\rangle_{\no}=\no\Bigl (\mx m+\delta\,\langle\Qo\rangle_{\no}\Bigr )\,,
\eeq
\bq
\delta\langle{\,\rm Tr\,}\Qo\rangle_{\no}=\frac{1}{\no}\langle\frac{\partial}{\partial {\hat a}_0} \Bigl (\delta \w_{SU(N_c-\no)}^{\,SYM}+\delta \w_{a_0,a}\Bigr )\rangle\approx -2 \mx\frac{\langle\Lambda^{SU(N_{c}-\no)}_{{\cal N}=2\,\,SYM}\rangle^2}{m}\,,\label{(A.13)}
\eq
while on the other hand,
\bbq
\delta\langle{\,\rm Tr\,}\Qo\rangle_{\no}=\frac{1}{\no}\langle\,\frac{\partial}{\partial {\hat a}}\Bigl (\delta \w_{SU(N_c-\no)}^{\,SYM}+\delta \w_{a_0,a}\Bigr )\approx -2 \mx\frac{\langle\Lambda^{SU(N_{c}-\no)}_{{\cal N}=2\,\,SYM}\rangle^2}{m} \,,
\eeq
as it should be.

Therefore, on the whole for these $U(N_c)$ br1 vacua, see \eqref{(A.6)} for $\langle\Qo\rangle^{U(N_c)}_{N_c}$,
\bq
\langle\Qo\rangle^{U(N_c)}_{\no}=\langle{\ov Q}^1_1\rangle\langle Q^1_1\rangle\approx\mx m\Bigl [\,1-2\frac{\langle\Lambda^{SU(N_{c}-\no)}_{{\cal N}=2\,SYM}\rangle^2}{m^2}\,\Bigr ]\approx\mx m \Bigl [\, 1-2\Bigl (\frac{\lm}{m}\Bigr )^{\frac{2N_c-N_F}{N_c-\no}}\,\Bigr ]\,,\label{(A.14)}
\eq
\bbq
\langle\Qo\rangle^{U(N_c)}_{N_c}\approx\mx m\Bigl [\,1-\frac{\langle\Lambda^{SU(N_{c}-\no)}_{{\cal N}=2\,\,SYM}\rangle^2}{m^2}\,\Bigr ].
\eeq

We can compare also the result for $\langle\Qo\rangle^{U(N_c)}_{\no}$ in \eqref{(A.14)} with those from \eqref{(2.2.10)}.
\footnote{\,
In these simplest br1 vacua of $U(N_c)$ the charges and multiplicities of particles massless at $\mx\ra 0$ are evident, these are original electric particles from $SU(\no)\times U^{(0)}(1)$ and $N_c-\no-1$ magnetic monopoles from $SU(N_c-\no)$ SYM.
}
To obtain definite predictions from \eqref{(2.2.10)} one has to find first the values of roots entering \eqref{(2.2.10)}. In the case  considered, to obtain definite values of all roots of the curve \eqref{(1.2)} in the br1 vacua of $U(N_c)$ theory one needs only one additional relation \cite{CIV,CSW} for the two single roots $e^{\pm}$ of the curve \eqref{(1.2)}
\bq
e^{\pm}=\pm 2\sqrt{\langle S\rangle_{N_c}/\mx}\,,\quad e^{\pm}_c=\frac{1}{2} (e^{+}+e^{-})=0\,.\label{(A.15)}
\eq
One can obtain then from the $U(N_c)$ curve \eqref{(1.2)} the values of $\no$ equal double roots $e^{(Q)}_i$ of original electric quarks from $SU(\no)$ and $N_c-\no-1$ unequal double roots $e^{(M)}_k$ of magnetic monopoles from $SU(N_c-\no)$ SYM. We find from \eqref{(1.2)},\eqref{(A.15)} with our accuracy, see \eqref{(A.6)},\eqref{(A.8)},
\bbq
e^{+}=-e^{-}\approx 2 m\Bigl (\frac{\lm}{m}\Bigr )^{\frac{2N_c-N_F}{2(N_c-\no)}}\Biggl (1+O\Bigl (\frac{\langle\Lambda^{SU(N_{c}-\no)}_{{\cal N}=2\,\,SYM}\rangle^2}{m^2} \Bigr ) \Biggr )
\approx 2\langle\Lambda^{SU(N_{c}-\no)}_{{\cal N}=2\,\,SYM}\rangle,\quad e^{(Q)}_i= - m,\,\, i=1...\no\,,
\eeq
\bq
e^{(M)}_k\approx 2\cos (\frac{\pi k}{N_c-\no})\langle\Lambda^{SU(N_{c}-\no)}
_{{\cal N}=2\,\,SYM}\rangle-\frac{\nt-\no}{N_c-\no-1}\frac{\langle\Lambda^{SU(N_{c}-\no)}
_{{\cal N}=2\,\,SYM}\rangle^2}{m}\,,\quad k=1...(N_c-\no-1)\,.\label{(A.16)}
\eq
Let us note that these roots satisfy the sum rule, see \eqref{(A.7)},
\bq
\sum_{n=1}^{N_c}\phi_n=\frac{1}{2}\sum_{n=1}^{2 N_c}(-e_n)=N_c\langle a_0\rangle\,, \label{(A.17)}
\eq
as it should be.

Therefore, from \eqref{(2.2.10)},\eqref{(A.16)}
\bbq
\langle\Qo\rangle^{U(N_c)}_{\no}= - \mx\sqrt{(e^{(Q)}_i-e^+)(e^{(Q)}_i-e^-)}\approx
\eeq
\bq
\approx\mx m \Bigl (1-2\frac{\langle\Lambda^{SU(N_{c}-\no)}_{{\cal N}=2\,\,SYM}\rangle^2}{m^2} \Bigr)\approx
\mx m \Bigl [\, 1-2\Bigl (\frac{\lm}{m}\Bigr )^{\frac{2N_c-N_F}{N_c-\no}}\,\Bigr ]\,,\quad i=1...\no\,,\,\label{(A.18)}
\eq
this agrees with \eqref{(A.14)}.\\

Besides, we can use the knowledge of roots of the curve \eqref{(1.2)} for the $U(N_c)$ theory and of $\langle a_0\rangle$, see \eqref{(A.7)},\eqref{(A.16)}, to obtain the values of $\langle\Qo\rangle^{SU(N_c)}_{\no}$ condensates in the $SU(N_c)$ theory. For this, the curve \eqref{(1.2)} for the $U(N_c)$ theory
\bq
y^2=\prod_{i=1}^{N_c}(z+\phi_i)^2-4\lm^{\bb}(z+m)^{N_F}\,,\quad \sum_{i=1}^{N_c}\phi_i=\frac{1}{2}\sum_{n=1}^{2 N_c}(-e_n)=N_c\langle a_0\rangle{\rm\,\,\, in\,\,\, U(N_c)}\,,\label{(A.19)}
\eq
we rewrite in the form of the $SU(N_c)$ theory
\bq
y^2=\prod_{i=1}^{N_c}({\hat z}+{\hat \phi}_i)^2-4\lm^{\bb}({\hat z}+{\hat m})^{N_F}\,,\,\, \sum_{i=1}^{N_c}{\hat\phi}_i=\frac{1}{2}\sum_{n=1}^{2 N_c}(-{\hat e}_n)=0\,,
\,\, {\hat m}=m\, (1-m^{-1}\langle a_0\rangle)\,.\,\label{(A.20)}
\eq

But clearly, it remained the same $U(N_c)$ theory with the same condensates \eqref{(A.18)}. Therefore, for the $SU(N_c)$ curve \eqref{(1.2)} the quark condensates will be as in \eqref{(A.18)} with $m$ replaced by
${m^\prime}=m/(1-m^{-1}\langle a_0\rangle)$, both in the leading terms and in all power corrections, i.e., see \eqref{(A.7)},
\bbq
\frac{\langle a_0\rangle}{m}\,\,\ra\,\,\approx \Biggl (\,\frac{\no}{N_c}+\frac{\nt-\no}{N_c}\Bigl (\frac{\lm}{m_3}\Bigr )^{\frac{2N_c-N_F}{N_c-\no}}\,\Biggr )\,,
\eeq
\bq
m\ra {m^\prime}\approx m_3\Biggl (1+\frac{\nt-\no}{N_c-\no}\Bigl (\frac{\lm}{m_3}\Bigr )^{\frac{\bb}{N_c-\no}}\Biggr )\,,\quad m_3=\frac{N_c}{N_c-\no}\,m\,.\label{(A.21)}
\eq
Then, from \eqref{(A.18)},\eqref{(A.21)},
\bq
\langle\Qo\rangle^{SU(N_c)}_{\no}\approx \mx m_3 \Biggl (1+\frac{\nt-\no}{N_c-\no}\Bigl (\frac{\lm}{m_3}\Bigr )^{\frac{\bb}{N_c-\no}} \Biggr )\Bigl [\, 1-2\Bigl (\frac{\lm}{m_3}\Bigr )^{\frac{2N_c-N_F}{N_c-\no}}\,\Bigr ]\approx \label{(A.22)}
\eq
\bbq
\approx \mx m_3 \Bigl [\, 1-\frac{\bb}{N_c-\no}\Bigl (\frac{\lm}{m_3}\Bigr )^{\frac{2N_c-N_F}{N_c-\no}}\,\Bigr ]\,,
\eeq
this agrees with \eqref{(A.5)}.\\

{\bf 3) br2 vacua of $SU(N_c),\,\, m\ll\lm$ in section 2.1}

The calculations of leading power corrections to $\langle\,\delta \w_{\rm tot}^{\,\rm low}\rangle$ and $\langle\,\delta \w^{\,\rm eff}_{\rm tot}\rangle$ have been presented in \eqref{(2.1.15)}. Therefore, we present here the calculation of the leading power correction $\delta\,\langle\Qo\rangle_{\no}\sim \mx \langle\Lambda^{SU(N_{c}-\no)}_{{\cal N}=2\,\,SYM}\rangle^2/m$ to the quark condensate. For this, see \eqref{(2.1.4)},\eqref{(2.1.15)}, we write ($a_{1,2}=\langle a_{1,2}\rangle+{\hat a}_{1,2}$)
\bbq
\delta \w^{(SYM)}_{SU(\nd-\no)}\approx (\nd-\no)\wmu\, \delta\,\Bigl (\Lambda^{SU(\nd-\no)}_{{\cal N}=2\,\,SYM}\Bigr )^2\approx (\nd-\no)\wmu\,\delta\,\Biggl (\frac{\Lambda_{SU(\nd)}^{2\nd-N_F}
(m_2-{\hat a}_1-c_2\,{\hat a}_2)^{N_F}}{[\,m_2+(1-c_2)\,{\hat a}_2\,]^{\,2\no}}\Biggr )^{\frac{1}{\nt-N_c}}\approx
\eeq
\bq
\approx \wmu\frac{\langle\Lambda^{SU(N_{c}-\no)}_{{\cal N}=2\,\,SYM}
\rangle^2}{m_1}\,\Bigl [\,N_F{\hat a}_1+\frac{\no(2N_c-N_F)}{N_c-\nt}\,{\hat a}_2 \Bigr ],
\,\, m_1= \frac{N_c}{N_c-\nt}m= -\, m_2,\,\, \wmu=-\,\mx,\label{(A.23)}
\eq
\bbq
\langle{\,\rm Tr\,}\qq\rangle_{\no}=\no\mx m_1+\delta\langle{\,\rm Tr\,}\qq\rangle_{\no}=\no\Bigl (\mx m_1+\delta\,\langle\Qo\rangle_{\no}\Bigr )\,,\quad \Lambda_{SU(\nd)}=-\lm\,,
\eeq
From this, see \eqref{(2.1.5)}, and \eqref{(2.1)} for $\langle\Qt\rangle^{SU(N_c)}_{N_c}$,
\bbq
\delta\,\langle\Qo\rangle_{\no}=\frac{1}{\no}\langle\frac{\partial}{\partial {\hat
a}_2}\w^{(SYM)}_{SU(\nd-\no)}\rangle\approx\wmu m_1\Bigl [\,\frac{2N_c-N_F}{N_c-\nt}\frac{\langle\Lambda^
{SU(\nd-\no)}_{{\cal N}=2\,\,SYM}\rangle^2}{m^2_1}\,\Bigr ],\,\,\frac{\langle\Lambda^{SU(\nd-\no)}_{{\cal N}=2\,\,SYM}\rangle^2}{m^2_1}\approx\Bigl (\frac{m_1}{\lm}\Bigr )^{\frac{2N_c-N_F}{\nt-N_c}},
\eeq
\bq
\langle\Qo\rangle^{SU(N_c)}_{\no}=\langle{\ov Q}^1_1\rangle\langle Q^1_1\rangle\approx\mx m_1\Bigl [\,1+\frac{2N_c-N_F}{\nt-N_c}\,
\frac{\langle\Lambda^{SU(\nd-\no)}_{{\cal N}=2\,\,SYM}\rangle^2}{m^2_1}\,\Bigr ]\approx \label{(A.24)}
\eq
\bbq
\approx\mx m_1\Bigl [\,1+\frac{2N_c-N_F}{\nt-N_c}\,\Bigl (\frac{m_1}{\lm}\Bigr )^{\frac{2N_c-N_F}{\nt-N_c}}\Bigr ]\,,\quad m_1= \frac{N_c}{N_c-\nt}m\,,
\eeq
\bbq
\langle\Qt\rangle^{SU(N_c)}_{N_c}=\mx m_1-\frac{N_c-\no}{N_c-\nt}\,\langle\Qo\rangle_{N_c}
\approx\mx m_1\Bigl [\, 1+\frac{N_c-\no}{\nt-N_c}\frac{\langle\Lambda^{SU(\nd-\no)}_{{\cal N}=2\,\,SYM}\rangle^2}{m^2_1}\,\Bigr ]\,.
\eeq
It is seen that, as in \eqref{(A.5)}, the leading corrections to $\langle\Qo\rangle_{\no}$ and $\langle\Qt\rangle_{N_c}$ are different. \\

{\bf 4) br2 vacua of $U(N_c),\,\, m\ll\lm$ in section 2.2}

Keeping the leading order corrections $\sim\langle\Lambda^{SU(\nd-\no)}_{{\cal N}=2\,\, SYM}\rangle^2/m$, from \eqref{(2.2.2)},\eqref{(2.2.4)},\eqref{(2.1.15)} (put attention that the quark and gluino condensates, and $\Lambda^{SU(\nd-\no)}_{{\cal N}=2\,\, SYM}$ are different in br2 vacua of $SU(N_c)$ or $U(N_c)$ theories, see \eqref{(A.24)},\eqref{(A.25)},
\bq
\langle\Qt\rangle_{N_c}=\mx m-\langle\Qo\rangle_{N_c}\,,\quad \langle\Qo\rangle_{N_c}\approx\mx m\Bigl ( \frac{m}{\lm}\Bigr )^{\frac{2N_c-N_F}{\nt-N_c}}\approx \mx\frac{\langle\Lambda^{SU(\nd-\no)}_{{\cal N}=2\,\, SYM}\rangle^2}{m} \,,\label{(A.25)}
\eq
\bbq
\frac{\langle a_0\rangle}{m}=\frac{1}{N_c\mx m}\langle{\rm Tr\,}\qq\rangle_{N_c}\approx \frac{\nt}{N_c}+\frac{\no-\nt}{N_c}\,
\Bigl ( \frac{m}{\lm}\Bigr )^{\frac{2N_c-N_F}{\nt-N_c}}\,.
\eeq
\bq
\delta \w_{SU(\nd-\no)}^{\,SYM}=(\nd-\no)\wmu\,\delta\,\Biggl [\Bigl (\Lambda^{SU(\nd-\no)}_{{\cal N}=2\,\, SYM}\Bigr )^{2}=\Biggl (\frac{\Lambda_{SU(\nd)}^{2\nd-N_F}(-m-{\hat a}_0-{\hat a}_1-c_2\,{\hat a}_2)^{N_F}}
{[\,-m+(1-c_2)\,{\hat a}_2\,]^{\,2\no}}\Biggr )^{\frac{1}{\nt-N_c}}\Biggr ]\,\,\,\label{(A.26)}
\eq
\bbq
\approx\wmu\,\frac{\langle\Lambda^{SU(\nd-\no)}_{{\cal N}=2\,\, SYM}\rangle^2}{m}\, \Bigl [\,N_F({\hat a}_0+{\hat a}_1)-\no\frac{ 2N_c-N_F}{\nt-N_c}\,{\hat a}_2\,\Bigr ],\,\, \frac{\langle\Lambda^{SU(\nd-\no)}_{{\cal N}=2\,\,SYM}\rangle^2}{m^2}=
\Bigl (\frac{m}{\lm}\Bigr )^{\frac{2N_c-N_F}{\nt-N_c}},\,\, \Lambda_{SU(\nd)}=-\,\lm\,,
\eeq
\bbq
\delta \w_{a}\approx\mx\frac{\langle\Lambda^{SU(\nd-\no)}_{{\cal N}=2\,\, SYM}\rangle^2}{m}\,\Bigl [(\no-\nt) {\hat a}_0-\frac{N_F}{2N_c-N_F} {\hat a}_1-\no\frac{\nt-\no}{\nt-N_c}\,{\hat a}_2\,\Bigr ]\,,\quad a_i=\langle a_i\rangle+{\hat a}_i,\, i=0,\,1,\,2\,,\,
\eeq
\bbq
\delta\langle\Qo\rangle^{(U(N_c)}_{\no}=\frac{1}{\no}\langle\frac{\partial}{\partial {\hat a}_2}\Bigl (\delta\w_{SU(\nd-\no)}^{\,SYM}+\delta \w_{a}\Bigr )\rangle \approx - 2\mx m \frac{\langle\Lambda^{SU(\nd-\no)}_{{\cal N}=2\,\, SYM}\rangle^2}{m^2}\approx - 2\mx m\Bigl (\frac{m}{\lm}\Bigr )^{\frac{2N_c-N_F}{\nt-N_c}}\,.
\eeq
Therefore, on the whole for these $U(N_c)$ br2 vacua after accounting for the leading power corrections, see \eqref{(A.25)},\eqref{(A.26)},
\bq
\langle\Qo\rangle^{U(N_c)}_{\no}=\langle{\ov Q}^1_1\rangle\langle Q^1_1\rangle\approx\mx m\Bigl [1-2\Bigl (\frac{m}{\lm}\Bigr )^{\frac{2N_c-N_F}{\nt-N_c}}
\Bigr ],\,\, \langle\Qt\rangle^{U(N_c)}_{N_c}\approx\mx m\Bigl [1-\Bigl (\frac{m}{\lm}\Bigr )^{\frac{2N_c-N_F}{\nt-N_c}}\Bigr ].\,\,\,\,\label{(A.27)}
\eq

It is seen that, as above in $SU(N_c)$ br2 vacua \eqref{(A.24)}, the non-leading terms in $\langle\Qo\rangle_{\no}$ and $\langle\Qt\rangle_{N_c}$ are different.\\

We can compare now the value of $\langle\Qo\rangle^{U(N_c)}_{\no}$ from \eqref{(A.27)} with those from
\eqref{(2.2.10)} (see however the footnote \ref{(f7)}, in these br2 vacua the charges of massless at $\mx\ra 0$ particles are not trivial and not obvious beforehand, as well as their multiplicities). But we know the charges and multiplicities of massless at $\mx\ra 0$ particles from sections 2.1 and 2.2. To obtain definite predictions from \eqref{(2.2.10)} for the quark condensates $\langle\Qo\rangle_{\no}$ we need the values of $\no$ double roots $e^{(Q)}_k$ corresponding to original electric quarks from $SU(\no)$ and the values of two single roots $e^{\pm}$. Similarly to \eqref{(A.15)},\eqref{(A.16)}, these look here as
\bq
e^{(Q)}_k= - m\,,\quad k=1...\no\,,\quad  e^{+}=-e^{-}\approx 2\langle\Lambda^{SU(\nd-\no)}_{{\cal N}=2\,\, SYM}\rangle\,,\quad \frac{\langle\Lambda^{SU(\nd-\no)}_{{\cal N}=2\,\, SYM}\rangle^2}{m^2}\approx\Bigl (\frac{m}{\lm}\Bigr )^{\frac{2N_c-N_F}{\nt-N_c}}\ll 1\,.\,\,\,\label{(A.28)}
\eq
From this
\bq
\langle\Qo\rangle^{U(N_c)}_{\no}= - \mx\sqrt{(e^{(Q)}_k-e^+)(e^{(Q)}_k-e^-)}\approx\mx m \Bigl [\,1-2\Bigl (\frac{m}{\lm}\Bigr )^{\frac{2N_c-N_F}{\nt-N_c}}\,\Bigr ]\,,\label{(A.29)}
\eq
this agrees with \eqref{(A.27)}.

Besides, proceeding similarly to br1 vacua in \eqref{(A.19)}-\eqref{(A.22)}, we obtain for $\langle\Qo\rangle^{SU(N_c)}_{\no}$ in br2 vacua of $SU(N_c)$, see \eqref{(A.25)},
\bbq
\frac{\langle a_0\rangle}{m}\,\,\ra\,\,\approx \Biggl (\,\frac{\nt}{N_c}+\frac{\no-\nt}{N_c}\Bigl (\frac{m_1}{\lm}\Bigr )^{\frac{2N_c-N_F}{\nt-N_c}}\,\Biggr )\,,
\eeq
\bq
m\ra {m^\prime}\approx m_1\Biggl (1+\frac{\no-\nt}{N_c-\nt}\Bigl (\frac{m_1}{\lm}\Bigr )^{\frac{\bb}{\nt-N_c}} \Biggr )\,,\quad m_1=\frac{N_c}{N_c-\nt}\,m\,,\label{(A.30)}
\eq
and from \eqref{(A.29)},\eqref{(A.30)}
\bq
\langle\Qo\rangle^{SU(N_c)}_{\no}\approx \mx m_1 \Biggl (1+\frac{\no-\nt}{N_c-\nt}\Bigl (\frac{m_1}{\lm}\Bigr )^{\frac{\bb}{\nt-N_c}}\Biggr )\Bigl [\, 1-2\Bigl (\frac{m_1}{\lm}\Bigr )^{\frac{2N_c-N_F}{\nt-N_c}}\,\Bigr ]\approx \label{(A.31)}
\eq
\bbq
\approx \mx m_1 \Bigl [\, 1+\frac{\bb}{\nt-N_c}\Bigl (\frac{m_1}{\lm}\Bigr )^{\frac{2N_c-N_F}{\nt-N_c}}\,\Bigr ]\approx \mx m_1\Biggl [\,1+\frac
{2N_c-N_F}{\nt-N_c}\frac{\langle\Lambda^{SU(\nd-\no)}_{{\cal N}=2\,\,SYM}\rangle^2}{m^2_1}\,\Biggr ]\,,
\eeq
this agrees with \eqref{(A.24)}.

\section{Power corrections to dyon  condensates}

Consider first the br2 vacua of $\mathbf{U(N_c)}$ with $m\ll\lm$ in section 2.2. Because, unlike the br1 vacua of $U(N_c)$ theory in section 4.2, there are now in addition massless dyons in br2 vacua, to obtain with our accuracy the definite values of roots from the $U(N_c)$ curve \eqref{(1.2)}, one needs to use not only \eqref{(A.15)}, but also the sum rule \eqref{(A.17)}. The sum rule \eqref{(A.17)} looks in these br2 vacua as, see \eqref{(A.25)},
\bq
N_c\langle a_0\rangle\approx \Bigl (\nt m+(\no-\nt)\frac{\langle\Lambda^{SU(\nd-\no)}_{{\cal N}=2\,\, SYM}\rangle^2}{m} \Bigr )\approx \Biggl (\nt m+(\no-\nt) m\Bigl (\frac{m}{\lm}\Bigr )^{\frac{\bb}{\nt-N_c}}\Biggr )\approx \label{(B.1)}
\eq
\bbq
\approx\Bigl (\no m+(\nd-\no-1)A^{(M)}_c\frac{\langle\Lambda^{SU(\nd-\no)}_{{\cal N}=2\,\, SYM}\rangle^2}{m}-\frac{1}{2}(e^{+}+e^{-}=0)+(2N_c-N_F)C^{(D)}_c \Bigr )\,,
\eeq
where $\no(-e^{(Q)}_i)=\no m$ is the contribution of $\no$ equal double roots of original electric quarks from $SU(\no)$, the term with $A^{(M)}_c$ is the contribution from the centre of $\nd-\no-1$ unequal double roots of magnetic monopoles, and the term with $C^{(D)}_c$ is the contribution from the centre of $2N_c-N_F$ unequal double roots of dyons. We obtain then from the curve \eqref{(1.2)} together with \eqref{(B.1)} (with our accuracy)
\bq
A^{(M)}_c=\frac{\nt-\no}{\nd-\no-1}\,,\quad C^{(D)}_c=\frac{\nt-\no}{2N_c-N_F} m \Bigl ( 1-2 \frac{\langle\Lambda^{SU(\nd-\no)}_{{\cal N}=2\,\, SYM}\rangle^2}{m^2} \Bigr )\,,\,\,\,\label{(B.2)}
\eq
\bbq
e^{(M)}_k\approx 2\cos (\frac{\pi k}{\nd-\no})\langle\Lambda^{SU(\nd-\no)}
_{{\cal N}=2\,\,SYM}\rangle-\frac{\nt-\no}{\nd-\no-1}\frac{\langle\Lambda^{SU(\nd-\no)}
_{{\cal N}=2\,\,SYM}\rangle^2}{m}\,,\quad k=1...(\nd-\no-1)\,,
\eeq
\bbq
e^{(D)}_j\approx \omega^{j-1}\lm\Biggl (1+O\Bigl (\frac{m}{\lm}\Bigr )^{\bb}\Biggr )-\frac{\nt-\no}{2N_c-N_F}\,m\Biggl (1-2\Bigl
(\frac{m}{\lm}\Bigr )^{\frac{\bb}{\nt-N_c}}\Biggr )\,,\,\, j=1...(2N_c-N_F)\,,
\eeq
while $e^{\pm}$ are given in \eqref{(A.28)}.

From \eqref{(2.2.10)},\eqref{(B.2)} we obtain now for the dyon condensates in these br2 vacua of $U(N_c)$ at $m\ll\lm$ (with the same accuracy)
\bq
\langle{\ov D}_j D_j\rangle_{U(N_c)}=\langle{\ov D}_j\rangle\langle D_j\rangle\approx \mx\Biggl [\,-\, \omega^{j-1}\lm+\frac{\nt-\no}{2N_c-N_F}\,m
\Biggl (1-2\Bigl (\frac{m}{\lm}\Bigr )^{\frac{\bb}{\nt-N_c}}\Biggr ) \Biggr ]\,, \label{(B.3)}
\eq
\bbq
\langle\Sigma^{U(N_c)}_D\rangle\equiv\sum_{j=1}^{2N_c-N_F}\langle{\ov D}_j D_j\rangle_{U(N_c)}=\sum_{j=1}^{2N_c-N_F}\langle{\ov D}_j\rangle\langle D_j\rangle
\approx \mx m\,(\nt-\no)\Biggl (1-2\Bigl (\frac{m}{\lm}\Bigr )^{\frac{\bb}{\nt-N_c}}\Biggr ), \,\,\, j=1...\bb\,.
\eeq

Now, using only the leading terms $\sim \mx m$ of \eqref{(B.3)},\eqref{(2.2.3)},\eqref{(2.2.5)},
\eqref{(2.2.8)}, we can find the value of $\delta_3=O(1)$ in \eqref{(2.1.2)},\eqref{(2.1.5)},\eqref{(2.2.4)},\eqref{(6.2.3)}. From \eqref{(2.2.4)}
\bq
\langle\frac{\partial}{\partial a_0}\w^{\,\rm low}_{\rm tot}\rangle=0= - \langle{\rm Tr\,}\qq\rangle_{\no}-\langle\Sigma^{U(N_c)}_D\rangle+\mx N_c\langle a_0\rangle-\mx N_c\delta_3\langle a_1\rangle= - \mx N_c\delta_3\langle a_1\rangle \ra {\boldsymbol\delta}{\mathbf{_3=0}}.\,\,\quad\label{(B.4)}
\eq
(The same result follows from $\langle\partial \w^{\,\rm low}_{\rm tot}/{\partial a_1}\rangle=0$ in \eqref{(2.2.4)}\,).

As for the non-leading terms $\delta\langle\Sigma^{U(N_c)}_D\rangle\sim \mx m (m/\lm)^{\frac{\bb}{\nt-N_c}}$ of $\langle\Sigma^{U(N_c)}_D\rangle$ in \eqref{(B.3)}, with $\delta_3=0$ from \eqref{(B.4)}, these can now be also calculated {\it independently} from \eqref{(2.2.4)},\eqref{(A.25)},\eqref{(A.26)}, i.e.
\bbq
\langle\frac{\partial}{\partial {\hat a}_0}\delta\w^{\,\rm low}_{\rm tot}\rangle=0\ra \delta\langle\Sigma^{U(N_c)}_D\rangle\approx\Bigl [-\delta\langle{\rm Tr}\qq\rangle_{\no}=2\no\mx m\Bigl (\frac{m}{\lm}\Bigr )^{\frac{\bb}{\nt-N_c}}\Bigr ]_Q+\Bigl [(\no-\nt)\mx m\Bigl (\frac{m}{\lm}\Bigr )^{\frac{\bb}{\nt-N_c}}\Bigr ]_{\delta\w_{a_0}}\,\,\,
\eeq
\bq
+\Bigl [\langle\frac{\partial}{\partial {\hat a}_0}\delta \w_{SU(\nd-\no)}^{\,SYM}\rangle= - N_F\mx m\Bigl (\frac{m}{\lm}\Bigr )^{\frac{\bb}{\nt-N_c}}\Bigr ]_{\delta\w_{SYM}}\approx -2\mx m\,(\nt-\no)
\Bigl (\frac{m}{\lm}\Bigr )^{\frac{\bb}{\nt-N_c}}\,,\label{(B.5)}
\eq
this agrees with \eqref{(B.3)}.\\

As for the value of $\langle\Sigma^{SU(N_c)}_D\rangle$ in br2 vacua of $\mathbf{SU(N_c)}$, with $\delta_3=0$ from \eqref{(B.4)}, it can be found now directly from \eqref{(2.1.5)},\eqref{(2.1.12)},\eqref{(A.23)},\eqref{(A.24)}
\bq
\langle\frac{\partial}{\partial {\hat a}_1}\w^{\,\rm low}_{\rm tot}\rangle=0\ra \langle\Sigma^{SU(N_c)}_D\rangle\approx \mx m_1(\nt-\no)\Biggl [1+\frac{\bb}
{\nt-N_c}\Bigl (\frac{m_1}{\lm}\Bigr )^{\frac{\bb}{\nt-N_c}} \Biggr ]\,,\,\, m_1=\frac{N_c}{N_c-\nt}\,m\,.\,\,\,\label{(B.6)}
\eq
(Really, the power suppressed term in \eqref{(B.6)} can be found {\it independently of the value of $\delta_3$} because it is parametrically different and, unlike \eqref{(B.5)} in $U(N_c)$ theory, there are no power corrections to $\langle a_1\rangle$ in $SU(N_c)$ theory, see \eqref{(2.1.6)}).

It can also be obtained from \eqref{(B.3)} in br2 vacua of $U(N_c)$ proceeding similarly to \eqref{(A.30)},\eqref{(A.31)} for the quark condensates. We obtain from \eqref{(A.30)} and \eqref{(B.3)}
\bbq
\langle\Sigma^{SU(N_c)}_D\rangle=\sum_{j=1}^{2N_c-N_F}\langle{\ov D}_j\rangle\langle D_j\rangle\approx \mx m_1\Biggl (1+\frac{\no-\nt}{N_c-\nt}\Bigl (\frac{m_1}{\lm}\Bigr )^{\frac{\bb}{\nt-N_c}}\Biggr )(\nt-\no)\Bigl [\, 1-2\Bigl (\frac{m_1}{\lm}\Bigr )^{\frac{2N_c-N_F}{\nt-N_c}}\,\Bigr ]\approx
\eeq
\bq
\approx\mx m_1(\nt-\no)\Biggl [1+\frac{\bb}{\nt-N_c}\Bigl (\frac{m_1}{\lm}\Bigr )^{\frac{\bb}{\nt-N_c}} \Biggr ]\,,\label{(B.7)}
\eq
this agrees with \eqref{(B.6)}.

On the whole, the dyon condensates in br2 vacua of $SU(N_c)$ look at $m\ll\lm$ as, $j=1,...,\bb$,
\bq
\hspace*{-4mm}\langle{\ov D}_j D_j\rangle^{SU(N_c)}\approx \mx\Biggl [-\omega^{j-1}\lm\Biggl (1+O\Bigl (\frac{m}{\lm}\Bigr )^{\bb}\Biggr )+\frac{\nt-\no}{2N_c-N_F}\,m_1\Biggl (1+\frac{\bb}{\nt-N_c}\Bigl (\frac{m_1}{\lm}\Bigr )^{\frac{\bb}{\nt-N_c}}\Biggr ) \Biggr ].\,\,\,\label{(B.8)}
\eq
\vspace*{2mm}

Finally, not going into details, we present below the values of roots of the $\mathbf{SU(N_c)}$ spectral curve \eqref{(1.2)} in br1 vacua of section 4.1 and br2 vacua of section 2.1.

{\bf 1) br1 vacua}. There are at $\mx\ra 0$ $\,\no$ equal double roots $e^{(Q)}_i$ of massless original electric quarks $Q_i$, then $N_c-\no-1$ unequal double roots $e^{(M)}_k$ of massless pure magnetic monopoles $M_k$ and two single roots $e^{\pm}$ from ${\cal N}=2\,\, SU(N_c-\no)$ SYM. They look (with our accuracy) as, see \eqref{(A.3)},
\bbq
e^{(Q)}_i= - m\,,\,\,i=1...\no\,,\quad e^{\pm}\approx e^{\pm}_{c}\pm 2\langle\Lambda^{SU(N_{c}-\no)}_{{\cal N}=2\,\,SYM}\rangle,\quad \langle\Lambda^{SU(N_{c}-\no)}_{{\cal N}=2\,\,SYM}\rangle\approx  m_3\Bigl (\frac{\lm}{m_3}\Bigr )^{\frac{2N_c-N_F}{2(N_c-\no)}},
\eeq
\bq
e^{\pm}_{c}=\frac{1}{2}(e^{+}+e^{-})\approx m_3\Biggl [\,\frac{\no}{N_c} +\frac{\nt-\no}{N_c-\no}\frac{\langle\Lambda^{SU(N_{c}-\no)}_{{\cal N}=2\,\,SYM}\rangle^2}{m^2_3} \,\Biggr ], \quad m_3=\frac{N_c}{N_c-\no} m\,,\label{(B.9)}
\eq
\bbq
e^{(M)}_k\approx \Biggl [\,e^{\pm}_{c}-\frac{\nt-\no}{N_c-\no-1}\frac{\langle\Lambda^{SU(N_{c}-\no)}_{{\cal N}=2\,\,SYM}\rangle^2}{m_3}\,\Biggr ]+ 2\cos (\frac{\pi k}{N_c-\no})\langle\Lambda^{SU(N_{c}-\no)}
_{{\cal N}=2\,\,SYM}\rangle,\quad \frac{1}{2}\sum_{n=1}^{2 N_c}(-e_n)=0\,.
\eeq
It should be emphasized that instead of $e^{\pm}_{c}=0$ in $\mathbf{U(N_c)}$ theory, see \eqref{(A.15)}, it looks in $\mathbf{SU(N_c)}$ theory as, see \eqref{(A.21)},\eqref{(B.9)},
\bbq
e^{\pm}_{c}\approx \Bigl [\frac{\langle a_0\rangle}{m}\Bigr ]\Bigl [\,m\,\Bigr ]\,\ra\,\Biggl [\,\frac{\langle a_0\rangle}{m}\,\,\ra\,\,\, \approx\Biggl (\,\frac{\no}{N_c}+\frac{\nt-\no}{N_c}\Bigl (\frac{\lm}{m_3}\Bigr )^{\frac{2N_c-N_F}{N_c-\no}}\,\Biggr )\,\Biggr ]\Biggl [m\,\ra\, {m^\prime}
\approx m_3\Biggl (\,1+\frac{\nt-\no}{N_c-\no}
\eeq
\bq
\cdot\Bigl (\frac{\lm}{m_3}\Bigr )^{\frac{\bb}{N_c-\no}}\Biggr )\,\Biggr ]\approx m_3\Biggl [\frac{\no}{N_c}+\frac{\nt-\no}{N_c-\no}\Bigl (\frac{\lm}{m_3}\Bigr )^{\frac{2N_c-N_F}{N_c-\no}} \Biggr ]\approx m_3\Biggl [\,\frac{\no}{N_c}+ \frac{\nt-\no}{N_c-\no}\frac{\langle\Lambda^{SU(N_{c}-\no)}_{{\cal N}=2\,\,SYM}\rangle^2}{m^2_3} \,\Biggr ]\,.\label{(B.10)}
\eq

{\bf 2) br2 vacua}. There are at $\mx\ra 0$ $\,\no$ equal double roots $e^{(Q)}_i$ of massless original electric quarks $Q_i$, then $\nd-\no-1$ unequal double roots $e^{(M)}_k$ of massless pure magnetic monopoles $M_k$ and two single roots $e^{\pm}$ from ${\cal N}=2\,\, SU(\nd-\no)$ SYM, and $2N_c-N_F$ unequal double roots of massless dyons $D_j$. They all look (with our accuracy) as, see \eqref{(A.24)},\eqref{(A.25)},\eqref{(A.30)},
\bbq
e^{(Q)}_i= - m\,,\,\,i=1...\no\,,\quad e^{\pm}\approx e^{\pm}_{c}\pm 2\langle\Lambda^{SU(\nd-\no)}_{{\cal N}=2\,\,SYM}\rangle,\quad \langle\Lambda^{SU(\nd-\no)}_{{\cal N}=2\,\,SYM}\rangle^2\approx  m^2_1\Bigl (\frac{m_1}{\lm}\Bigr )^{\frac{2N_c-N_F}{\nt-N_c}},
\eeq
\bbq
e^{\pm}_{c}=\frac{1}{2}(e^{+}+e^{-})\approx\Biggl [\,\frac{\langle a_0\rangle}{m}\,\ra\,\, \approx \Biggl (\,\frac{\nt}{N_c}+\frac{\no-\nt}{N_c}\Bigl (\frac{m_1}{\lm}\Bigr )^{\frac{2N_c-N_F}{\nt-N_c}}\,\Biggr )\,\Biggr ]\Biggl [\,m\,\ra\,\,{m^\prime}\approx m_1\Biggl (1+
\eeq
\bq
+\frac{\no-\nt}{N_c-\nt}\Bigl (\frac{m_1}{\lm}\Bigr )^{\frac{\bb}{\nt-N_c}}\Biggr )\,\Biggr ]\approx m_1\Biggl [\,\frac{\nt}{N_c}+\frac{\nt-\no}{\nt-N_c}
\frac{\langle\Lambda^{SU(\nd-\no)}_{{\cal N}=2\,\,SYM}\rangle^2}{m^2_1} \,\Biggr ], \quad m_1=\frac{N_c}{N_c-\nt} m\,,\label{(B.11)}
\eq
\bbq
e^{(M)}_k\approx \Biggl [\,e^{\pm}_{c}-\frac{\nt-\no}{\nd-\no-1}\frac{\langle\Lambda^{SU(\nd-\no)}_{{\cal N}=2\,\,SYM}\rangle^2}{m_1}\,\Biggr ]+ 2\cos (\frac{\pi k}{\nd-\no})\langle\Lambda^{SU(\nd-\no)}_{{\cal N}=2\,\,SYM}\rangle,\quad k=1...(\nd-\no-1)\,,
\eeq
\bq
e^{(D)}_j\approx \omega^{j-1}\lm+\frac{N_F}{2N_c-N_F}\,m\,,\quad j=1...2 N_c-N_F\,,
\quad \frac{1}{2}\sum_{n=1}^{2 N_c}(-e_n)=0\,.\label{(B.12)}
\eq

Now, with \eqref{(B.9)}-\eqref{(B.12)}, we can perform checks of possible applicability of \eqref{(2.2.10)} not only to softly broken $\mathbf{U(N_c)}$ $\,{\cal N}=2$ SQCD, but also directly to softly broken $\mathbf{SU(N_c)}$ $\,{\cal N}=2$ SQCD at appropriately small $\mx$ (clearly, this applicability is far not evident beforehand, see the footnote \ref{(f7)}).

{\bf a)} From \eqref{(B.9)} for the quark condensates in br1 vacua of section 4.1
\bq
\langle\Qo\rangle^{SU(N_c)}_{\no}= - \mx\sqrt{(e^{(Q)}_k-e^+)(e^{(Q)}_k-e^-)}\approx\mx m_3\Biggl [\,1-\frac{2N_c-N_F}{N_c-\no}
\frac{\langle\Lambda^{SU(N_{c}-\no)}_{{\cal N}=2\,\,SYM}\rangle^2}{m^2_3}\, \Biggr ]\,,\label{(B.13)}
\eq
this agrees with the {\it independent} calculation in \eqref{(A.5)}.\\

{\bf b)} From \eqref{(B.11)} for the quark condensates in br2 vacua of section 2.1
\bq
\langle\Qo\rangle^{SU(N_c)}_{\no}= - \mx\sqrt{(e^{(Q)}_k-e^+)(e^{(Q)}_k-e^-)}\approx\mx m_1\Biggl [\,1+\frac
{2N_c-N_F}{\nt-N_c}\frac{\langle\Lambda^{SU(\nd-\no)}_{{\cal N}=2\,\,SYM}\rangle^2}{m^2_1}\,\Biggr ]\,,\label{(B.14)}
\eq
this agrees with the {\it independent} calculation in \eqref{(A.24)}.\\

{\bf c)} From \eqref{(B.11)},\eqref{(B.12)} for the dyon condensates in br2 vacua of section 2.1
\bbq
\langle{\ov D}_j D_j\rangle^{SU(N_c)}=\langle{\ov D}_j\rangle^{SU(N_c)}\langle D_j\rangle^{SU(N_c)}= - \mx\sqrt{(e^{(D)}_j-e^+)(e^{(D)}_j-e^-)}\approx
\eeq
\bq
\approx\mx\Biggl [-\omega^{j-1}\lm+\frac{\nt-\no}{2N_c-N_F}\,m_1\Biggl (1+\frac{\bb}{\nt-N_c}\frac{\langle\Lambda^{SU(\nd-\no)}_{{\cal N}=2\,\,SYM}\rangle^2}{m^2_1}\Biggl )\,\Biggr ]\,,\quad j=1...2 N_c-N_F\,,\label{(B.15)}
\eq
this agrees with the {\it independent} calculation in \eqref{(B.8)}.\\

\end{document}